\documentclass[useAMS,usenatbib]{mn2e}
\usepackage{txfonts,graphicx}
\usepackage{longtable,url}

\newcommand{\Ti}[5]{\mbox{$#1\,^#2{\rm #3}^{{\rm #4}}_{\rm #5}$}}
\newcommand\ion[2]{#1$\;${\scshape{#2}}} 
\newcommand{\opd}{\log \tau_{\rm 5000}}    
\newcommand{\SH}{S\!_{\rm H}}             
\newcommand{\Se}{S\!_{\rm e}}             
\newcommand{\Te}{T_{\rm e}}               
\newcommand{\mA}{{\rm m\AA}}              
\newcommand{\Elow}{E_{\rm low}}           
\newcommand{\EW}{W_{\lambda}}
\newcommand{\loge}{\log\varepsilon}       

\newcommand{\Teff}{\ensuremath{T_{\mathrm{eff}}}}     
\newcommand{\kms}{km s$^{-1}$}
\newcommand{\Vmic}{\xi_{\rm{t}}}          
\newcommand{\Vmac}{\xi_{\rm{RT}}}         

\newcommand{\loggfestar}{\log (gf\varepsilon)_{\ast}}
\newcommand{\loggfesun}{\log (gf\varepsilon)_{\odot}}

\newcommand{\logeFesun}[1]{\log\varepsilon_{\rm Fe, \odot}}
\newcommand{\logeFe}[1]{\log\varepsilon_{\rm Fe, \star}}
\newcommand{\logemean}[1]{\log\varepsilon_{\rm mean}}

\title[Ionization balance of Ti in late-type stars]{Ionization balance of
Ti in the photospheres of the Sun and four late-type stars}
\author[M. Bergemann]{Maria Bergemann\thanks{E-mail:
mbergema@mpa-garching.mpg.de}\\
Max-Planck Institute for Astrophysics, Karl-Schwarzschild Str. 1, 85741,
Garching, Germany \\}

\begin{document}

\date{Accepted Date. Received Date; in original Date}

\pagerange{\pageref{firstpage}--\pageref{lastpage}} \pubyear{2010}

\maketitle

\label{firstpage}

\begin{abstract}
In this paper we investigate statistical equilibrium of Ti in the atmospheres
of late-type stars. The Ti I/Ti II level populations are computed with available
experimental atomic data, except for photoionization and collision induced
transition rates, for which we have to rely on theoretical approximations. For
the Sun, the NLTE line formation with adjusted H I inelastic collision rates and
MAFAGS-OS model atmosphere solve the long-standing discrepancy between Ti I and
Ti II lines. The NLTE abundances determined from both ionization stages agree
within $0.01$ dex with each other and with the Ti abundance in C I meteorites.
The Ti NLTE model does not perform similarly well for the metal-poor stars,
overestimating NLTE effects in the atmospheres of dwarfs, but underestimating
overionization for giants. Investigating different sources of errors, we find
that only [Ti/Fe] ratios based on Ti II and Fe II lines can be safely used in
studies of Galactic chemical evolution. To avoid spurious abundance trends with
metallicity and dwarf/giant discrepancies, it is strongly recommended to
disregard Ti I lines in abundance analyses, as well as in determination of
surface gravities.

\end{abstract}

%
%
\section{Introduction}{\label{sec:intro}}
Most of abundance analyses in metal-poor stars focus on few key elements, among
others, Ti. This has several reasons. The main nucleosynthesis site of Ti has
not yet been unambiguously identified and all models of Galactic chemical
evolution completely fail to describe observational trend of [Ti/Fe] with
metallicity \citep{1995AJ....109.2757M}, which increases with decreasing
[Fe/H]. In spectra of late-type stars, Ti is represented by a large number
of spectral lines in both neutral and ionized stages. Therefore, Ti can be
used to verify basic stellar parameters ($\Teff, \log g$) and microturbulence,
which are determined, for example, by means of photometric methods or from the
excitation and ionization balance of Fe\footnote{In this context the notation of
excitation and ionization equilibrium refers to the equality of abundances
determined from all detected spectral lines of an element, which is represented
by several ionization stages in a stellar atmosphere}.

Traditionally, Ti abundances in stellar atmospheres are determined assuming that
distributions of atoms among excitation states and ionization stages derive from
the formulas of Saha-Boltzmann for the local values of electron temperature and
density in models of atmospheres, so-called local thermodynamic equilibrium
(LTE). Important evidence against this assumption comes, in return, from LTE
analyses of high-resolution spectra. It is known that low-excitation Ti I lines
deliver systematically lower abundances compared to the high-excitation lines
and lines of Ti II for different stellar parameters. LTE studies of the Sun with
1D static \citep{1987A&A...180..229B} and 3D hydrodynamical model atmospheres
\citep{2009ARA&A..47..481A} demonstrate that these discrepancies are larger
than $0.1$ dex, which is unacceptable given the precisely known fundamental
parameters of the Sun. The same problem was reported for other nearby stars with
well-known parameters, such as Pollux \citep{1980A&A....92...70R}.

Analyses of metal-poor stars also point to pronounced NLTE effects on Ti I in
their atmospheres. Systematically higher [Ti/Fe] abundance ratios in metal-poor
dwarfs compared to giants with similar metallicites were reported by
\citet{2009A&A...501..519B}. Positive offsets between abundances based on Ti II
and Ti I lines were reported for horizontal branch stars
\citep{1995AJ....110.2319C} and for giants \citep{1983ApJ...265L..93B,
1991A&A...241..501G, 2002ApJS..139..219J, 2010arXiv1008.3721T}.
\citet{2008ApJ...681.1524L} found a trend of [Ti/Fe] abundances ratios with
stellar $\Teff$, which was supported by the data of \citet{2006AJ....132...85P},
\citet{2004A&A...416.1117C}, and \citet{2004ApJ...612.1107C}.

Theoretical studies of NLTE effects in Ti are very sparse.
\citet*{1999ApJ...512..377H} investigated statistical equilibrium of Ti in
M-type dwarfs and giants. They found small effect of Ti I overionization on the
number density of TiO molecules, the main opacity agent in the atmospheres of M
stars. However, they demonstrated that the NLTE effects on the Ti I line
formation are important and grow with increasing model $\Teff$ for dwarfs and
decreasing $\Teff$ for giants. \citet*{2002A&A...394.1093M} and
\citet*{2009ASPC..405..275S} used NLTE populations of Ti to model polarization
in the Ti I lines of multiplet $42$. Their goal was to study weak solar magnetic
fields. None of the investigations focused on the NLTE effects on Ti abundances
in stellar atmospheres.

We present first results of a study aimed at the NLTE modelling of Ti line
formation in the atmospheres of late-type stars. In Sect. \ref{sec:methods}, we
give a detailed description of the methods developed for NLTE and spectrum
synthesis calculations. Statistical equilibrium of Ti for a restricted range of
stellar parameters is discussed in Sect. \ref{sec:nlte_effect}. Ti abundances
for the Sun and four metal-poor stars are given in Sect. \ref{sec:results}. The
results are summarized in Sect. \ref{sec:summary}.
%
%
\section{Methods}{\label{sec:methods}}
In order to carry out the abundance analysis, we use a unique program
package developed at the University Observatory Munich (the group of T.
Gehren). The package consists of three codes, which have been developed
consistently and are adapted for spectroscopic analysis of A-,F- and G-type
stars. MAFAGS is a 1D LTE model atmosphere code. Detail is a code for solving
multi-level non-LTE radiative transfer problems with a \emph{given} static 1D
model atmosphere. SIU is an interactive non-LTE spectrum synthesis code that
makes use of MAFAGS model atmospheres and level populations from DETAIL. Below,
we give some details for each of the programs. The general procedure of
abundance determination is the same as in our previous work on Mn, Co, and Cr
\citep*{2008A&A...492..823B,2010MNRAS.401.1334B,2010arXiv1006.0243B}.

\subsection{Model atmospheres}
\begin{figure}
\resizebox{\columnwidth}{!}{
{\includegraphics[scale=1]{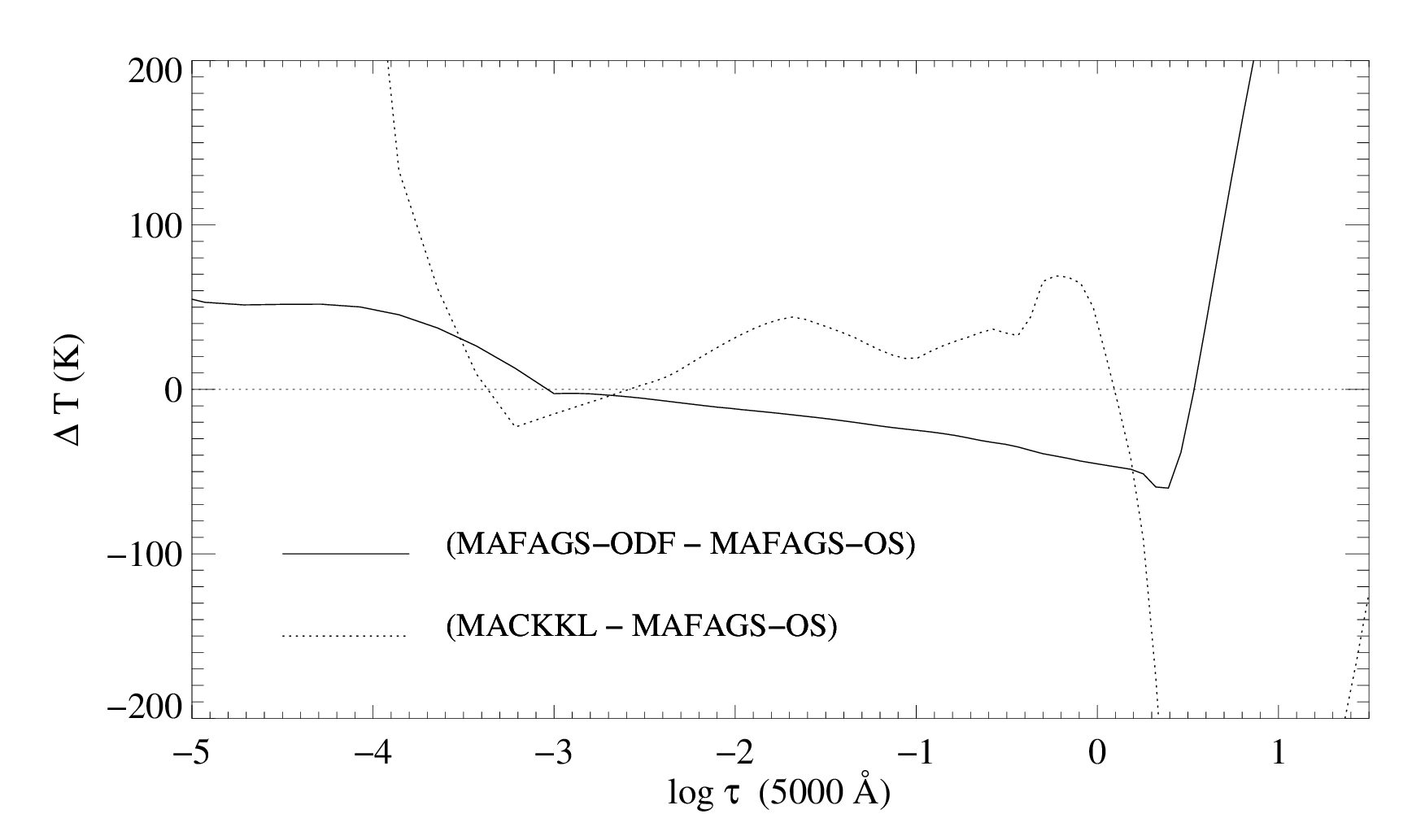}}}
\resizebox{\columnwidth}{!}{
{\includegraphics[scale=1]{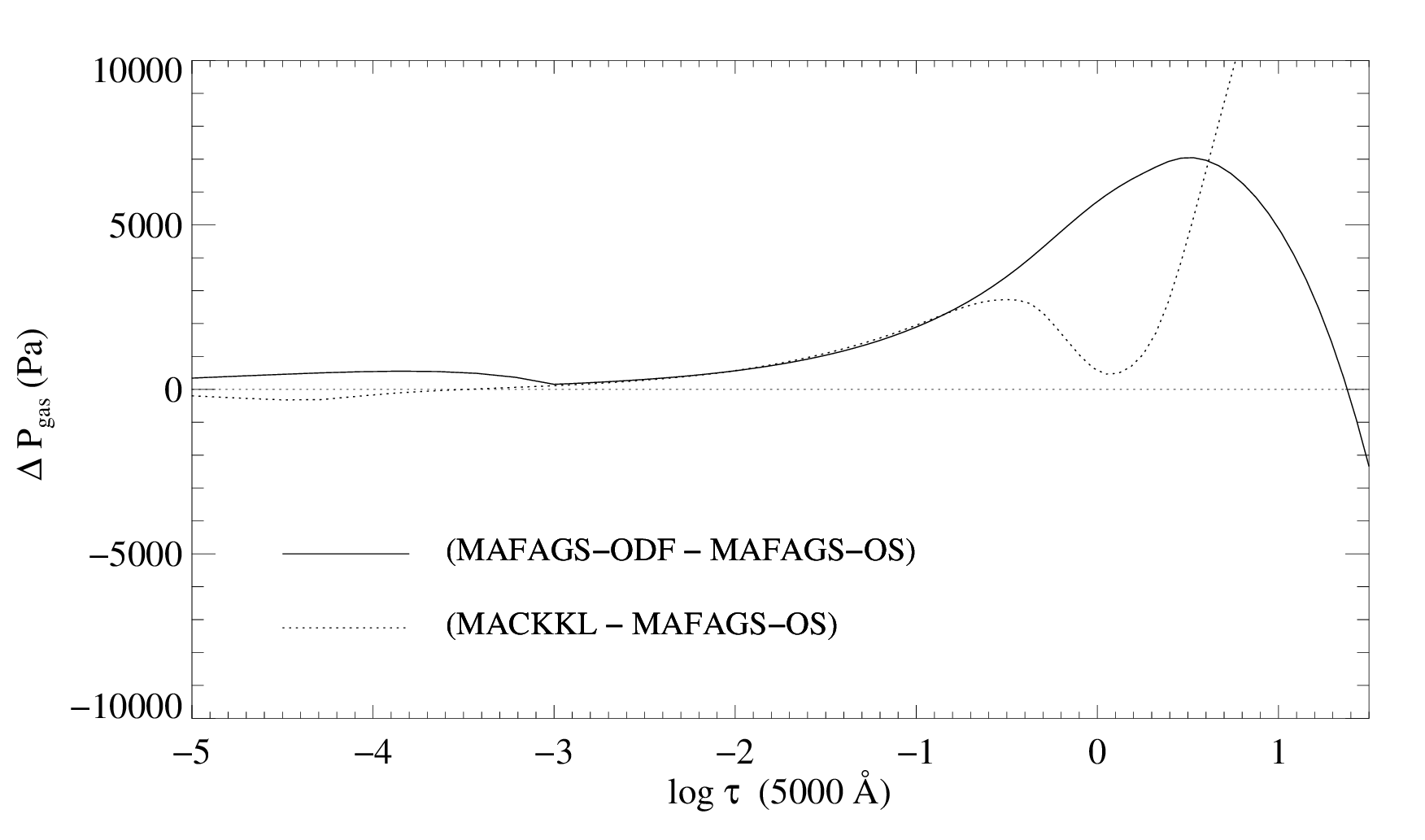}}}
\caption[]{Difference in temperature and gas pressure stratifications of the
solar theoretical MAFAGS-OS, MAFAGS-ODF, and semi-emprirical MACKKL models.}
\label{temp}
\end{figure}

The static plane-parallel LTE model atmospheres were kindly generated by
F. Grupp with the MAFAGS code. In the most recent version, MAFAGS-OS, line
blanketing is treated with an opacity sampling method, with line absorption
sampled at $\sim 86\,000$ wavelength points. All opacity sources relevant to
modelling A0-K0 stars including line absorption due to various diatomic
molecules are taken into account. Full description of these models, including
the reference set of solar element abundances, as well as selection criteria of
the wavelength grid and lines, is given in \citet{2004A&A...420..289G}. In
particular, $\log \epsilon_{\rm Fe} = 7.50$ was adopted, which is consistent
with the NLTE solar abundance of Fe determined by \citet{2001A&A...380..645G}.
Convection is included in the formulation of \citet*{1991ApJ...370..295C}, which
differs from the mixing-length theory of \citet*{1958ZA.....46..108B} in that it
assumes a full spectrum of turbulent eddies. Compressible turbulence is
accounted for using the mixing length, $\Lambda = \alpha_{\rm cm}/\rm{H_p}$,
where $H_p$ is the pressure scale height and the free parameter $\alpha_{\rm
cm}$ is set to $0.82$. The latter value was derived by
\citet{1998A&A...332..127B} from stellar evolution calculations for the Sun and
from the requirement that the MAFAGS models fit the observed Balmer line
profiles.

We also performed abundance calculations with the model atmospheres computed
with older version of the MAFAGS code \citep{1997A&A...323..909F}; these models
were used to determine spectroscopic stellar parameters for the selected stars
(Sect. \ref{sec:stars}). This version is based on opacity distribution functions
(ODF) from \citet*{1992RMxAA..23...45K}. Convection was taken into account with
the mixing-length theory \citep{1958ZA.....46..108B} and the mixing length was
set to $0.5$ pressure scale heights, the latter value was also calibrated on the
solar Balmer lines by \citet*{1993A&A...271..451F}. A detailed comparison of the
solar MAFAGS-ODF and MAFAGS-OS models is given in \citet{2004A&A...426..309G}.

MAGAFS models do not include chromospheres that is the case with all other
theoretical 1D models so far. Clearly, this is a crude approximation for the
Sun. The existence of chromospheres has also been demonstrated for metal-poor
dwarfs \citep*{1998MNRAS.297..388S} and giants \citep*{1990ApJ...353..623D}. To
discuss this deficiency of the models, we will also present some results for the
semi-empirical solar model atmosphere with chromosphere \citep[][hereafter
MACKKL]{1986ApJ...306..284M}. The MACCKL model was interpolated to $80$ depth
points to increase numerical accuracy in line formation calculations
\citep[][PhD thesis]{reetz}. Differences between temperature and pressure
stratifications of the solar MACCKL, MAFAGS-ODF, and MAFAGS-OS models as a
function of continuum optical depth at $5000$ \AA, $\opd$, are shown in
Fig.\ref{temp}.

We emphasize that the semi-empirical model atmosphere is used here only to
demonstrate the shortcomings of the theoretical atmospheres, which are still
the main tool to model any star other than the Sun. Thus, the solar abundance
analysis, as well as the differential analysis of the metal-poor stars, are
performed only with the MAFAGS models.

\subsection{Statistical equilibrium and line formation codes}
The NLTE level populations for Ti were computed with an updated version of the
DETAIL code \citep*{Butler85}. In calculations of statistical
equilibrium, each Ti line was treated with a Gaussian profile with $9$ frequency
points. The code solves a restricted NLTE problem, i.e. coupled statistical
equilibrium and radiative transfer equations are solved for a fixed input model
atmosphere. This may not be a bad approximation for Ti in the range of stellar
parameters we are interested in. Ti I, which is, in fact, affected by NLTE, is
not an important opacity source in the atmospheres of FG stars and does not
contribute significantly to the free electron pool due to the low element
abundance. Ti II is the dominant ionization stage and its number densities are
well described in LTE (see Sect. \ref{sec:nlte_effect}).
\citet{1999ApJ...512..377H} computed full NLTE line-blanketed model atmospheres
for M dwarfs and giants with solar metallicity including Ti as one of the NLTE
species. They showed that the NLTE effects in Ti influence the atmospheres of
cool stars only in the upper layers by changing the concentration of TiO$^+$
molecules, and the effect on the number density of TiO molecules is negligible.
Also, the effect of Ti on the atmospheric structure, even if present, is much
smaller than that of Fe, which is an important source of the bound-free opacity
in the UV and has, by far, the largest number of lines all over the spectrum of
a typical F-type star \citep[][Fig. 1]{2009A&A...503..177G}.

Still one has to keep in mind that, in general, the assumption of LTE in
modelling stellar atmospheres is not realistic. As demonstrated by
\citet{2005ApJ...618..926S}, the \emph{integral} effect of NLTE in Fe-group
elements (Ti, Mn, Fe, Co, Ni) on the structure of the solar model atmosphere is
significant. The differences in the temperature structure of their NLTE PHOENIX
model with respect to LTE models are as large as $\pm 200$ K at different
depths. These differences are comparable with the temperature fluctuations
in the line forming regions in the 3D hydrodynamical solar model
\citep{2005ARA&A..43..481A}. Since NLTE effects in the presence of convective
inhomogeneities are amplified \citep[Fe:][]{2001ApJ...550..970S}, one can expect
that self-consistent 3D NLTE hydrodynamical model atmospheres will have
radically different structure compared to existing 1D LTE, 1D NLTE, or 3D LTE
models. On the other side, such physically realistic models do not exist yet,
and there is little chance they will appear in the near future.

Emergent flux spectra were computed with an updated version of the code
SIU \citep{reetz}, using MAFAGS model atmospheres and departure coefficients of
Ti I/Ti II levels from the DETAIL code. Ti lines were computed with full
Voigt profiles taking into account various broadening processes (see Sect.
\ref{sec:sun}). Ti abundances were determined by visually fitting the LTE and
NLTE synthetic line profiles to the observed flux spectra including LTE
modelling of blending features. This method is more reliable than $\chi$-square
fitting or equivalent width (EW) measurements for the solar-type stars, because
a majority of Ti I and Ti II lines are blended and/or display an asymmetry in
the core and red wing, which is a typical signature of atmospheric temperature
and velocity inhomogeneities \citep[e.g.][]{2000A&A...359..729A} not taken into
account by static plane-parallel model atmospheres.

%
%
\subsection{Ti model atom}
\begin{figure}
\resizebox{\columnwidth}{!}{\rotatebox{90}
{\includegraphics[scale=1]{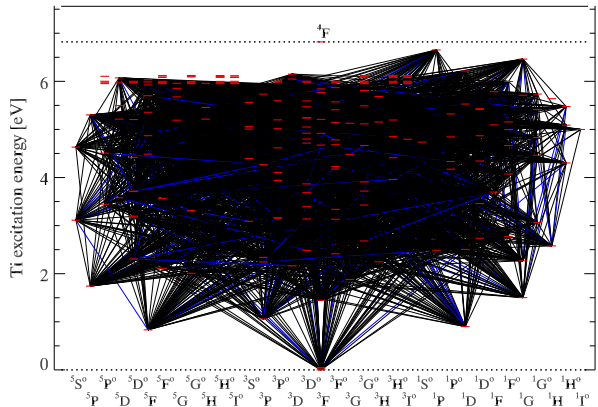}}}
\caption[]{Ti grotrian diagram. Strong transitions with $\log gf > -1$ are shown
with blue trace.}
\label{ti_grotrian}
\end{figure}

{\label{sec:modelatom}}
A model of the Ti atom was constructed with energy levels, wavelengths of
transitions, and transition probabilities from the Kurucz'
database\footnote{http://kurucz.harvard.edu/atoms.html}, which includes all 
laboratory data. The number of energy levels is $216$ for Ti I and $77$ for
Ti II, with uppermost excited levels located at $0.17$ eV and $1.1$ eV below the
respective ionization limits, $6.82$ eV and $13.58$ eV. The model is closed by
the Ti III ground state. The total number of radiatively-allowed transitions is
$4671$ ($3435$ Ti I and $1236$ Ti II). Fine structure was neglected in the
statistical equilibrium calculations, except for the ground state of Ti
I \Ti{a}{3}{F}{}{} (configuration $1s^2 2s^2 2p^6 3s^2 3p^6 3d^2 4s^2$).
Excitation energy of each LS term was computed as a weighted mean of statistical
weights and excitation energies of fine structure sub-levels. Also, transitions
between fine structure levels were combined, with the total transition
probability of each term being the weighted mean of $\log gf$'s of their fine
structure components. A Grotrian diagram for Ti I is shown in Fig.
\ref{ti_grotrian}.

The electron collision cross-sections from states with allowed bound-bound and 
bound-free transitions were computed with the formulas of
\citet*{1962ApJ...136..906V} and \citet*{1962amp..conf..375S}, respectively.
Electron collision cross-sections from states connected only by forbidden
transitions were treated with the formula of \citet*{1973asqu.book.....A}.
According to \citet{1996ASPC..108..140M} collision rates obtained with these
formulas have only an order of magnitude accuracy.

Threshold photoionization cross-sections for the levels \Ti{y}{3}{P}{\circ}{}
($3.93$ eV), \Ti{z}{3}{H}{\circ}{} ($3.96$ eV), and \Ti{v}{3}{F}{\circ}{}($4.2$
eV) were taken from \citet{2009ChJCP..22..615Y} and a hydrogenic approximation
was assumed for their frequency variation. The cross-sections were measured by
resonance ionization mass spectrometry and for the \Ti{v}{3}{F}{\circ}{} levels
they are consistent with Hartree-Fock calculations with relativistic corrections
\citep{1993AcSpe..48.1139S}. For the other levels, even quantum-mechanical
calculations are not available, thus the photoionization cross-sections were
derived from the hydrogenic approximation \citep{1978stat.book.....M}. The
experimental threshold cross-sections from \citet{2009ChJCP..22..615Y} are lower
than the hydrogenic cross-sections with effective principal quantum numbers by
one order of magnitude. For example, the measured values for the
\Ti{v}{3}{F}{\circ}{} fine structure levels are $0.6 - 1.2$ Mb, whereas the
hydrogenic cross-section is roughly $18$ Mb.

In many aspects including the atomic structure, our model atom is very similar
to that of \citet{1999ApJ...512..377H} with the main
difference\footnote{\citet{1999ApJ...512..377H} also used different
photoionization cross-section for the Ti I ground state.} that inelastic
collisions with \ion{H}{i} atoms are included in our model. The corresponding
bound-bound and bound-free rates were computed from the formulas of
\citet{1969ZPhy..225..483D} in the version of \citet*{1984A&A...130..319S}.

Since the atomic data are of a low accuracy, we perform calculations of
ionization equilibria for several Ti NLTE model atoms, which differ in the
efficiency of $e^-$ and H I collisions (Sect. \ref{sec:solarab}). Thus, for the
total $e^-$ collision rates we use the scaling factors $\Se = 0.01, 1, 10$, and
for the H I collision rates $\SH = 0.05, 3$. The final adopted values are
$\SH = 3$ and $\Se = 1$, which give the smallest abundance scatter in the solar
abundance analysis (see Sect. \ref{sec:sun}).
\subsection{Line parameters}
A careful selection of Ti lines is essential, because ionization equilibrium of
Ti in the solar photosphere is used to calibrate the efficiency of inelastic
collisions with H I.

Ti I and Ti II lines were selected by the critical inspection of the KPNO
atlas of solar fluxes \citep{1984sfat.book.....K} and of the disk-center
intensity spectrum \citep*{1972kpsa.book.....B}. For Ti I, we rejected all lines
that showed obvious blends in the profile, i.e. the contribution of the
theoretically computed blend to the total $\EW$ of a line was more than $5$
\%. For Ti II, this threshold was somewhat higher, because of a very small
number of sufficiently strong Ti II lines with reliable transition
probabilities.

\begin{figure}
\resizebox{\columnwidth}{!}{{\includegraphics[scale=1]{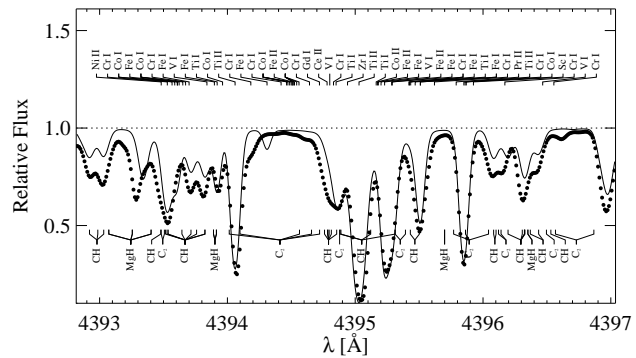}}}
\caption[]{A spectral window in the KPNO atlas of solar fluxes (filled circles)
compared to a synthetic spectrum (black trace) computed with the MAFAGS-OS
model atmosphere. All lines populating this spectral region are indicated.}
\label{cont}
\end{figure}

Some regions in the observed solar spectrum, such as shown in Fig. \ref{cont},
suffer from excessive line blanketing and are not well reproduced by our model
atmospheres and line lists. All Ti lines located in such regions were rejected
from the solar analysis. For the other regions, continuum is poorly defined.
Following \citet{2001A&A...366..981G}, we renormalized the continuum in
these regions if the depression extended over few tenths of \AA. The
re-adjustment of the continuum was never larger than $\sim 0.5\%$, which is
within the continuum placement uncertainty.

Tables \ref{ti_i} and \ref{ti_ii} contain important information for each line
including our evaluation of the line 'quality' in the solar spectrum, i.e.
whether the line is blended or has an asymmetric profile. Term designations,
wavelengths and excitation potentials of lower levels were taken from the NIST
database, whenever available \citep*[otherwise from][]{1988atps.book.....F}. The
equivalent widths given were computed from the best-fitting synthetic profiles,
excluding contribution of blends. Equivalent widths are not used in the
abundance calculations and serve only as a demonstration of line strength. For
example, lines with $\EW < 50~ \mA$ are not detectable in our spectra of
metal-poor stars.
%
%
\begin{table*}
\begin{minipage}{\linewidth}
\renewcommand{\footnoterule}{}
\tabcolsep1.2mm
\caption{Parameters of the Ti I lines used for the spectrum synthesis. The
abundances in columns $10-13$ are given for $gf$-values from Blackwell-Whitehead
et al. (2006), if the latter are specified in column $7$ (otherwise for the
$gf$-values measured by the Oxford group, which are given in column $8$, and
exact references to the latter are given in column $9$).}
\label{ti_i}
\begin{center}
\begin{tabular}{lll|rcrc|ccccc|llll}
\noalign{\smallskip}\hline\noalign{\smallskip} ~~~$\lambda$ & Mult. & Rem.$^a$ &
Lower & Upper & $\Elow$ & $\log gf$ & $\log gf^b$ & Source$^b$ & $W_\lambda$ & 
\multicolumn{2}{c}{MAFAGS-ODF}  & 
\multicolumn{2}{c}{MAFAGS-OS}  \\
  & &  & level & level & [eV] & & &  & [\mA] & 
$\log \epsilon_{\rm LTE}$ & $\log \epsilon_{\rm NLTE}$ &
$\log \epsilon_{\rm LTE}$ & $\log \epsilon_{\rm NLTE}$ \\
\noalign{\smallskip}\hline\noalign{\smallskip}
4281.37 & 17 & b & \Ti{a}{5}{F}{}{1} & \Ti{x}{5}{D}{\circ}{2} & 0.81 & & -1.303
& 2 & 33 & 4.78 & 4.81 & 4.88 & 4.91 \\
4512.73 & 16 & a & \Ti{a}{5}{F}{}{4} & \Ti{y}{5}{F}{\circ}{5} & 0.84 & &
-0.424 & 2 & 67 & 4.81 & 4.84 & 4.92 & 4.95 \\
4518.03	& 16 & b & \Ti{a}{5}{F}{}{3} & \Ti{y}{5}{F}{\circ}{4} & 0.83 & & -0.269
& 2 & 74 & 4.82 & 4.85 & 4.92 & 4.95 \\
4533.24 & 16 & b & \Ti{a}{5}{F}{}{5} & \Ti{y}{5}{F}{\circ}{5} & 0.85 & &  0.532
& 2 & 113 & 4.8 & 4.82 & 4.89 & 4.9  \\
4534.78 & 16 & b & \Ti{a}{5}{F}{}{4} & \Ti{y}{5}{F}{\circ}{4} & 0.84 & &  0.336
& 2 & 101 & 4.8 & 4.82 & 4.89 & 4.9  \\
4548.77 & 16 & s & \Ti{a}{5}{F}{}{3} & \Ti{y}{5}{F}{\circ}{2} & 0.83 & & -0.298
& 2 & 73 & 4.82 & 4.85 & 4.92 & 4.95 \\
4555.49 & 16 & a & \Ti{a}{5}{F}{}{5} & \Ti{y}{5}{F}{\circ}{4} & 0.85 & & -0.432
& 2 & 67 & 4.82 & 4.85 & 4.92 & 4.95 \\
4562.63 & 5 & b & \Ti{a}{3}{F}{}{3} & \Ti{z}{1}{D}{\circ}{2} & 0.02 & & -2.600 &
1 & 13 & 4.83 & 4.87 & 4.92 & 4.96 \\
4617.28	& 57 & a & \Ti{a}{5}{P}{}{3} & \Ti{w}{5}{D}{\circ}{4} & 1.75 & &  0.445
& 4 & 63 & 4.75 & 4.76 & 4.86 & 4.87 \\
4639.93 & 57 & b & \Ti{a}{5}{P}{}{1} & \Ti{w}{5}{D}{\circ}{1} & 1.73 & & -0.136
& 4 & 38 & 4.76 & 4.79 & 4.88 & 4.9 \\
4681.92 & 4 & b & \Ti{a}{3}{F}{}{4} & \Ti{z}{3}{G}{\circ}{5} & 0.05 & -1.03 &
-1.01 & 1 & 75 & 4.8 & 4.84 & 4.9 & 4.94 \\
4758.12 & 74 & b & \Ti{a}{3}{H}{}{5} & \Ti{x}{3}{H}{\circ}{5} & 2.25 & & 0.481
& 4 & 46 & 4.77 & 4.79 & 4.86 & 4.89 \\
4759.27 & 74 & b & \Ti{a}{3}{H}{}{6} & \Ti{x}{3}{H}{\circ}{6} & 2.26 & & 0.570 &
4 & 49 & 4.76 & 4.78 & 4.85 & 4.87 \\
4820.41 & 55 & b & \Ti{a}{1}{G}{}{4} & \Ti{y}{1}{F}{\circ}{3} & 1.50 & & -0.385
& 4 & 44 & 4.8 & 4.84 & 4.88 & 4.9 \\ 
4840.88 & 23 & b & \Ti{a}{1}{D}{}{2} & \Ti{y}{1}{D}{\circ}{2} & 0.90 & & -0.453
& 2 & 64 & 4.82 & 4.81 & 4.92 & 4.93 \\
4964.7  & 66 & b & \Ti{z}{5}{G}{\circ}{2} & \Ti{e}{5}{F}{}{2} & 1.97 & & -0.819
& 4 & 9 & 4.77 & 4.8 & 4.87 & 4.89 \\
4981.74 & 13 & s & \Ti{a}{5}{F}{}{5} & \Ti{y}{5}{G}{\circ}{6} & 0.85 & &  0.560
& 2 & 120 & 4.78 & 4.8 & 4.89 & 4.91 \\
4991.07 & 13 & b,s & \Ti{a}{5}{F}{}{4} & \Ti{y}{5}{G}{\circ}{5} & 0.84 & & 
0.436 & 2 & 112 & 4.78 & 4.8 & 4.89 & 4.91 \\
4997.1 & 3 & a & \Ti{a}{3}{F}{}{2} & \Ti{z}{3}{D}{\circ}{2} & 0.00 & -2.07 &
-2.06 & 1 & 34 & 4.8 & 4.84 & 4.9 & 4.95 \\
4999.51 & 13 & b & \Ti{a}{5}{F}{}{3} & \Ti{y}{5}{G}{\circ}{4} & 0.83 & &  0.306
& 2 & 106 & 4.82 & 4.85 & 4.93 & 4.95 \\
5009.66 & 3 & b,uc & \Ti{a}{3}{F}{}{3}  &  \Ti{z}{3}{D}{\circ}{3}  &  0.02 &
-2.23 & -2.20 & 1 & 27 & 4.84 &  4.88  &   4.94 &  4.98 \\
5016.17 & 13 & uc & \Ti{a}{5}{F}{}{5}  &  \Ti{y}{5}{G}{\circ}{5}  &  0.85 &  &
-0.518 & 2 & 65 & 4.83 & 4.86  &   4.91 &  4.95 \\
5022.87 & 13 & b & \Ti{a}{5}{F}{}{3}  &  \Ti{y}{5}{G}{\circ}{3}  &  0.83 &  &
-0.38 & 2 & 74 & 4.83 & 4.87  &   4.92 &  4.96 \\
5039.96 & 3 & b,s & \Ti{a}{3}{F}{}{3}  &  \Ti{z}{3}{D}{\circ}{2}  &  0.02 &
-1.09 & -1.07 & 3 & 77 & 4.83 & 4.86  &   4.91 &  4.94 \\
5064.65 & 3 & b & \Ti{a}{3}{F}{}{4}  &  \Ti{z}{3}{D}{\circ}{3}  &  0.05 & -0.95
& -0.94 & 1 & 84 & 4.88 & 4.91  &   4.99 &  5.01 \\
5113.45 & 48 & b & \Ti{b}{3}{F}{}{3}  &  \Ti{v}{3}{D}{\circ}{2}  &  1.44 &  &
-0.727 & 4 & 27 & 4.77 & 4.79  &   4.86 &  4.88 \\
5147.48 & 2 & b,uc & \Ti{a}{3}{F}{}{2}  &  \Ti{z}{3}{F}{\circ}{3}  &  0.00 &
-1.89 & -1.95 & 1 & 38  & 4.73 &  4.78  &   4.83 &  4.88 \\
5152.19 & 2 & b & \Ti{a}{3}{F}{}{3}  &  \Ti{z}{3}{F}{\circ}{4}  &  0.02 & -1.90
& -1.96  & 1 &  38  & 4.74 &  4.78  &   4.83 &  4.88 \\
5173.75 & 2 & b & \Ti{a}{3}{F}{}{2}  &  \Ti{z}{3}{F}{\circ}{2}  &  0.00 & -1.07
& -1.06 & 1 &  79  & 4.78 &  4.82  &   4.88 &  4.91 \\
5192.98 & 2 & b & \Ti{a}{3}{F}{}{3}  &  \Ti{z}{3}{F}{\circ}{3}  &  0.02 & -0.96
& -0.95 & 1 &  85  & 4.78 &  4.82  &   4.88 &  4.91 \\
5210.39 & 2 & b & \Ti{a}{3}{F}{}{4}  &  \Ti{z}{3}{F}{\circ}{4}  &  0.05 & -0.85
& -0.82 & 1 &  88   & 4.84 & 4.88  &   4.94 &  4.97 \\
5219.7  & 2 & b & \Ti{a}{3}{F}{}{3}  &  \Ti{z}{3}{F}{\circ}{2}  &  0.02 & -2.26
& -2.23 & 1 & 28   & 4.88 & 4.92  &   4.97 &  5.01 \\
5295.78 & 31 & b & \Ti{a}{3}{P}{}{2}  &  \Ti{x}{3}{D}{\circ}{3}  &  1.07 & &
-1.577 & 3 & 13 & 4.82 &  4.86  &   4.92 &  4.95 \\
5490.15 & 45 & b & \Ti{b}{3}{F}{}{4}  &  \Ti{x}{5}{D}{\circ}{3}  &  1.46 & &
-0.877 & 4 & 24 & 4.8  &  4.82  &   4.89 &  4.91 \\
5662.15 & 78 & b & \Ti{z}{5}{D}{\circ}{4}  &  \Ti{e}{5}{F}{}{5}  &  2.32 & &
-0.053 & 4 & 24 & 4.81 & 4.83  &   4.9  &  4.93 \\ 
5866.46 & 30 & b & \Ti{a}{3}{P}{}{2}  &  \Ti{y}{3}{D}{\circ}{3}  &  1.07 & &
-0.784 & 2 & 48 & 4.81 &  4.82  &   4.9 &  4.91 \\
5922.12 & 27 & b,uc & \Ti{a}{3}{P}{}{0}  &  \Ti{y}{3}{D}{\circ}{1}  &  1.05 & &
-1.410 & 2 & 21   & 4.84 &  4.87  &   4.93 &  4.96 \\
5965.83 & 62 & b,uc & \Ti{a}{3}{G}{}{4}  &  \Ti{z}{3}{H}{\circ}{5}  &  1.88 &
&-0.353 & 4 & 28 & 4.8 &  4.82  &   4.89 &  4.91 \\
5978.54 & 62 & a & \Ti{a}{3}{G}{}{3}  &  \Ti{z}{3}{H}{\circ}{4}  &  1.87 &
&-0.440 & 4 & 25   & 4.79 &  4.81  &   4.87 &  4.89 \\
6064.63 & 25 & b,uc & \Ti{a}{3}{P}{}{0}  &  \Ti{z}{3}{S}{\circ}{1}  &  1.05 &
&-1.888 & 3 & 9  & 4.85 &  4.88  &   4.96 &  5.00 \\
6092.8  & 61 & uc & \Ti{a}{3}{G}{}{5}  &  \Ti{w}{3}{G}{\circ}{5}  &  1.89 &
&-1.323 & 4 & 5  & 4.82 &  4.86  &   4.92 &  4.96 \\
6126.22 & 25 & a & \Ti{a}{3}{P}{}{2}  &  \Ti{z}{3}{S}{\circ}{1}  &  1.07 &
&-1.369 & 3 & 23   & 4.85 &  4.87  &   4.95 &  4.97 \\
6258.71 & 43 & a & \Ti{b}{3}{F}{}{4}  &  \Ti{y}{3}{G}{\circ}{5}  &  1.46 &
&-0.299 & 4 & 53   & 4.79 &  4.79  &   4.88 &  4.89 \\
6261.1  & 43 & b,uc & \Ti{b}{3}{F}{}{2}  &  \Ti{y}{3}{G}{\circ}{3}  &  1.43 &
&-0.423 & 4 & 49   & 4.83 &  4.84   &   4.92 &  4.93 \\
6303.76 & 43 & b,rc & \Ti{b}{3}{F}{}{3}  &  \Ti{y}{3}{G}{\circ}{3}  &  1.44 &
&-1.510 & 4 & 9    & 4.83 &  4.86  &   4.93 &  4.97 \\
6312.24 & 43 & b,rc & \Ti{b}{3}{F}{}{4}  &  \Ti{y}{3}{G}{\circ}{4}  &  1.46 &
&-1.496 & 4 & 9    & 4.81 &  4.84  &   4.9  &  4.94 \\
6554.24 & 41 & b & \Ti{b}{3}{F}{}{3}  &  \Ti{x}{3}{F}{\circ}{3}  &  1.44 &
&-1.162 & 4 & 11 & 4.83 &  4.86  &   4.91 &  4.94 \\
6556.08 & 41 & a & \Ti{b}{3}{F}{}{4}  &  \Ti{x}{3}{F}{\circ}{4}  &  1.46 &
&-1.018 & 4 & 11 & 4.81 &  4.83  &   4.89 &  4.91 \\
7357.74 & 35 & b,rc & \Ti{b}{3}{F}{}{3}  &  \Ti{y}{3}{F}{\circ}{3}  &  1.44 &
&-1.066 & 4 & 23 & 4.85 &  4.87  &   4.94 &  4.96 \\
8426.51 & 7 & b & \Ti{a}{5}{F}{}{3}  &  \Ti{z}{5}{D}{\circ}{2}  &  0.83 &
&-1.197 & 2 & 51 & 4.89 &  4.92  &   4.96 &  5    \\
8435.66	& 7 & b & \Ti{a}{5}{F}{}{4}  &  \Ti{z}{5}{D}{\circ}{3}  &  0.84 &
&-0.967 & 2 & 62 &  4.88 &  4.91  &   4.96 &  4.98 \\
8682.99	& 24 & uc & \Ti{a}{3}{P}{}{1}  &  \Ti{z}{3}{D}{\circ}{2}  &  1.05 &
-1.69 & -1.82 & 3 & 15 &  4.74 &  4.77  &   4.83 &  4.86 \\
\noalign{\smallskip}\hline\noalign{\smallskip}
\end{tabular}
\end{center}
\end{minipage}
$^a$ Line identification: a - unblended; b - blended; uc - uncertain continuum;
s - saturated; rc - renormalized continuum
$^b$ References to $gf$-values: (1) \citet[][a]{1982MNRAS.199...21B}; (2)
\citet[][b]{1982MNRAS.201..611B}; (3) \citet{1983MNRAS.204..883B}; (4)
\citet{1986MNRAS.220..289B}
\end{table*}
%
%
\begin{table*}
\begin{minipage}{\linewidth}
\renewcommand{\footnoterule}{}  
\tabcolsep1.2mm
\caption{Parameters of Ti II lines used for the spectrum synthesis and
individual abundances. The lines at $4394$, $4395$, $4443$ \AA\ were not used
in the solar abundance analysis. For some transitions
\citet{2001ApJS..132..403P} do not provide accuracies of $gf$-values.}
\label{ti_ii}
\begin{center}
\begin{tabular}{rrl|rr|ccccc|llll}
\noalign{\smallskip}\hline\noalign{\smallskip} ~~~$\lambda$ & Mult. & Rem.$^a$ &
Lower & Upper & $\Elow$ & $W_\lambda$ & $\log gf$ & Unc. & Source$^b$ & 
\multicolumn{2}{c}{MAFAGS-ODF}  & 
\multicolumn{2}{c}{MAFAGS-OS}  \\
   & & & level & level & [eV] & [\mA] &  &  &  &
$\log \epsilon_{\rm LTE}$ & $\log \epsilon_{\rm NLTE}$ &
$\log \epsilon_{\rm LTE}$ & $\log \epsilon_{\rm NLTE}$ \\ 
\noalign{\smallskip}\hline\noalign{\smallskip}
  4394.05 & 12 & b & \Ti{a}{2}{P}{}{0.5} & \Ti{z}{4}{D}{\circ}{1.5} & 1.22 &
77 & -1.78 & 0.04 & 1 &  &  &  &  \\
  4395.85 & 3 & b & \Ti{b}{4}{P}{}{2.5} & \Ti{z}{4}{D}{\circ}{3.5} & 1.24 & 71 
& -1.93 & 0.04 & 1 &  &  &  &  \\
  4443.8  & 3 & b & \Ti{a}{2}{D}{}{1.5} & \Ti{z}{2}{F}{\circ}{2.5} & 1.08 & 138
& -0.72 & 0.02 & 1 &  &  &  &  \\
  4444.56 & 6 & b & \Ti{a}{2}{G}{}{3.5} & \Ti{z}{2}{F}{\circ}{3.5} & 1.12 &  61
& -2.24 &      & 1 & 4.97 & 4.97 & 4.99 & 4.99 \\
  4468.5  & 6 & b & \Ti{a}{2}{G}{}{4.5} & \Ti{z}{2}{F}{\circ}{3.5} & 1.13 & 139
& -0.6  & 0.02 & 2 & 4.94 & 4.92 & 4.98 & 4.96 \\
  4488.32 & 26 & b,uc & \Ti{c}{2}{D}{}{2.5} & \Ti{x}{2}{F}{\circ}{3.5} & 3.12 & 
50 & -0.51 & 0.03 & 1 & 4.86 & 4.84 & 4.91 & 4.90  \\
  4493.52 & 2 & b & \Ti{a}{2}{D}{}{1.5} & \Ti{z}{4}{F}{\circ}{2.5} & 1.08 &  33
& -2.83 & 0.05 & 1 & 4.89 & 4.89 & 4.90  & 4.90  \\
  4563.77 & 11 & b & \Ti{a}{2}{P}{}{0.5}  & \Ti{z}{2}{D}{\circ}{1.5} & 1.22 &
129 & -0.69 & 0.02 & 1  & 4.98 & 4.98 & 5.01 & 5.01 \\
  4583.41 & 8 & b & \Ti{a}{4}{P}{}{1.5}  & \Ti{z}{2}{F}{\circ}{2.5} & 1.17 & 31 
& -2.92 &     & 1  & 5.03 & 5.03 & 5.04 & 5.03 \\
  4636.32 & 7 & b &  \Ti{a}{4}{P}{}{1.5}   & \Ti{z}{4}{F}{\circ}{2.5} &
1.16 & 20  &  -3.02 &     & 1  & 4.81 & 4.8  & 4.85 & 4.84 \\
  4657.2  & 13 & b & \Ti{b}{4}{P}{}{2.5}  & \Ti{z}{2}{F}{\circ}{3.5} & 1.24 &
52 & -2.32 & 0.03 & 2  & 4.98 & 4.97 & 4.99 & 4.98 \\
  4708.66 & 10 & uc & \Ti{a}{2}{P}{}{1.5} & \Ti{z}{2}{F}{\circ}{2.5} & 1.24 & 51
& -2.34 & 0.05 & 1 & 4.96 & 4.95 &  4.97 & 4.96 \\
  4798.53 & 1 & a & \Ti{a}{2}{D}{}{1.5} & \Ti{z}{4}{G}{\circ}{2.5} & 1.08 & 44 &
-2.68 & 0.08 & 1 & 4.97 & 4.97 & 5.00 & 5.00  \\
  5129.15 & 20 & b & \Ti{b}{2}{G}{}{4.5} & \Ti{z}{2}{G}{\circ}{4.5} & 1.89 & 72
& -1.24 & 0.03 & 1 & 4.96 & 4.93 &  4.99 & 4.96 \\
  5336.81 & 16 & b & \Ti{b}{2}{D}{}{2.5} & \Ti{z}{2}{F}{\circ}{3.5} & 1.58 & 71
& -1.63 & 0.03 & 2 & 5.05 & 5.02 & 5.03 & 5.00 \\
  5381.01 & 16 & b & \Ti{b}{2}{D}{}{1.5} & \Ti{z}{2}{F}{\circ}{2.5} & 1.57 & 57
& -1.92 & 0.05 & 1 & 4.95 & 4.94 & 4.95  & 4.94 \\
  5418.77 & 16 & b &  \Ti{b}{2}{D}{}{2.5}  &  \Ti{z}{2}{F}{\circ}{2.5} & 1.58 &
51 & -2   & 0.05 & 1 & 4.90  & 4.89 & 4.90  & 4.89 \\
\noalign{\smallskip}\hline\noalign{\smallskip}
\end{tabular}
\end{center}
\end{minipage}
$^a$  Line identification: a - unblended; b - blended; uc - uncertain continuum
$^b$ References to $gf$-values: (1) \citet{2001ApJS..132..403P}; (2)
\citet{1993A&A...273..707B}
\end{table*}
\subsection{Isotopic shift}
Ti is represented by five stable isotopes
$^{46}$Ti:$^{47}$Ti:$^{48}$Ti:$^{49}$Ti:$^{50}$Ti and their relative abundances
in the solar system are
$8:7.43:73.8:5.5:5.4$\footnote{http://www.nist.gov/pml/data/handbook/index.cfm},
respectively. It has been recognized long ago that Ti lines in the solar
spectrum are subject to an additional broadening due to the isotopic effect
\citep{1952AJ.....57..158A}. In analogue to hyperfine
splitting\footnote{Hyperfine splitting also affects atomic energy levels of
$^{47}$Ti and $^{49}$Ti, but the fraction of these odd-$Z$ isotopes is only
$13\%$ of the total element abundance. Thus, hyperfine splitting is ignored in
this work.}, isotopic shift of energy levels leads to splitting of spectral
lines into several components. For lighter multi-electron atomic systems, such
as Ti, the isotopic shift of energy levels is primarily due to the differences
in nuclear masses of isotopes giving rise to normal and specific mass shifts. As
shown below, non-negligible errors in the solar Ti abundance may arise if
isotopic structure is neglected in the line formation calculations.

Isotopic shifts in the UV and blue Ti II transitions were recently calculated
by \citet{2008JPhB...41w5702B} and measured by \citet{2010PhyS...81f5301N}. For
some Ti II transitions, the total separation of components due to the isotopic
effect is not larger than $\delta \nu \sim 900$ MHz or $\delta \lambda \sim 6$
\mA\ \citep{2010PhyS...81f5301N}. In contrast, very large isotopic shifts have
been observed for transitions from the unfilled $3$d shell (e.g. $3$d$^3 4$s$
\rightarrow 3$d$^2$ $4$p, or $3$d $4$s$^2 \rightarrow 3$d$^2$ $4$p). For the Ti
II line at $4488$ \AA, which we use in our abundance analysis,
\citet{2010PhyS...81f5301N} derive an isotope shift of $\sim 2400$ MHz with the
total separation of components of $\sim 4500$ MHz. This corresponds to a line
width of $\sim 30$ \mA\ that is comparable with thermal line broadening in the
solar atmosphere. There are also few measurements of the isotopic shift in Ti I,
which include several transitions from our sample.

In this study, isotopic structure was taken into account for two Ti I and three 
Ti II lines, which show large isotopic splittings, i.e. the separation of
components is $\sim 15 - 20$ m\AA. For the Ti I lines, we used the data from
\citet*{1995JPhB...28..957G}. The wavelengths and $gf$-values of individual
components are given in Table \ref{isocomp}. The total oscillator strengths for
the transitions are given in Tables \ref{ti_i} and \ref{ti_ii}.
\begin{table}
\begin{center}
\caption[]{Isotopic components of selected Ti II and Ti I transitions for the
solar mixture of the isotopes. See text.}
\label{isocomp}
\begin{tabular}{lclc}
\hline\noalign{\smallskip}
Transition &  Wavelength &  $\log gf_i$ & Isotope \\
           &  \AA        &              &       \\
\noalign{\smallskip}\hline\noalign{\smallskip}
Ti II & & & \\
\Ti{c}{2}{D}{}{2.5} -- \Ti{x}{2}{F}{\circ}{3.5} & 4488.346  & $-1.607$ & 46 \\
                                                & 4488.337  & $-1.638$ & 47 \\
                                                & 4488.330  & $-0.642$ & 48 \\
                                                & 4488.322  & $-1.77$  & 49 \\
                                                & 4488.315  & $-1.78$  & 50 \\
 & & & \\
\Ti{a}{2}{G}{}{4.5} -- \Ti{z}{2}{F}{\circ}{3.5} & 4468.498  & $-1.697$ & 46 \\
                                                & 4468.504  & $-1.729$ & 47 \\
                                                & 4468.510  & $-0.732$ & 48 \\
                                                & 4468.515  & $-1.860$ & 49 \\
                                                & 4468.521  & $-1.868$ & 50 \\
 & & & \\
\Ti{a}{2}{G}{}{3.5} -- \Ti{z}{2}{F}{\circ}{3.5} & 4444.548  & $-3.337$ & 46 \\
                                                & 4444.554  & $-3.369$ & 47 \\
                                                & 4444.560  & $-2.372$ & 48 \\
                                                & 4444.565  & $-3.500$ & 49 \\
                                                & 4444.571  & $-3.507$ & 49 \\
Ti I & & & \\
\Ti{a}{3}{P}{}{2} -- \Ti{y}{3}{D}{\circ}{3}     & 5866.440  & $-2.108$ & 46 \\
                                                & 5866.449  & $-2.100$ & 47 \\
                                                & 5866.459  & $-0.972$ & 48 \\
                                                & 5866.469  & $-1.977$ & 49 \\
                                                & 5866.480  & $-1.937$ & 49 \\
& & & \\
\Ti{a}{3}{P}{}{0} -- \Ti{y}{3}{D}{\circ}{1}     & 5922.100  & $-2.734$ & 46 \\
                                                & 5922.109  & $-2.726$ & 47 \\
                                                & 5922.119  & $-1.598$ & 48 \\
                                                & 5922.129  & $-2.603$ & 49 \\
                                                & 5922.140  & $-2.563$ & 49 \\
\noalign{\smallskip}\hline\noalign{\smallskip}
\end{tabular}
\end{center}
\end{table}
%
%
\section{Results}{\label{sec:results}}

In the following sections, we present the results of statistical equilibrium
calculations for Ti for a restricted range of stellar parameters and discuss
whether the LTE approximation can be adopted for different combinations of
$\Teff$, $\log g$, and [Fe/H]. We also compute NLTE abundances of Ti for the Sun
and four metal-poor stars (Sect. \ref{sec:stars}) and attempt to calibrate the
poorly-known cross-sections to inelastic H I and electron collisions following
the classical \citet{1968ZPhy..211..404D} formalism to satisfy ionization
equilibrium of Ti I/Ti II. Deviations from LTE and their effect on the
abundances are discussed in terms of NLTE abundance corrections defined as:
\[
\Delta_{\rm NLTE} = \log\varepsilon^{\rm NLTE} - \log\varepsilon^{\rm LTE}
\]
%
%
\subsection{NLTE effects on atomic number densities and line
formation}
\begin{figure}
\resizebox{\columnwidth}{!}{\rotatebox{0}
{\includegraphics[scale=1]{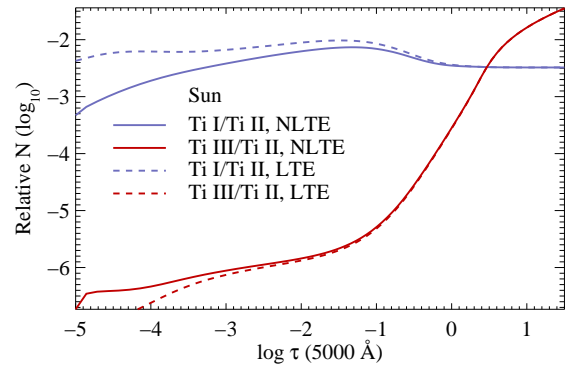}}}
\caption[]{Relative number densities N(Ti I)/N(Ti II) and N(Ti III)/N(Ti II) for
the solar model atmosphere with parameters $5780/4.44/0$ and [Ti/H] $ = 4.9$,
computed under LTE and NLTE.}
\label{numbers_ti}
\end{figure}

\begin{figure*}
\hbox{\includegraphics[scale=0.3]
{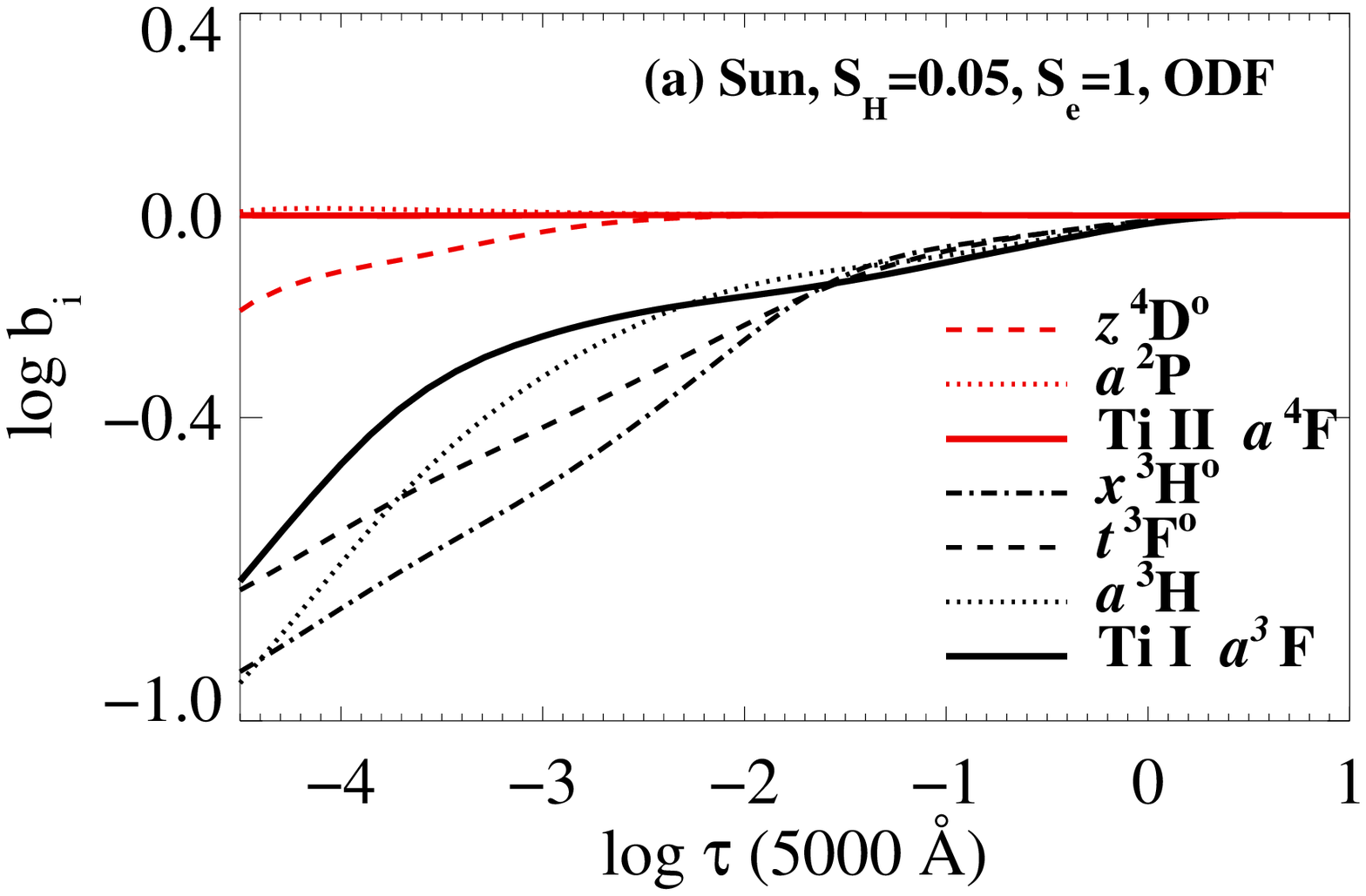}\hfill
\hspace{-5mm}
\includegraphics[scale=0.3]
{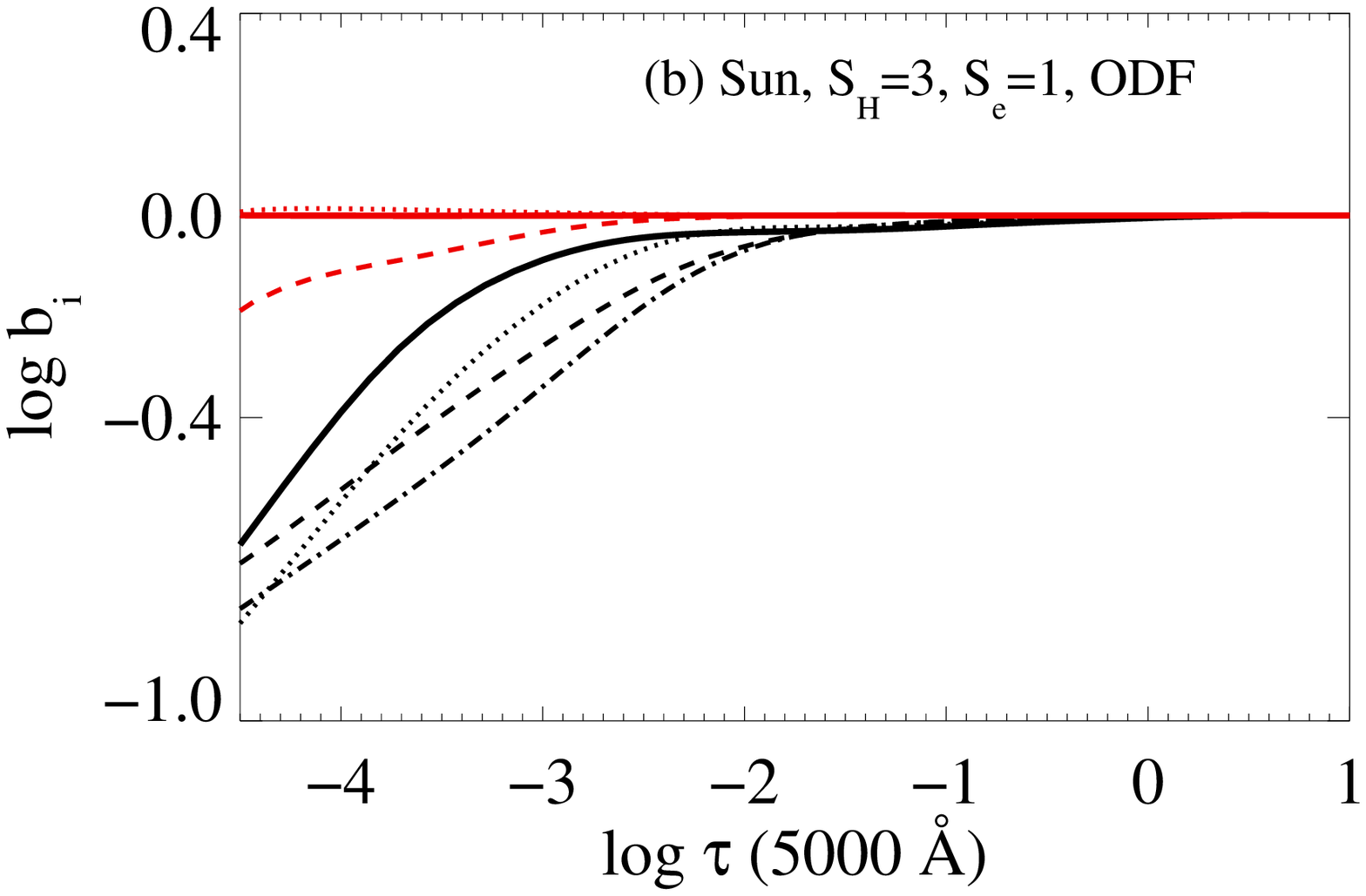}\hfill
\hspace{-5mm}
\includegraphics[scale=0.3]
{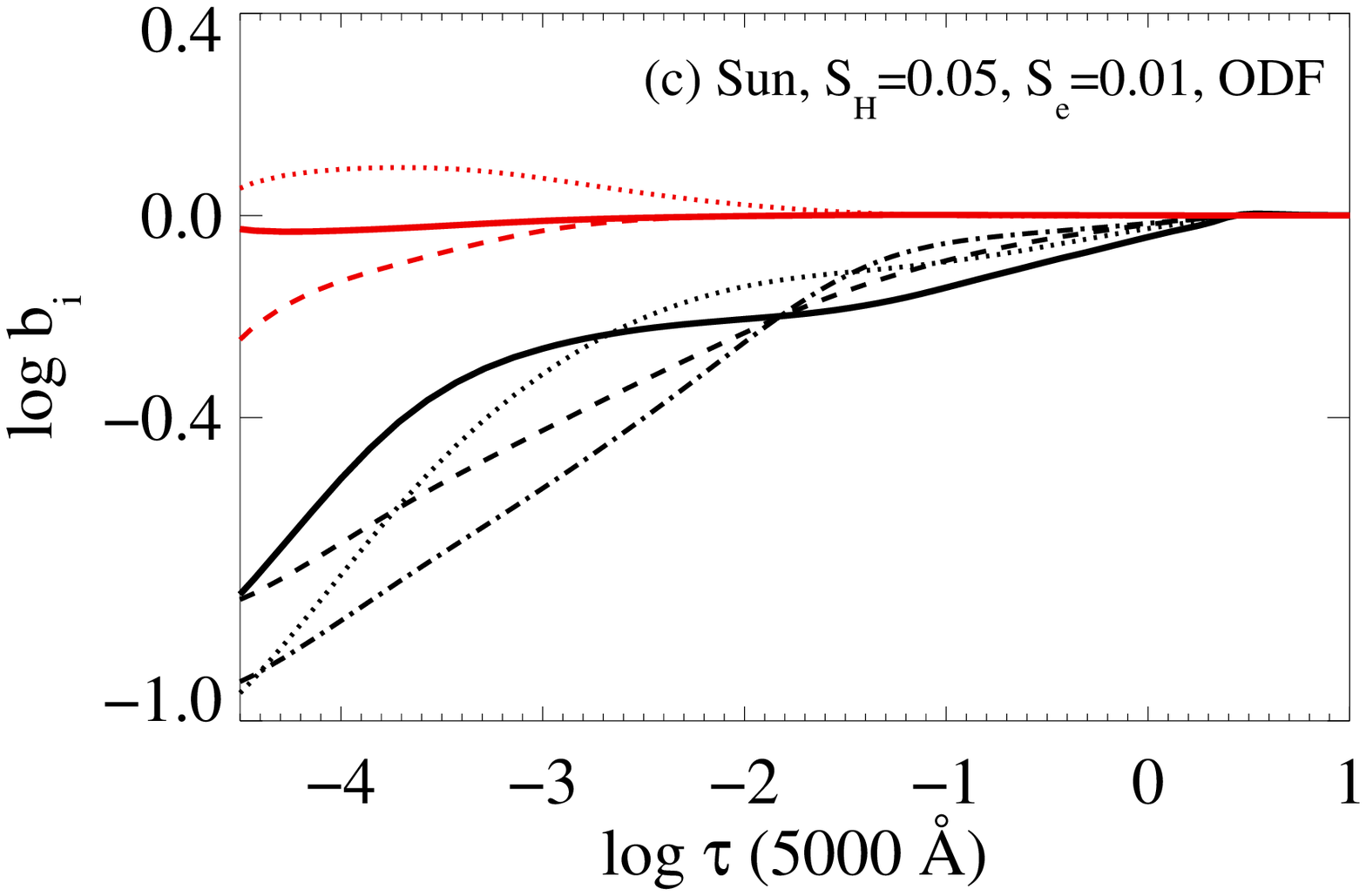}}
\vspace{-3mm}
\hbox{\includegraphics[scale=0.3]
{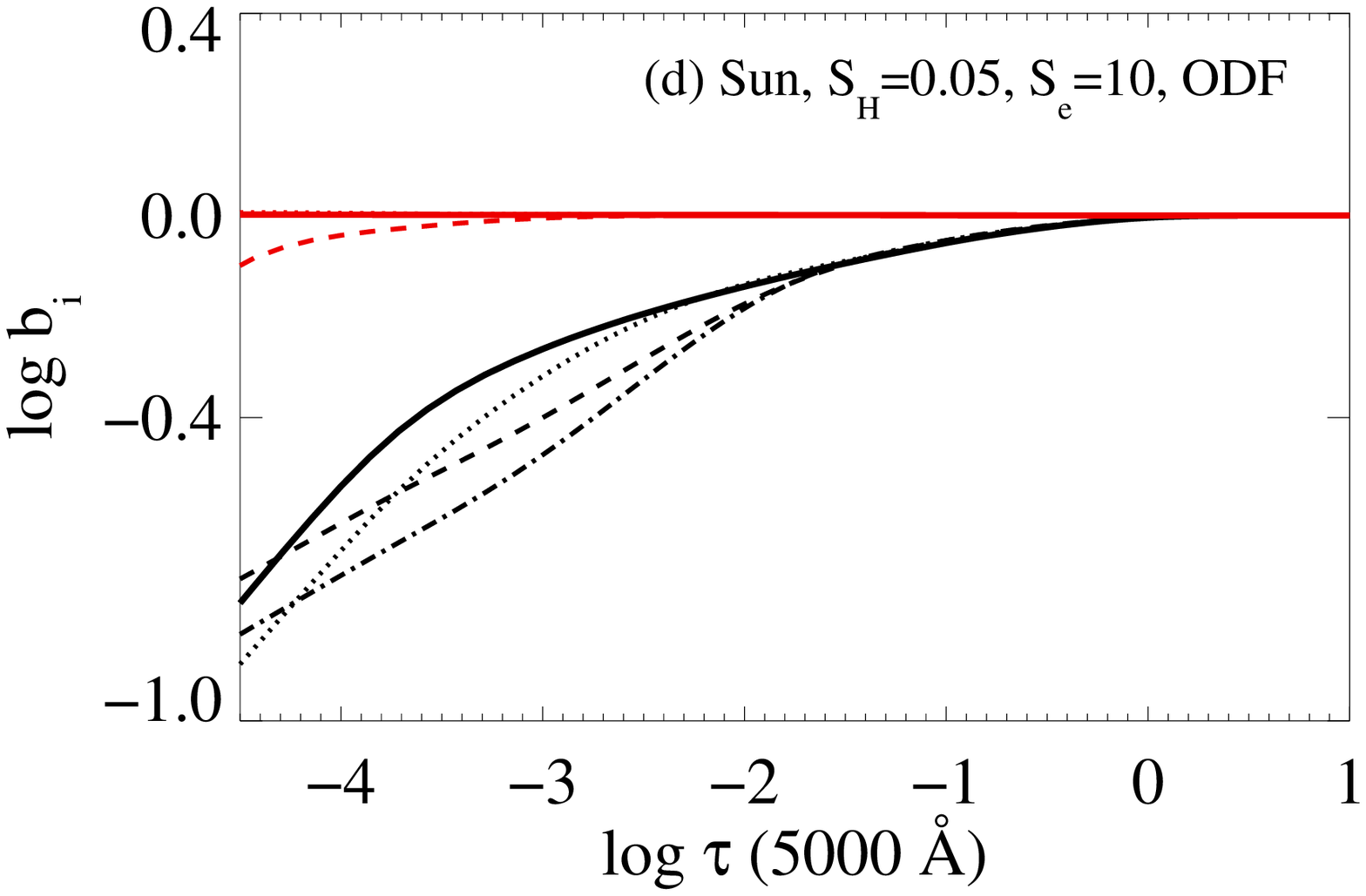}\hfill
\hspace{-5mm}
\includegraphics[scale=0.3]
{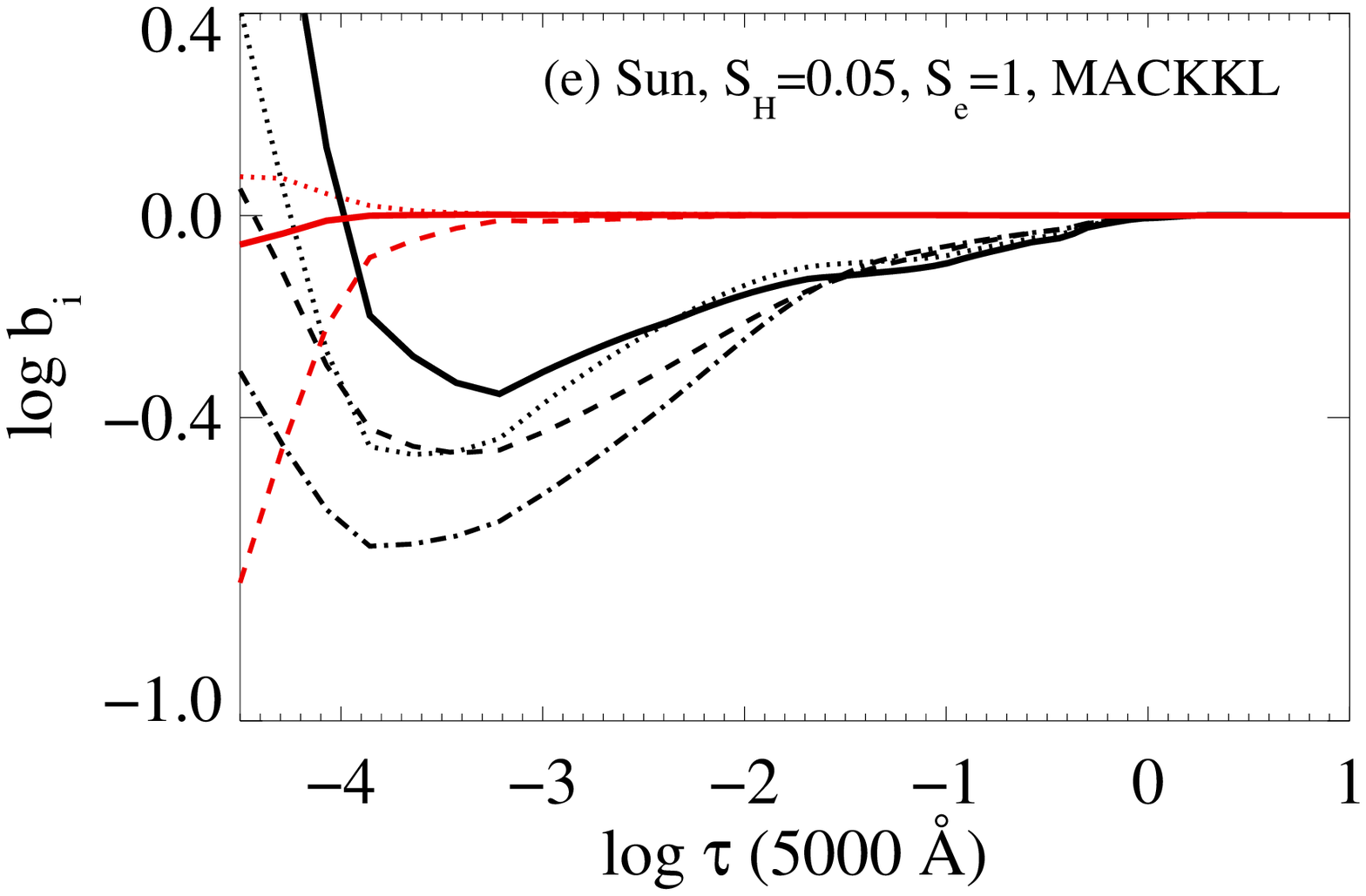}\hfill
\hspace{-5mm}
\includegraphics[scale=0.3]
{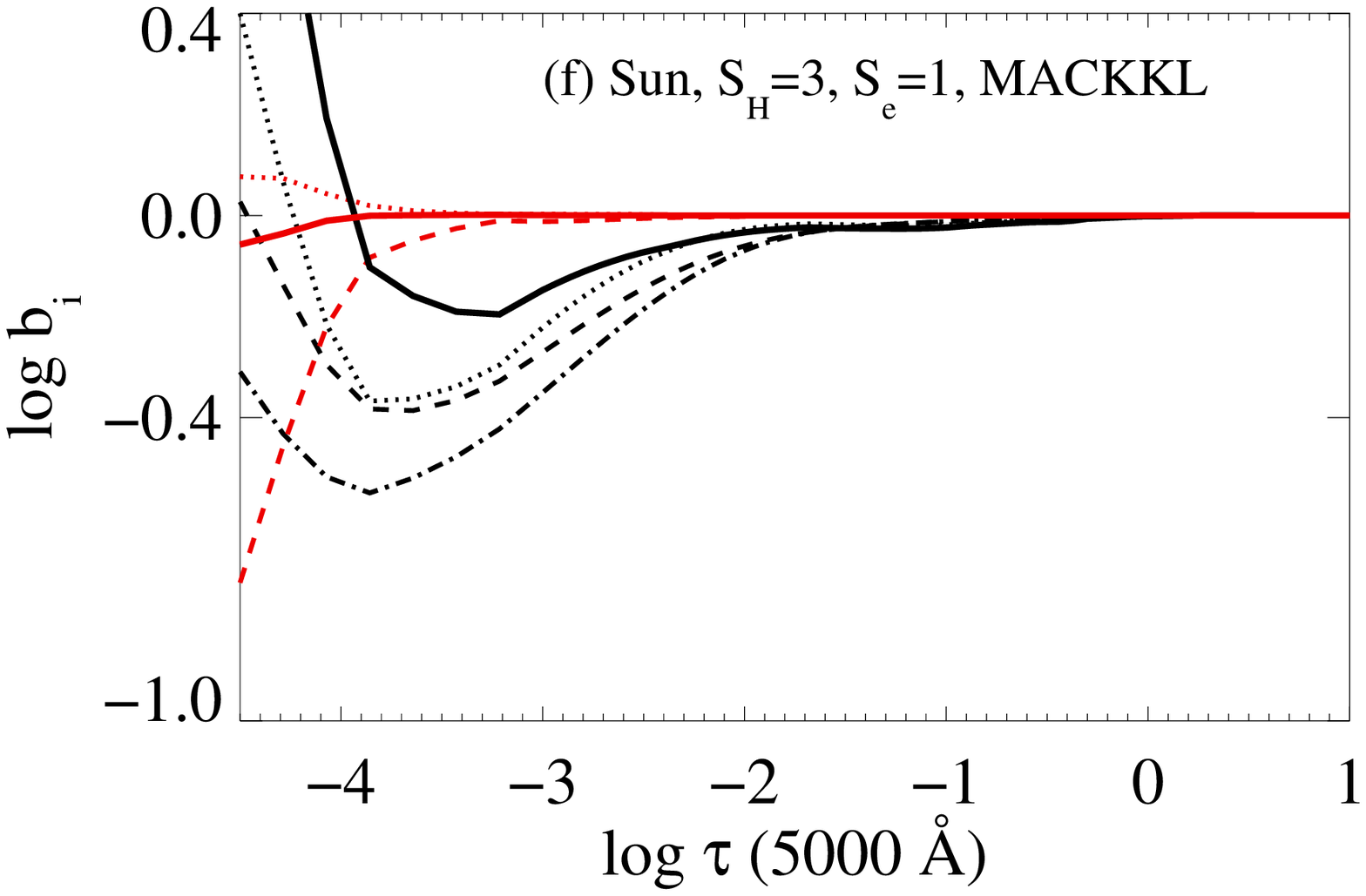}}
\vspace{-3mm}
\hbox{\includegraphics[scale=0.3]
{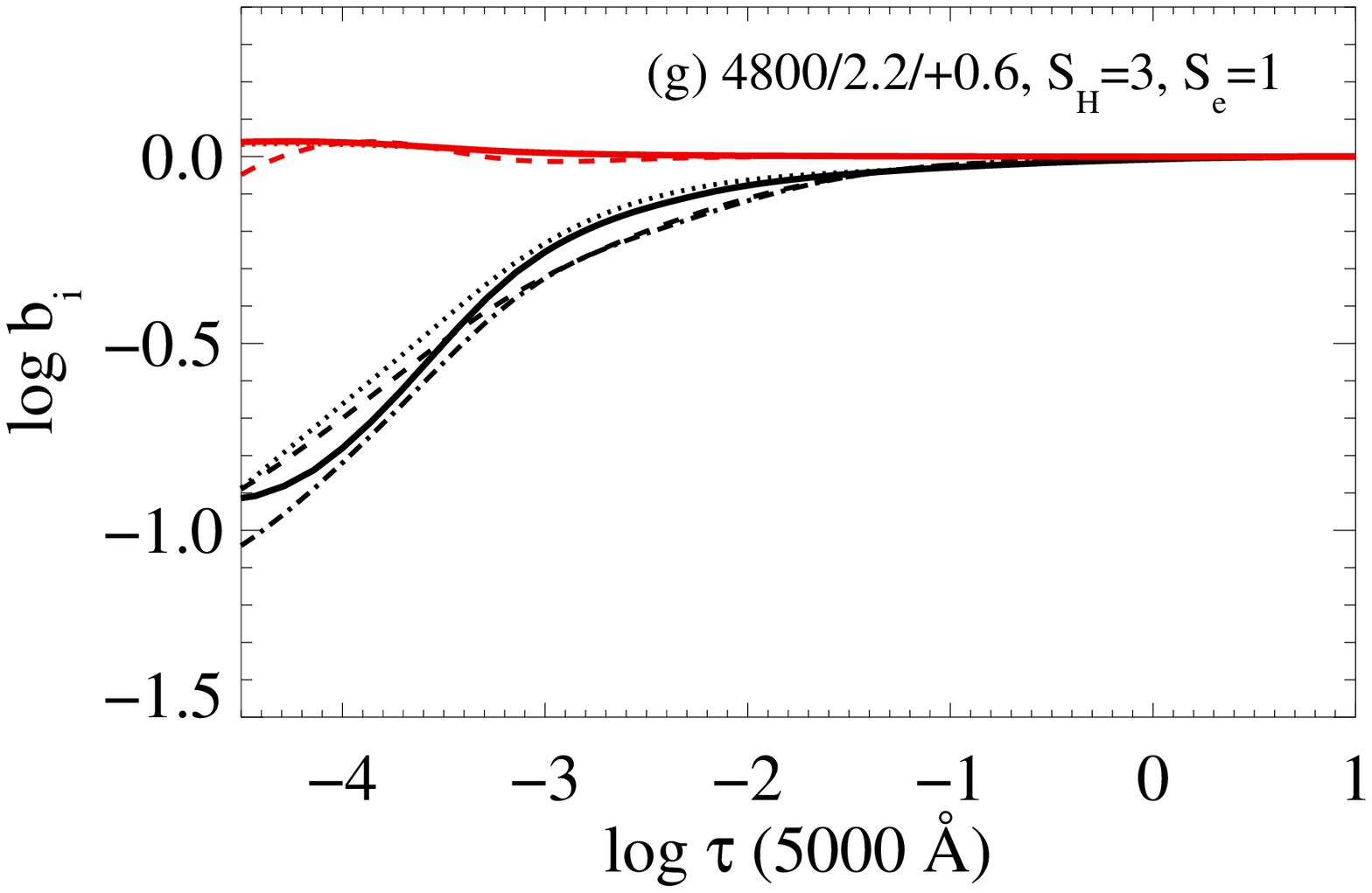}\hfill
\hspace{-5mm}
\includegraphics[scale=0.3]
{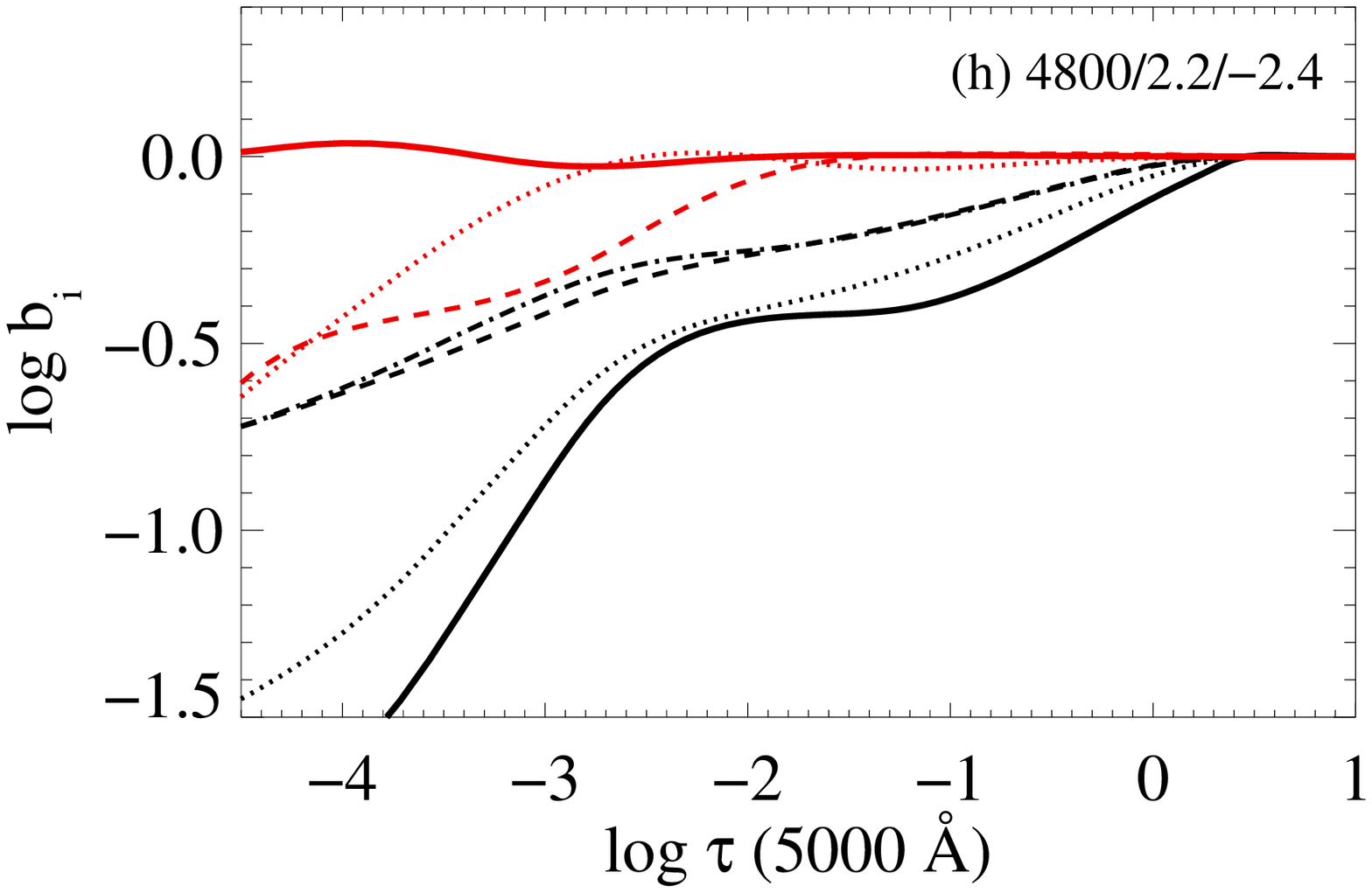}\hfill
\hspace{-5mm}
\includegraphics[scale=0.3]
{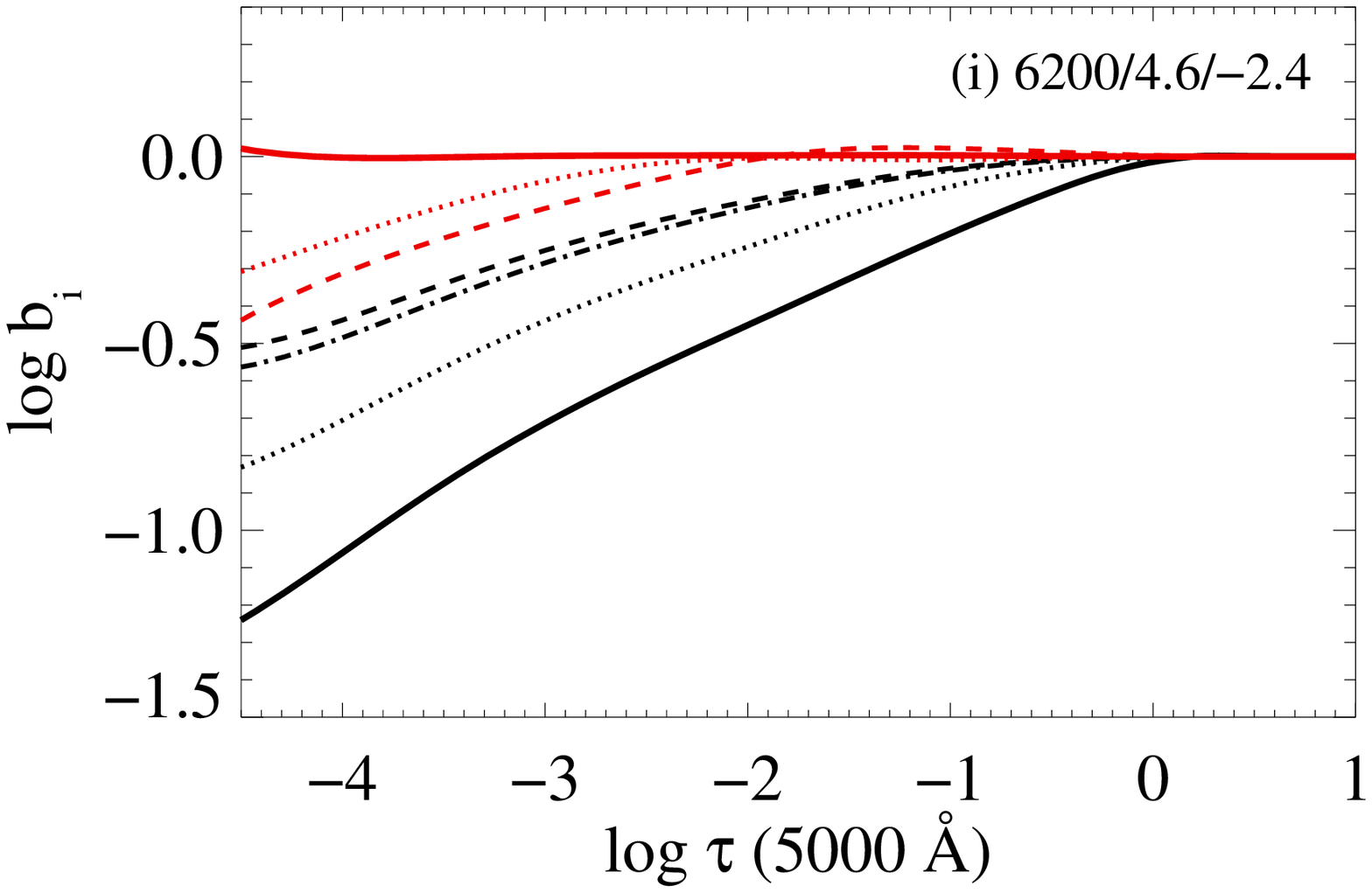}}
\vspace{-3mm}
\caption[]{Departure coefficients for selected Ti I and Ti II levels computed
for the MAFAGS-ODF (a)-(d) and MACKKL (e)-(f) solar model atmospheres with
different prescriptions for inelastic collision rates. (g) to (i): Same as (b),
but the model atmospheres are taken from the grid with stellar parameters
specified in each panel. Level designations in all panels are the same as in
(a). See text for details.}
\label{departures}
\end{figure*}

{\label{sec:nlte_effect}}
The qualitative analysis of level \emph{departure coefficients}\footnote{The
departure coefficients are defined as the ratio of NLTE to LTE number densities
of atoms in a certain excited level $i$, $b_i = n_i^{\rm NLTE}/n_i^{\rm LTE}$
\citep[definition according to][]{1972SoPh...23..265W}} helps to understand NLTE
effects in excitation and ionization balance of Ti. Departure coefficients
of selected atomic energy levels of Ti I computed for MAFAGS-ODF model
atmospheres\footnote{The choice of the ODF or OS model has no impact on the
conclusions drawn in this section} with different prescriptions for elastic
collisional rates are shown in Fig. \ref{departures}. Fig. \ref{numbers_ti}
shows relative number densities N(Ti I)/N(Ti II) and N(Ti III)/N(Ti II) for the
solar model atmosphere computed under LTE (dashed lines) and NLTE (solid
lines).

In general, most of the Ti I levels in the atmospheres of late-type stars are
underpopulated relative to their LTE number densities. The driving mechanism is
overionization by non-local UV radiation field with mean intensity $J_{\rm \nu}$
larger than the local Planck function $B_{\nu}(\Te)$. However, due to the
complexity of Ti atom, other NLTE mechanisms also play a role and population of
each level at each atmospheric depth is determined by its collisional and
radiative interaction with hundreds of other atomic states and, indirectly,
even with the levels of the next ionization stage, Ti II.
Thus, clear isolation of interaction processes leading to NLTE population of
each atomic level is not possible and we have chosen only few Ti I and Ti II
levels to illustrate the typical NLTE effects. Transitions between these levels
are strong and few of them, $\lambda 4758$ \AA\ and $\lambda 4759$ \AA\ 
(\Ti{a}{3}{H}{}{} $\leftrightarrow$ \Ti{x}{3}{H}{\circ}{}), were selected for
the stellar abundance analysis. Fig. \ref{departures} shows the Ti I ground
state \Ti{a}{3}{F}{}{}, even-parity metastable levels \Ti{a}{3}{H}{}{} ($\Elow =
2.2$ eV), odd-parity levels \Ti{t}{3}{F}{\circ}{} and \Ti{x}{3}{H}{\circ}{}
($\Elow = 4.8$ eV and $4.85$ eV, respectively). Also, few Ti II levels, which
give rise to the Ti II transitions visible in our stellar spectra, are shown.

Fig. \ref{departures}a illustrates departure coefficients computed for the
solar model atmosphere assuming a scaling factor $\SH = 0.05$ for the
\citet{1968ZPhy..211..404D} cross-sections for collisions with H I atoms and
non-scaled cross-sections for transitions due to electron collisions, $\Se = 1$
(Sect. \ref{sec:modelatom}).
LTE distribution functions for the Ti I are invalid already at $\opd \sim +0.3$,
which is where the atmosphere becomes transparent to UV continuum radiation. The
super-thermal radiation field, $J_{\rm \nu} > B_{\nu}(\Te)$, increases
photoionization rates from the Ti I levels with ionization edges at relevant
wavelengths, that is, at $\lambda \sim 300 - 400$ nm. Since this process is not
compensated by the recombination rates, which are fixed by the local kinetic
temperature $\Te$, the disbalance in number densities of well-populated Ti I
levels with excitation energy $2 - 3$ eV is transferred to the other levels via
collisions and strong UV radiative transitions. This is an interesting case,
when collisions inhibit deviations from LTE for some levels, but amplify them
for the others by thermalizing the relative populations of the levels with
different degree of underpopulation.

\begin{figure*}
\hbox{\includegraphics[scale=0.3]{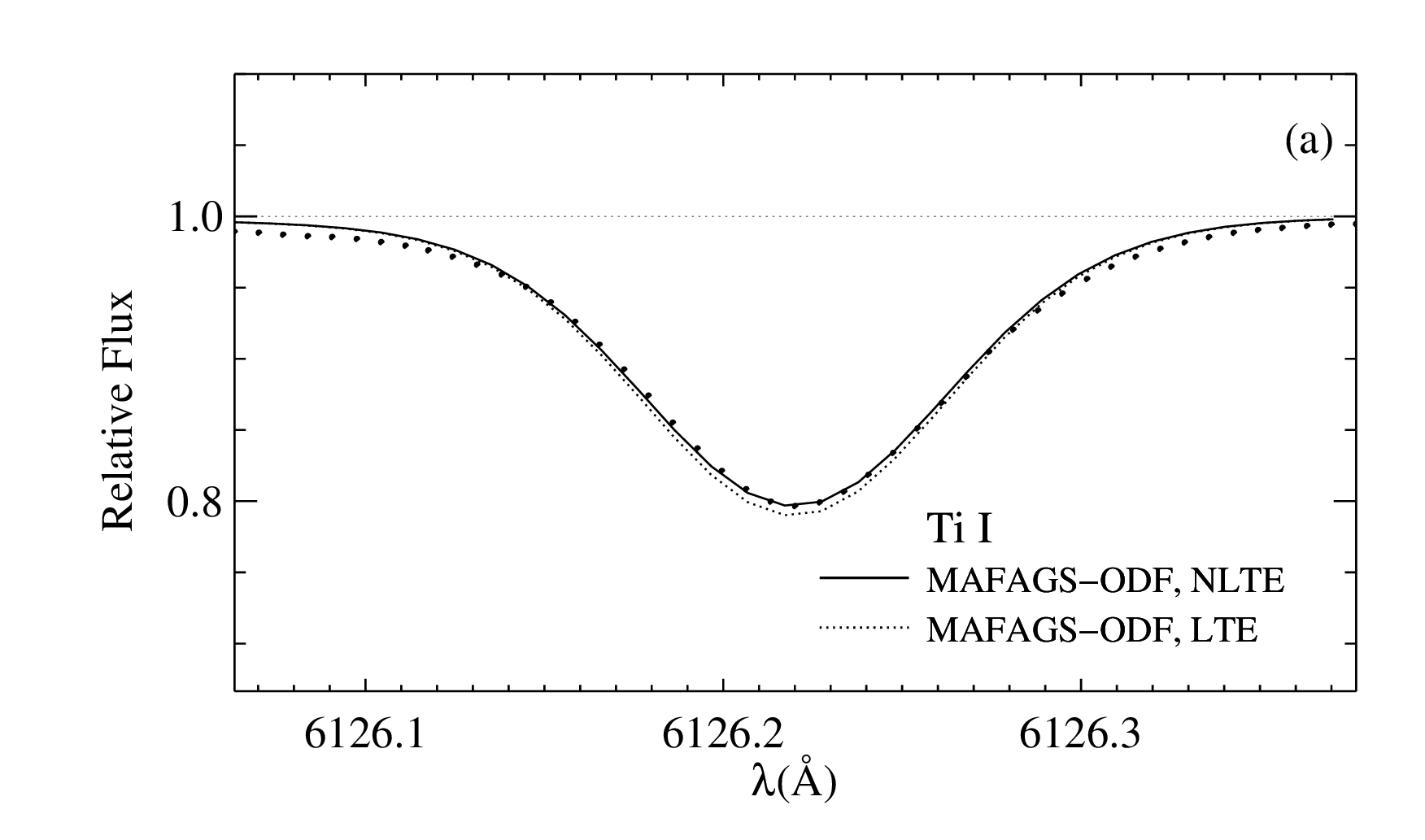}\hfill
\hspace{-5mm}
\includegraphics[scale=0.3]{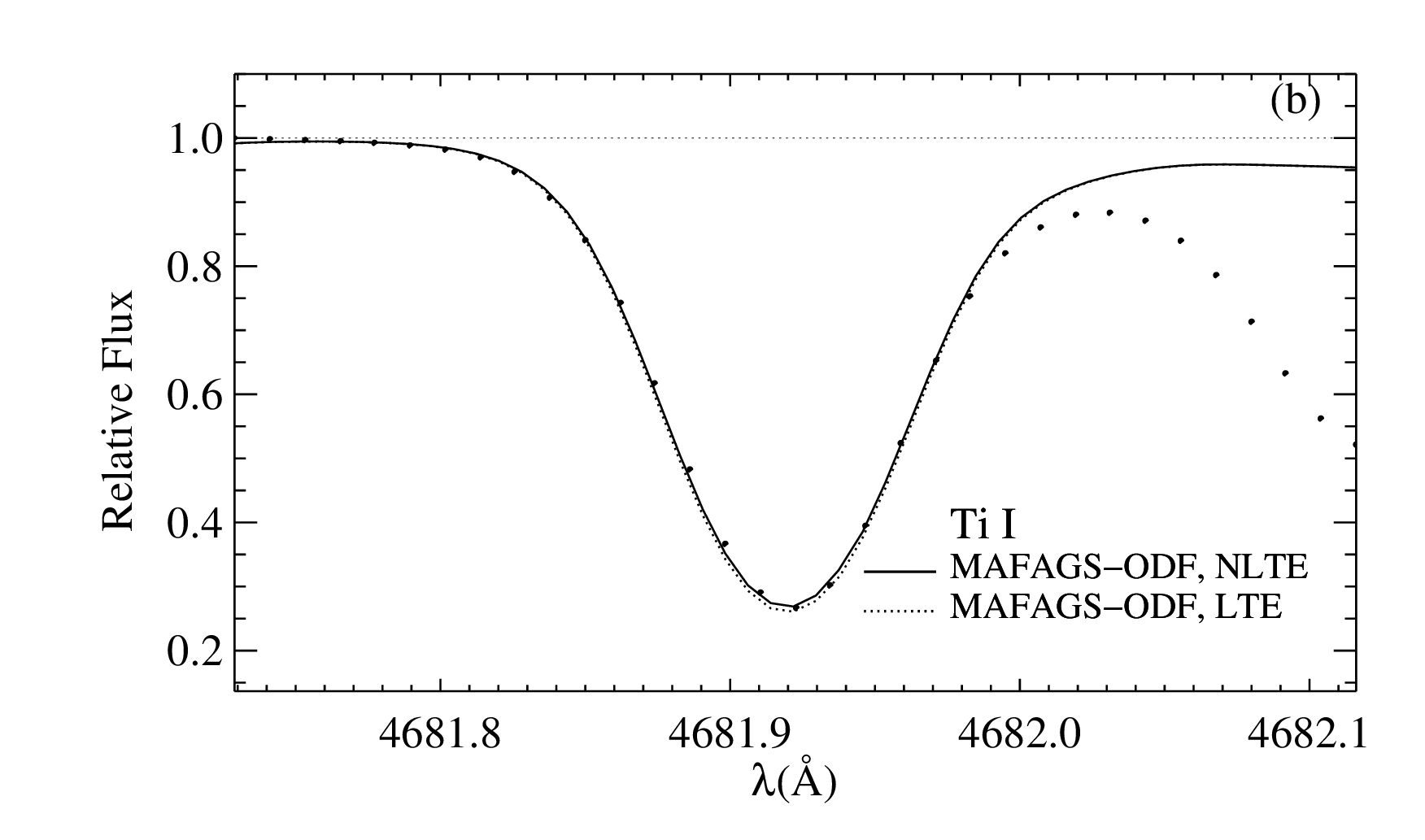}}
\vspace{-3mm}
\hbox{\includegraphics[scale=0.3]{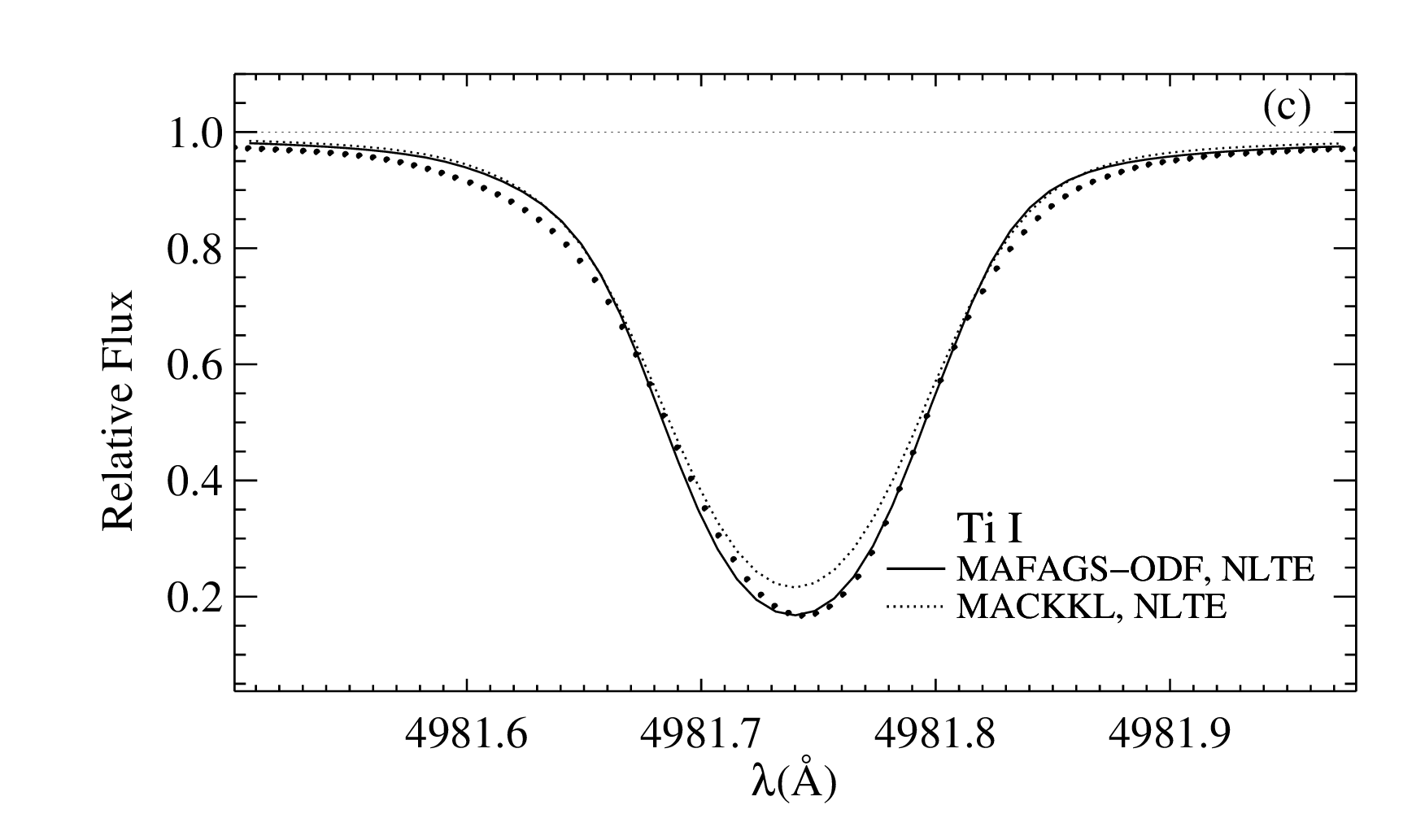}\hfill
\hspace{-5mm}
\includegraphics[scale=0.3]{fig18.ps}}
\vspace{-3mm}
\hbox{\includegraphics[scale=0.3]{fig19.ps}\hfill
\hspace{-5mm}
\includegraphics[scale=0.3]{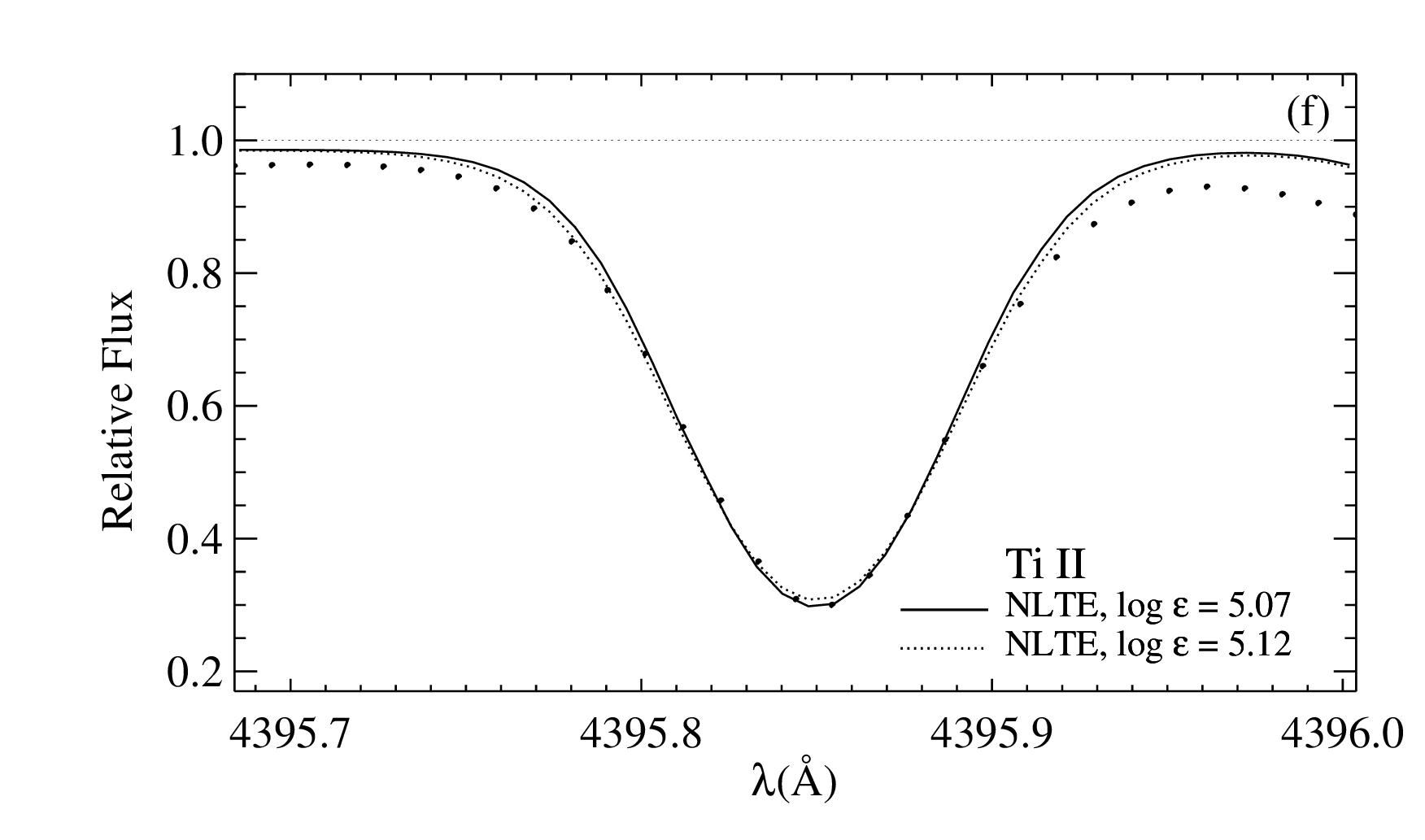}}
\vspace{-3mm}
\caption[]{Synthetic (black trace) and observed (symbols) profiles of selected
Ti I and Ti II lines in the solar spectrum. (d): for demonstration purposes, the
theoretical profile was synthesized excluding the contribution of blends in the
wings that makes the fit worse. (e): the line was not used for the solar
abundance calculations due to an unknown blend. See text for details.}
\label{profiles}
\end{figure*}

At $-1.5 \leq \opd \leq 0$ in the solar case, optical depth in the strong UV
transitions of Ti I is still larger than unity and photoionization
cross-sections of the levels in the hydrogenic approximation are a monotonic
function of frequency, $\sigma_{\rm ph-ion} \sim 1/\nu^3$, thus underpopulation
of all levels is very homogeneous. The weak Ti I lines, which are formed at
these optical depths, are only affected by the change in opacity, which scales
with the departure coefficient of their lower level. Since respective
$b_i$-factors are less than unity, NLTE model lines are weaker compared to LTE
and require larger Ti abundance to fit the observed spectrum. A few examples of
NLTE synthetic lines of Ti I computed with the MAFAGS-ODF solar model are given
in Fig. \ref{profiles}(a,b), where they are compared to the observed profiles.
The \emph{shapes} of line profiles computed with ODF or OS models are
very similar, thus profiles computed with the latter are not shown in the
Figure. The LTE profiles reproduce the NLTE profiles after adjustment of
the Ti abundance and macroturbulence.

Excitation balance in Ti I significantly deviates from thermal at $\opd \leq
-1.5$ for the MAFAGS solar model, where photon losses in numerous near-UV and
blue transitions influence the relative populations of the levels. The majority
of such transitions are resonance (multiplet numbers $1$ to $37$). For example,
the Ti I ground state, \Ti{a}{3}{F}{}{} is radiatively connected with the
\Ti{t}{3}{F}{\circ}{} level with the excitation energy $4.8$ eV (Fig.
\ref{departures} a). Although there is an ample pumping of the upper level by
super-thermal $J_{\rm \nu}$ at $-1 \leq \opd \leq 0$, photon escape in the lines
of the multiplet $30$ leads to a sudden underpopulation of \Ti{t}{3}{F}{\circ}{}
at $\opd \sim -1.3$. The latter process, combined with photon losses in other
resonance transitions, causes the outward flattening of the departure
coefficient for the Ti I ground state \Ti{a}{3}{F}{}{}, up to the depths where
its $b_i$-factor drops again due to the dominant overionization. The same
mechanism acts between the \Ti{a}{3}{H}{}{} and \Ti{x}{3}{H}{\circ}{} levels
(multiplet 114). The corresponding lines form at $\opd \sim -1.2$ (Fig.
\ref{departures}a).

In the outermost layers, $\opd \leq -3.5$, the theoretical solar model
atmosphere is not adequate: it lacks chromospheres. In the MACKKL semi-empirical
solar model, the temperature minimum occurs at $\opd \approx -3.4$. Below this
optical depth, the behaviour of departure coefficients is identical to the
MAFAGS model (Fig. \ref{departures}e,f). However, the outward rise of
temperature leads to the overpopulation of Ti I levels, contrary to the strong
underpopulation predicted by the MAFAGS models. As a result, NLTE abundance
corrections change in sign and magnitude, which is particularly important for
lines formed above $T_{\rm min}$, such as the strong resonance Ti I lines at
$3998.6$ \AA\ and $3989.76$ \AA. LTE abundances determined with the MACKKL and
MAFAGS models also differ because the former is slightly warmer than the latter
in the layers below $\opd = -3$ (Fig. \ref{temp}).

To illustrate the difference between the theoretical and semi-empirical models
with respect to abundance determinations, we consider two Ti I lines. The line
at $4981.73$ \AA\ (multiplet $44$) is the strongest in our list with $\EW \sim
120$ \mA. The NLTE line profiles computed with the MAFAGS-ODF and MACKKL solar
model atmospheres are compared with the observed solar flux profiles in the Fig.
\ref{profiles}c. The NLTE abundance corrections have different sings:
$\Delta_{\rm NLTE} \rm(MAFAGS-ODF) = +0.02$ dex and $\Delta_{\rm NLTE}
\rm(MACKKL) = -0.05$ dex. The NLTE Ti abundances deduced from such profile
fits are $4.86$ dex and $4.98$ dex for the MAFAGS-ODF and MACCKL models,
respectively. For the weak line at $4997$ \AA\ with $\EW \sim 34$ \mA, MACKKL
model also overestimates the abundance by $0.1$ dex compared to the MAFAGS-ODF
model, but the NLTE corrections for both model atmospheres are equal,
$\Delta_{\rm NLTE} = +0.04$ dex.

The influence of uncertainties in atomic data on the excitation and ionization 
balance of Ti was tested by changing the cross-sections for collision-induced
transitions by hydrogen atoms and free electrons. Scaling down the
cross-sections of transitions due to e$^-$ collisions by two orders of magnitude
(Fig. \ref{departures}c), $\Se = 0.01$, leads to decoupling of all levels at
$\opd \geq -1.5$, whereas increasing the cross-sections by a factor of ten (Fig.
\ref{departures}d) produces \emph{relative} thermalization of the levels with
excitation energy below $5$ eV in the same range of optical depths $\opd$.

Departure coefficients are very sensitive to the variation of the scaling factor
to inelastic H I collisions $\SH$. With $\SH = 3$, the Ti I levels nearly
thermalize at $-2 \leq \opd \leq 0$ (Fig. \ref{departures}b). It is only in the
uppermost layers with low densities, $\opd \leq -3$, that the influence of
collisions with H I is small. We note that $\SH = 3$ produces the smallest
abundance scatter between Ti I lines and simultaneously satisfies ionization
balance of Ti in the solar case (Sect. \ref{sec:sun}). Thus, the latter value
defines our reference model atom. Yet, it has to be kept in mind that
calibration of $\SH$ on the observed stellar spectrum may hide or compensate
other deficiencies in the modelling, such as those related to the use of the 1D
static model atmospheres with a certain prescription for convection, i.e.
mixing-length theory in our case, and the fixed microturbulence parameter. Thus,
our final value of $\SH$ may not be appropriate for other NLTE studies of Ti,
which are based on different model atmospheres and methods \citep*[see
e.g. discussion in][]{2005ARA&A..43..481A}.

We performed test NLTE calculations for several models with the following
parameters: (a) $\Teff = 4800$, $\log g = 2.2$, [Fe/H] $= +0.6$; (b) $\Teff =
4800$, $\log g = 2.2$, [Fe/H] $= -2.4$; (c) $\Teff = 6200$, $\log g = 4.6$,
[Fe/H] $= -2.4$. These parameters are representative of stars commonly used in
Galactic chemical evolution studies, i.e. metal-rich $K$-type giants in the
bulge, metal-poor giants and dwarfs in the halo. Departure coefficients for
the three models computed with $\SH = 3$ and $\Se = 3$ are shown in Fig.
\ref{departures}. A rather counter-intuitive result is that the Ti I and Ti II
level populations show departures from LTE even for the cool giant with
super-solar metallicity (Fig. \ref{departures}g). Even though UV radiation
field is weakened due to low photospheric temperatures and excessive line
blanketing, collisions at $\log g \approx 2$ are too weak to ensure LTE
conditions. Since the stellar flux maximum is shifted to longer wavelengths,
ionization from the Ti I levels with higher excitation potential, e.g.
\Ti{a}{3}{H}{}{}, becomes more important compared to the solar case.
Qualitatively, our results for cool stars of solar metallicity are very similar
to \citet[][their Fig. 4]{1999ApJ...512..377H}.

NLTE effects are particularly large for the metal-poor models with low gravity
(Fig. \ref{departures}h). NLTE number densities of the majority of Ti I levels
are reduced by a factor of $2$ already at $\opd \sim -0.5$. The undulations of
the $b_i$-curves with $\opd$ are due to strong radiative interaction between
levels, i.e. non-balanced excitations at the depths where photons escape in
the wings of strong Ti I lines and spontaneous de-excitations at the core
formation depths. In contrast, very homogeneous distribution of $b_i$-factors
with optical depth is characteristic of warm high-gravity models (Fig.
\ref{departures}i). This is an expected result, because collisions provide close
coupling of the levels and even the Ti I ground state with the largest
occupation number is subject to overionization caused by strong non-local UV
radiation field near its ionization threshold.

Deviations from LTE in Ti II are small for the combinations of stellar
parameters investigated in this work. Departure coefficients of low and
intermediate-excitation Ti II levels deviate from unity in the upper atmospheric
layers, at $\opd \leq -2$. Most of the Ti II lines form below this depth, where
$b_i$-factors are slightly larger than unity, and the corresponding effect on
abundances is never larger than $-0.05$ dex.
%
%
%
\subsection{The Sun}{\label{sec:sun}}
\subsubsection{Spectrum synthesis}
The solar abundance of Ti was determined by visually fitting synthetic
spectral lines computed with LTE or NLTE number densities to the Solar Flux
Atlas of \citet{1984sfat.book.....K}. We used both MAFAGS-ODF and MAFAGS-OS
model atmospheres. The standard line broadening mechanisms were taken into
account: solar rotation with $V_{\rm rot , \odot} = 1.8$ \kms, microturbulence
velocity $\Vmic = 0.9$ \kms, and a radial-tangential macroturbulence velocity
$\Vmac = 2.5 \ldots 4$ \kms. Line broadening caused by elastic collisions with H
I atoms was computed using the velocity parameters and temperature exponents
from \citet{1995MNRAS.276..859A}. In Tables \ref{ti_i} and \ref{ti_ii}, this
parameter is given in terms of the van der Waals damping constant $\log C_6$ for
each transition.
%
%
\begin{table*}
\caption{Solar abundance of Ti computed with different NLTE model atoms and
model atmospheres. The brackets specify a combination ($\SH, \Se$) used for each
model atom. See texts.} 
\label{abundances}
\begin{tabular}{cccccccccccc}
\noalign{\smallskip}\hline\noalign{\smallskip}
Ion & $N$ & $gf^a$ & \multicolumn{2}{c}{MAFAGS-ODF} & 
\multicolumn{5}{c}{MAFAGS-OS} \\
    &     &        & LTE    & $(3,1)$ & LTE &
$(3,0.01)$ & $(3,1)$ & $(3,10)$ &  $(0.05,0.01)$ & $(0.05,1)$ & $(0.05,10)$ \\
\noalign{\smallskip}\hline\noalign{\smallskip}
Ti I  & 12 & 1 & $4.80 \pm 0.05 $ & $4.84 \pm 0.05 $ & $4.90 \pm 0.05$ & 
$5.01 \pm 0.08$ & $4.94 \pm 0.05$ & $4.95 \pm 0.05$ & $5.11 \pm 0.06$ &
$5.02 \pm 0.05$ & $5.01 \pm 0.05$ \\
      & 52 & 2 & $4.81 \pm 0.03 $ & $4.84 \pm 0.03$ & $4.91 \pm 0.04$ &         
$4.95 \pm 0.06$ & $4.93 \pm 0.04$ & $4.93 \pm 0.05$ & $5.09 \pm 0.09$ & 
$5.00 \pm 0.05$ & $5.00 \pm 0.04$ \\
Ti II & 10 & 3 & $4.94 \pm 0.06$ & $4.93 \pm 0.07$ & $4.96 \pm 0.06$ &     &
$4.95 \pm 0.06$ &     &     &     &    \\
      & 11 & 4 & $4.98 \pm 0.04$ & $4.97 \pm 0.04$ & $4.99 \pm 0.04$ &     &
$4.98 \pm 0.04$ &     &     &     &    \\
\noalign{\smallskip}\hline\noalign{\smallskip}
\end{tabular}
$^a$ References: ~~~(1) \citet{2006MNRAS.373.1603B}; (2) Blackwell et al.
(1982a, 1982b, 1983, 1986), scaled by $+0.056$ dex following
\citet{1989A&A...208..157G}; (3) \citet{2001ApJS..132..403P}; (4)
\citet{1993A&A...273..707B}
\end{table*}

\subsubsection{Oscillator strengths}
Oscillator strengths for the Ti II transitions were adopted from
\citet*{2001ApJS..132..403P}, who combined their branching ratios obtained by
the methods of Fourier transform spectrometry with the radiative lifetimes
measured by means of time-resolved laser-induced fluorescence by
\citet{1993A&A...273..707B}. The accuracy of the latter is about $5 \%$. We
also used the $\log gf$ values of \citet{1993A&A...273..707B}. The oscillator
strengths from both sources agree well with each other for $\log gf > -1.5$
\citep[][their Fig. 3]{2001ApJS..132..403P}, but significant discrepancies, up
to $0.3$ dex, are present for the weaker transitions. The uncertainties of
$gf$-values from both references are typically less than $0.05$ dex,
but for some transitions experimental error was as large as $0.2$ dex.

For the Ti I transitions, we relied on the oscillator strengths of
\citet{2006MNRAS.373.1603B}, who applied the same experimental techniques
as \citet{2001ApJS..132..403P}. As a test case, we also used the $gf$-values of
Blackwell et al. (1982a, 1982b, 1983, 1986), which we corrected by $+0.056$ as
recommended by \citet*{1989A&A...208..157G}, who renormalized the relative
oscillator strengths measured with the Oxford furnace to more accurate lifetimes
of \citet*{1982JPhB...15L.599R}. The estimated errors of the absolute $f$-values
are less than $5\%$.

All parameters of the selected lines, including different sets of oscillator
strengths, are given in Tables \ref{ti_i} and \ref{ti_ii}. Given the
above-mentioned discrepancies between oscillator strengths from different
experimental sources, we decided to compute the average Ti abundance for each
set of $gf$-values.

\subsubsection{Abundance of Ti in the solar photosphere}{\label{sec:solarab}}
The mean Ti abundances obtained under LTE and NLTE with different model
atmospheres, model atoms, and sets of oscillator strengths are given in Table
\ref{abundances}. The individual abundances for each spectral line are listed in
Tables \ref{ti_i} and \ref{ti_ii}. NLTE abundances given in these tables were
derived assuming efficient inelastic H - Ti collisions in statistical
equilibrium calculations, $\SH = 3$. The choice of the scaling factor will be
described below.

In Table \ref{abundances}, it can be seen that the LTE abundances determined
from the Ti I lines are systematically lower than the NLTE abundances, with the
difference of $0.05 - 0.1$ dex, which depends on the treatment of inelastic
collisions in statistical equilibrium calculations. Also, the MAFAGS-ODF model
atmosphere underestimates the solar Ti abundance by $\sim 0.1$ dex compared to
the MAFAGS-OS model. This is due to differences in the temperature and pressure
structure of the models (Fig. \ref{temp}).
The NLTE values determined from the Ti I lines using the
\citet{2006MNRAS.373.1603B} and Blackwell et al. (1982a, 1982b, 1983, 1986) $gf$
values agree well with each other. Mean LTE abundances deduced with either of
the $gf$ sets are also consistent. \citet{2006MNRAS.373.1603B} noted that the
$gf$-values for several near-IR transitions of Ti I measured by
\citet{1983MNRAS.204..883B} are inaccurate due to non-equilibrium effects in the
Oxford furnace, and should be modified by $\sim 50\%$. Our results do not allow
to confirm or refute this statement, since in both cases the individual
abundances for the near-IR transition \Ti{a}{3}{P}{}{1} - \Ti{z}{3}{D}{\circ}{2}
($8683$ \AA) differ by more than $1 \sigma$ from the mean value computed for a
corresponding $\log gf$ set (Table \ref{ti_i}).

\begin{figure*}
\centering
\hbox{\includegraphics[scale=0.3]{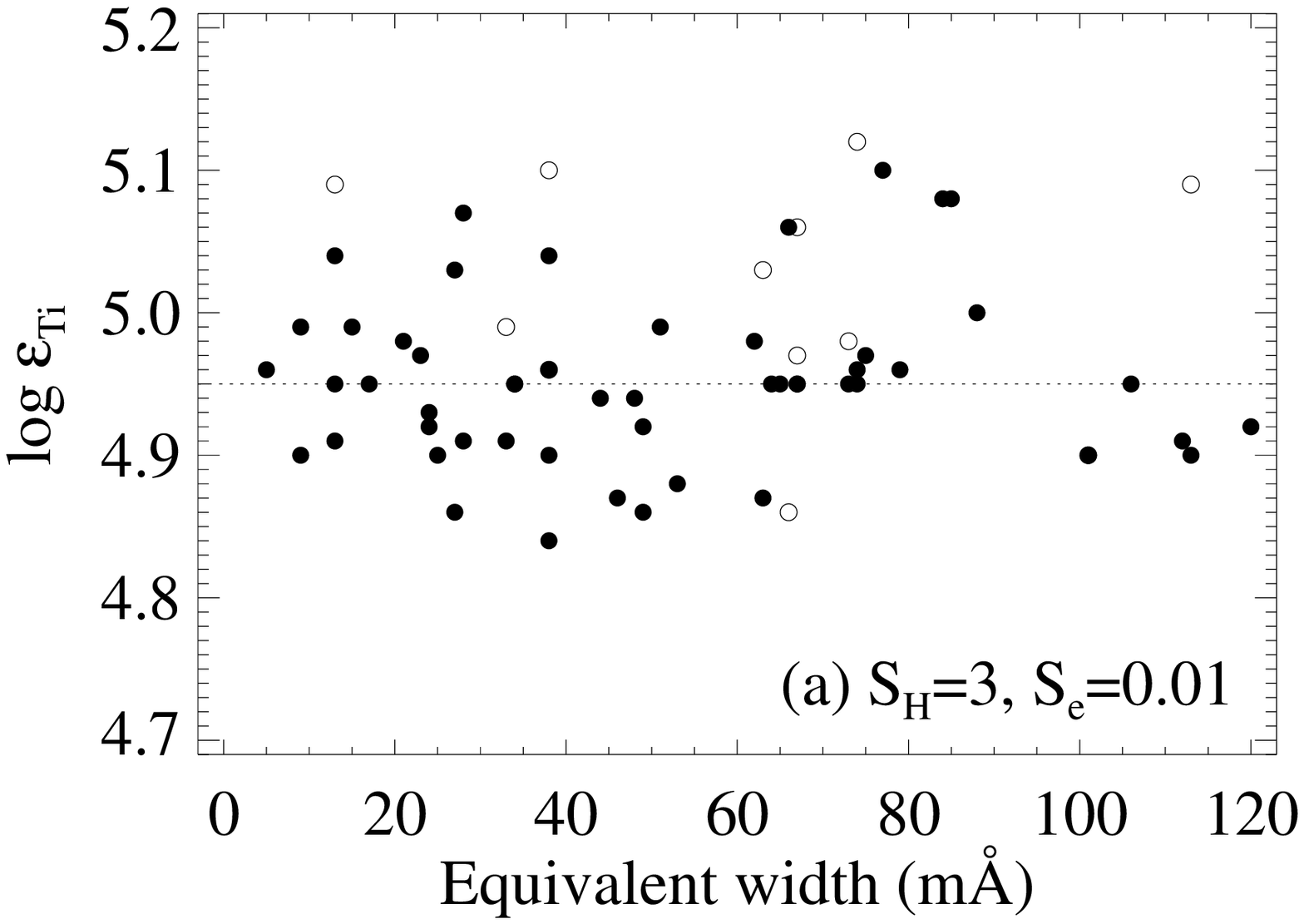}\hfill
\hspace{-6mm}
\includegraphics[scale=0.3]{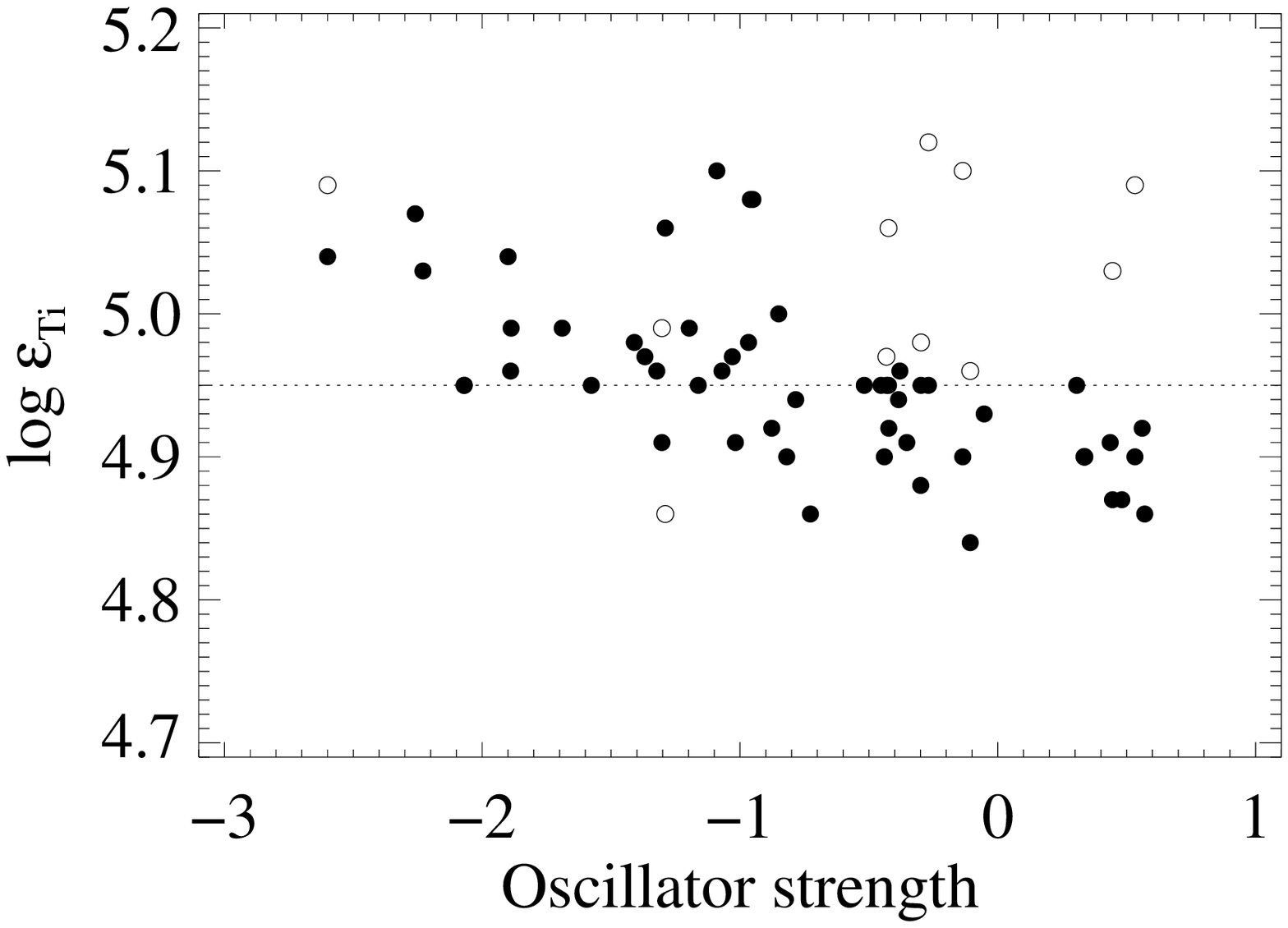}\hfill
\hspace{-6mm}
\includegraphics[scale=0.3]{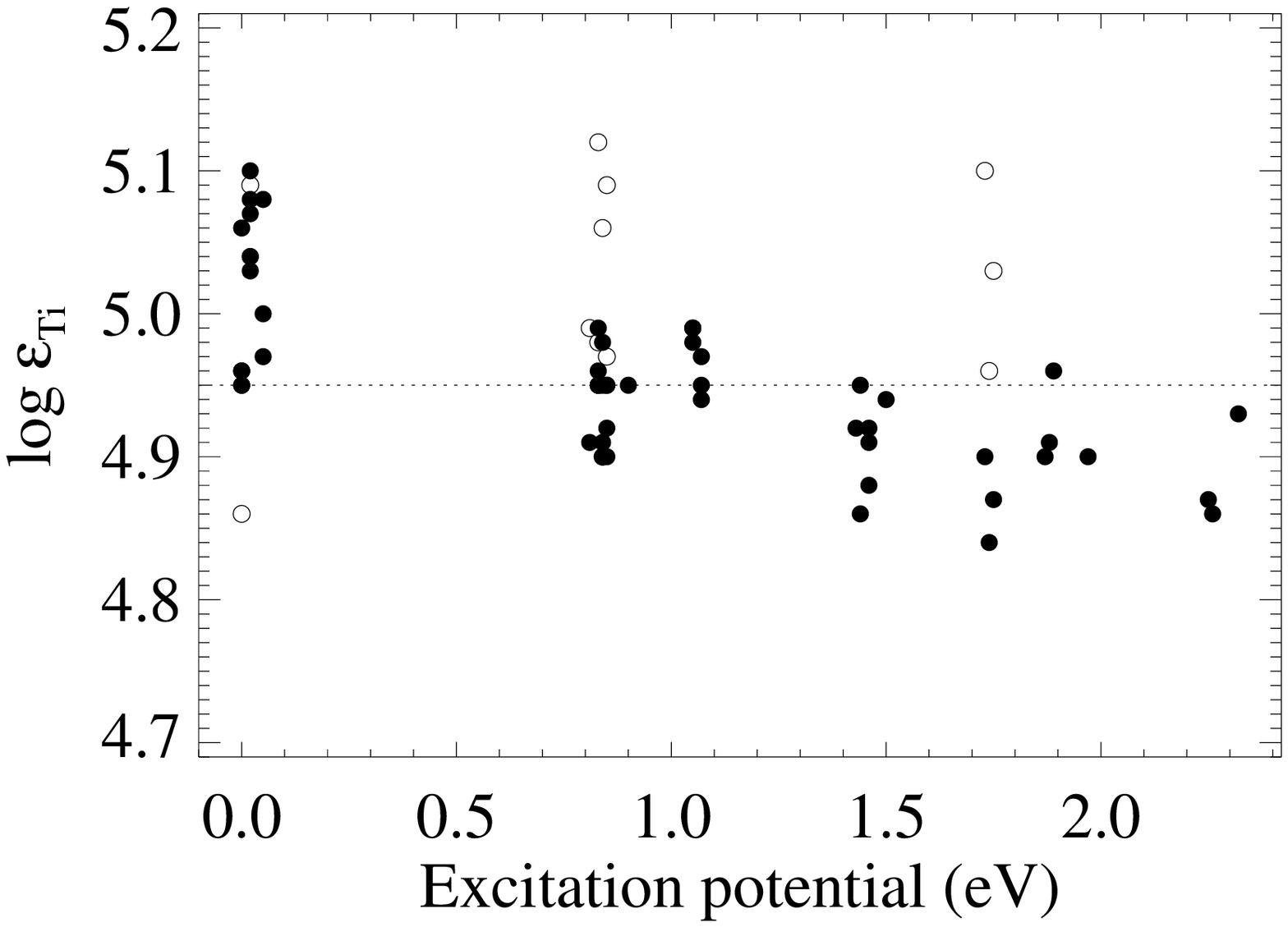}\hfill}
\vspace{-4mm}
\hbox{\includegraphics[scale=0.3]{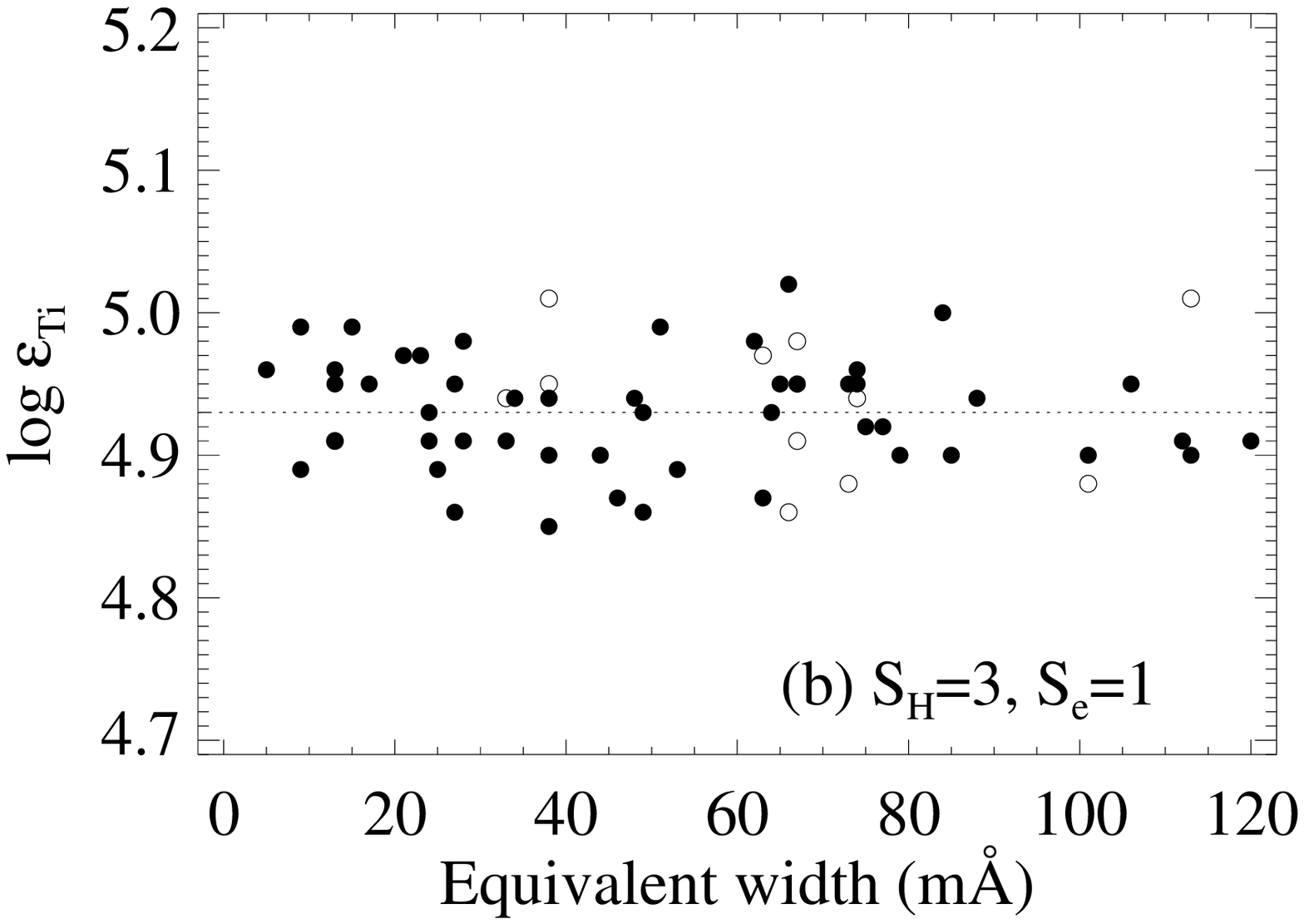}\hfill
\hspace{-6mm}
\includegraphics[scale=0.3]{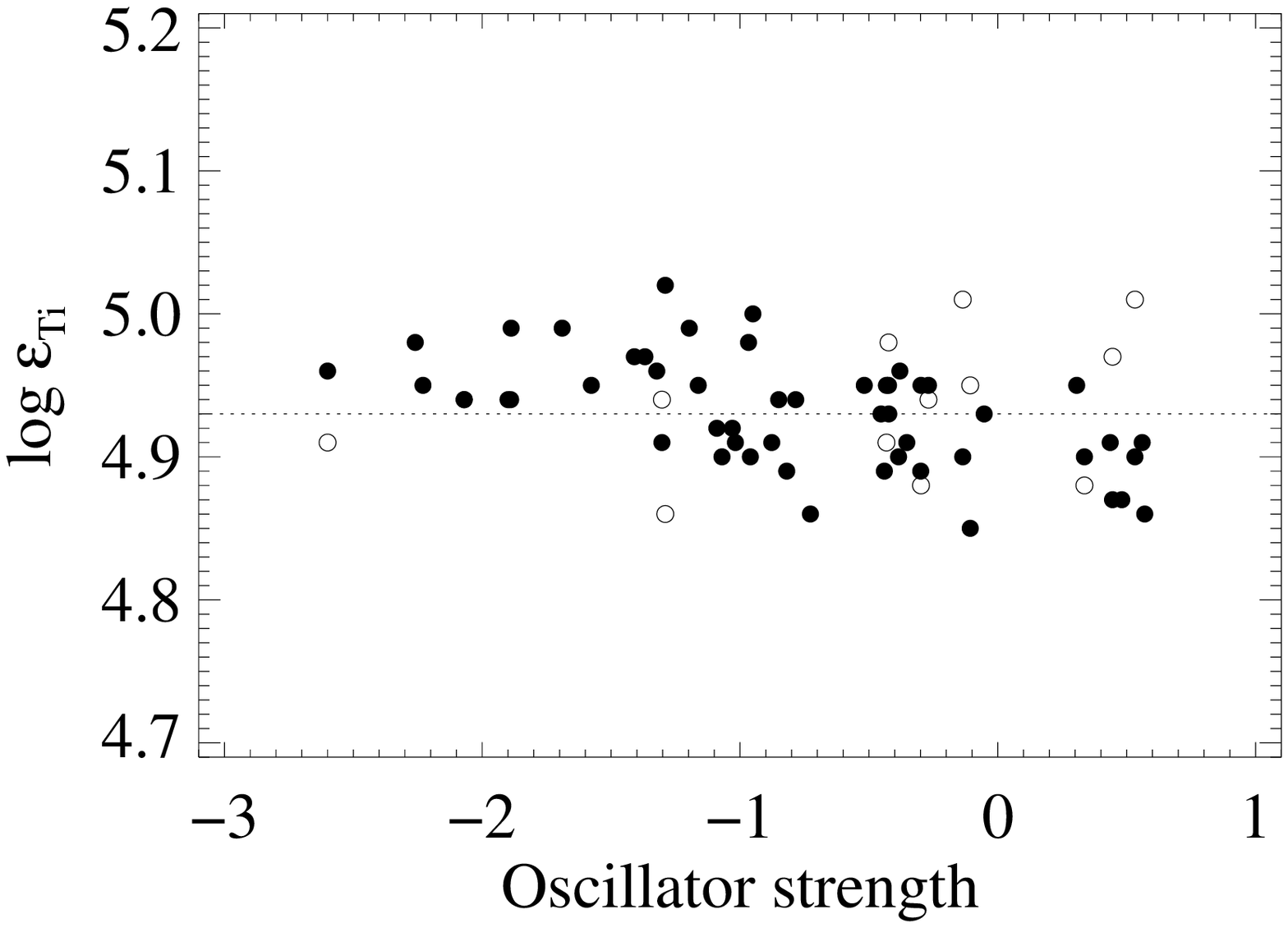}\hfill
\hspace{-6mm}
\includegraphics[scale=0.3]{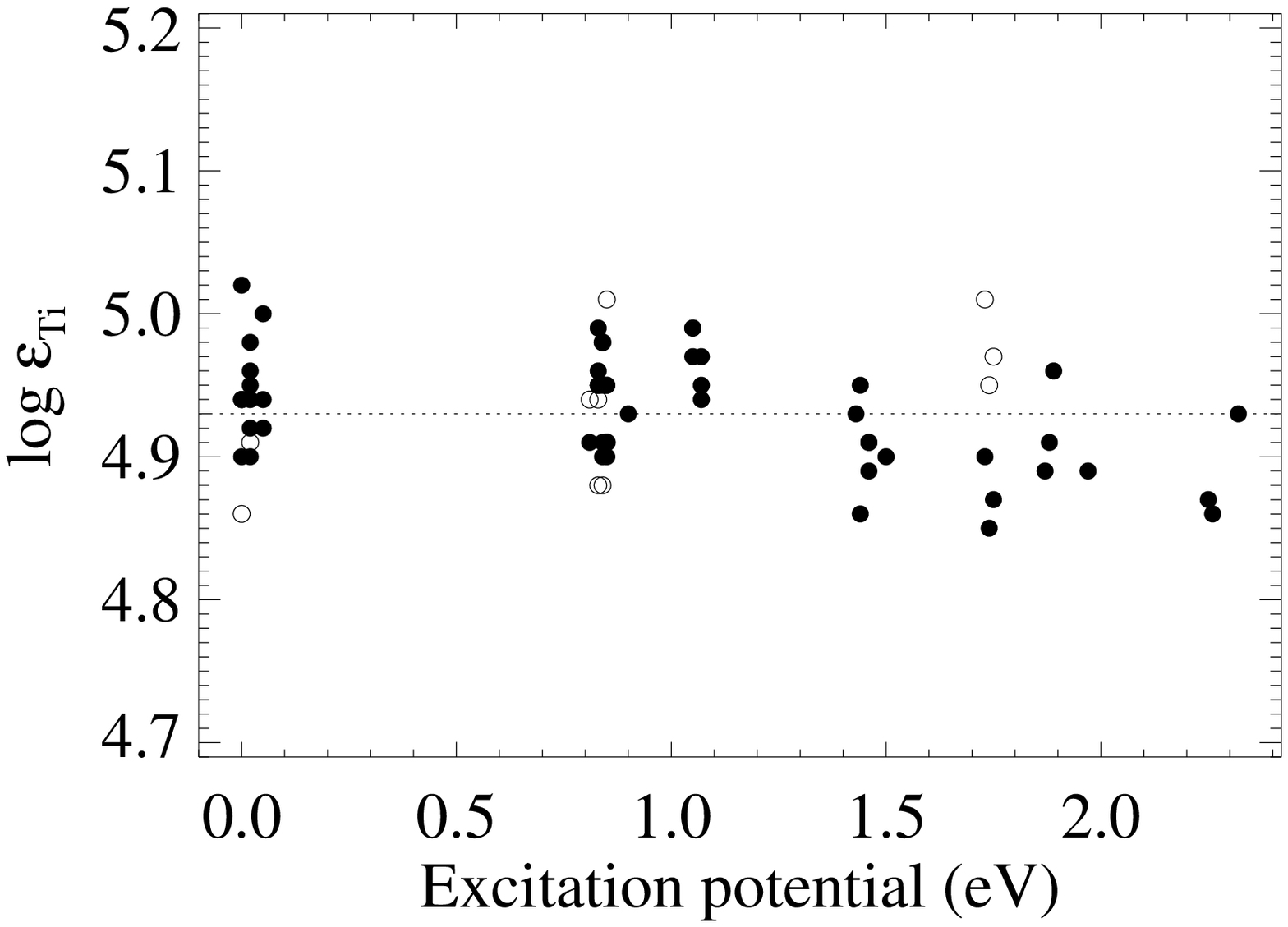}\hfill}
\vspace{-4mm}
\hbox{\includegraphics[scale=0.3]{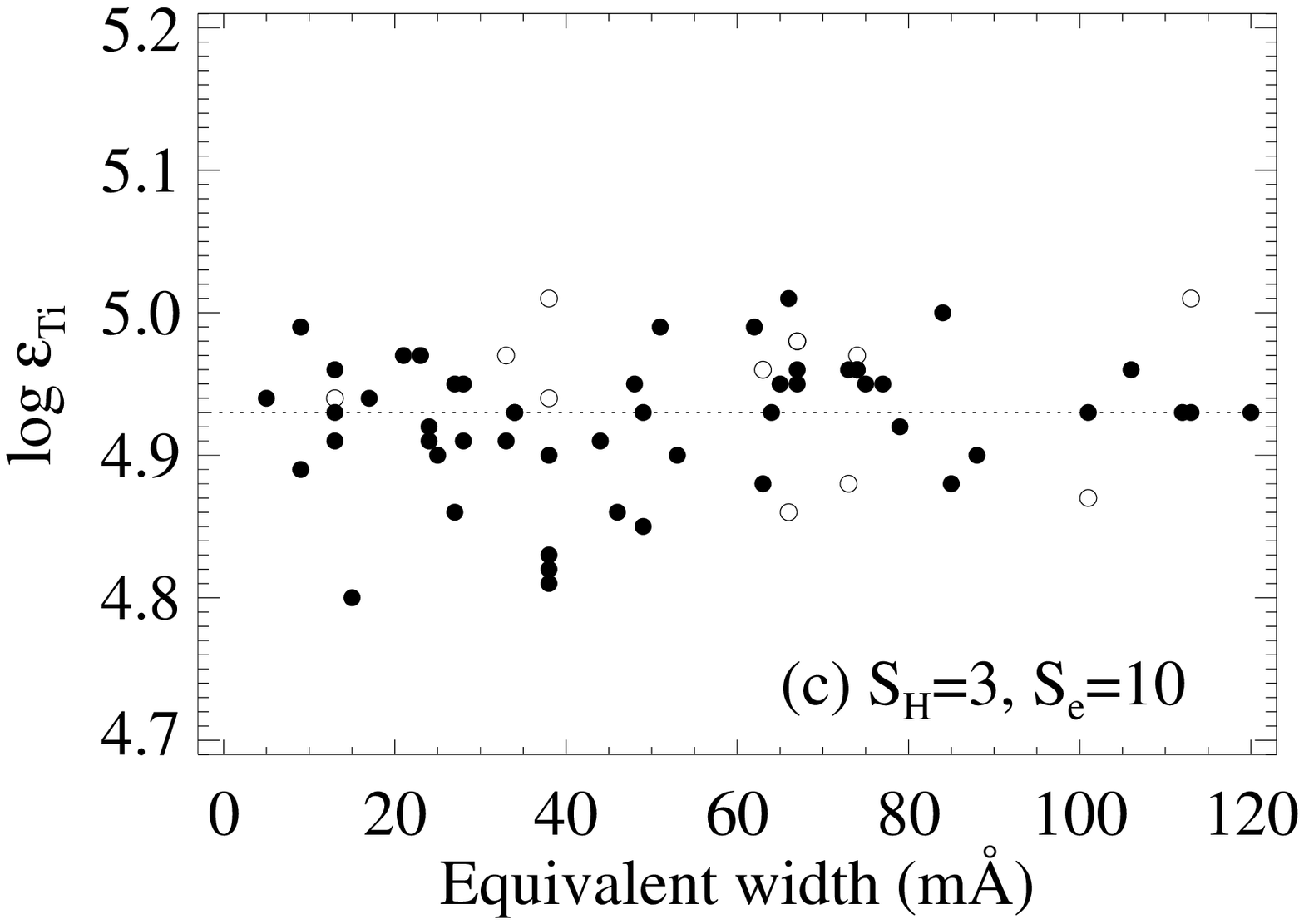}\hfill
\hspace{-6mm}
\includegraphics[scale=0.3]{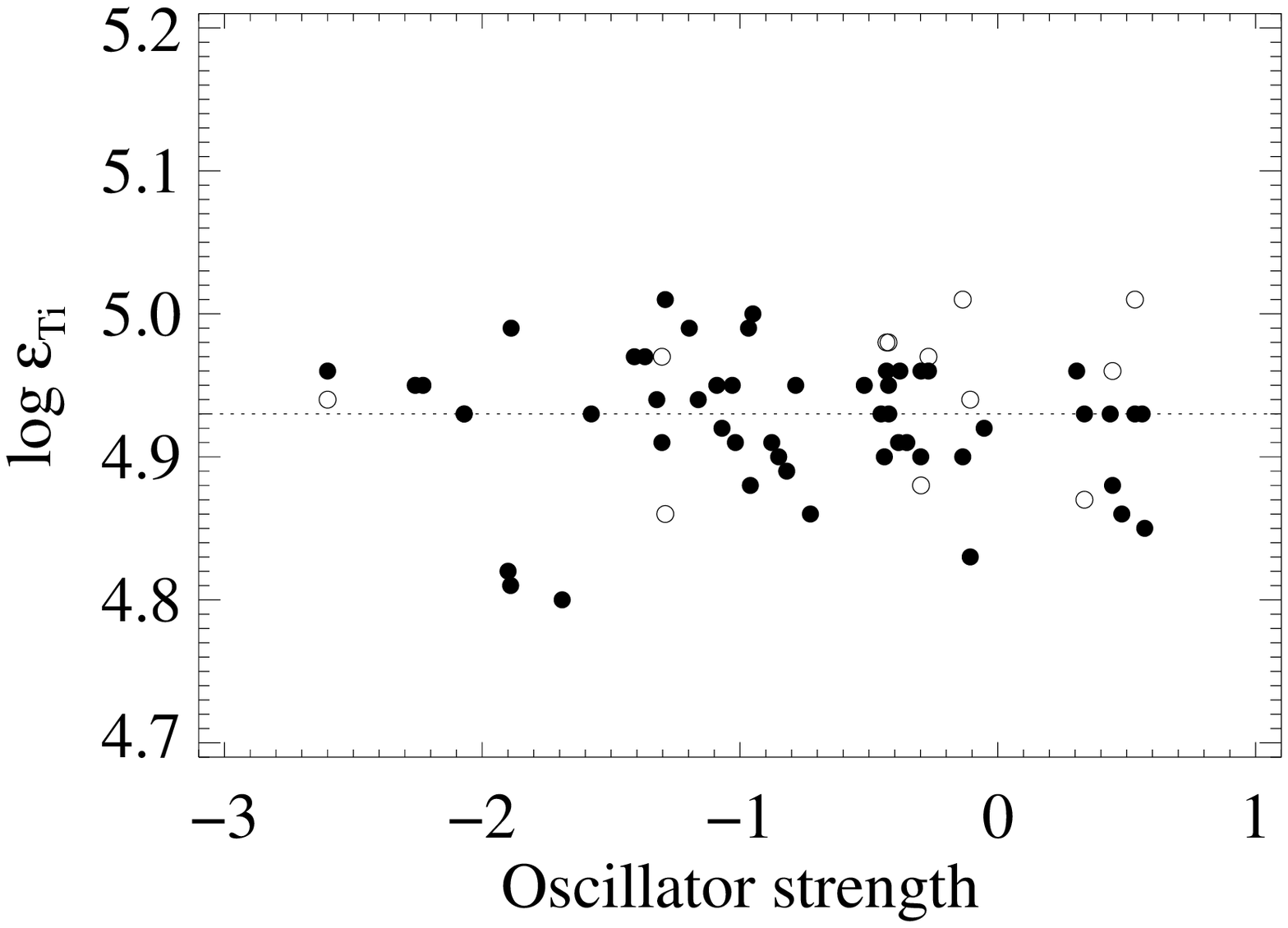}\hfill
\hspace{-6mm}
\includegraphics[scale=0.3]{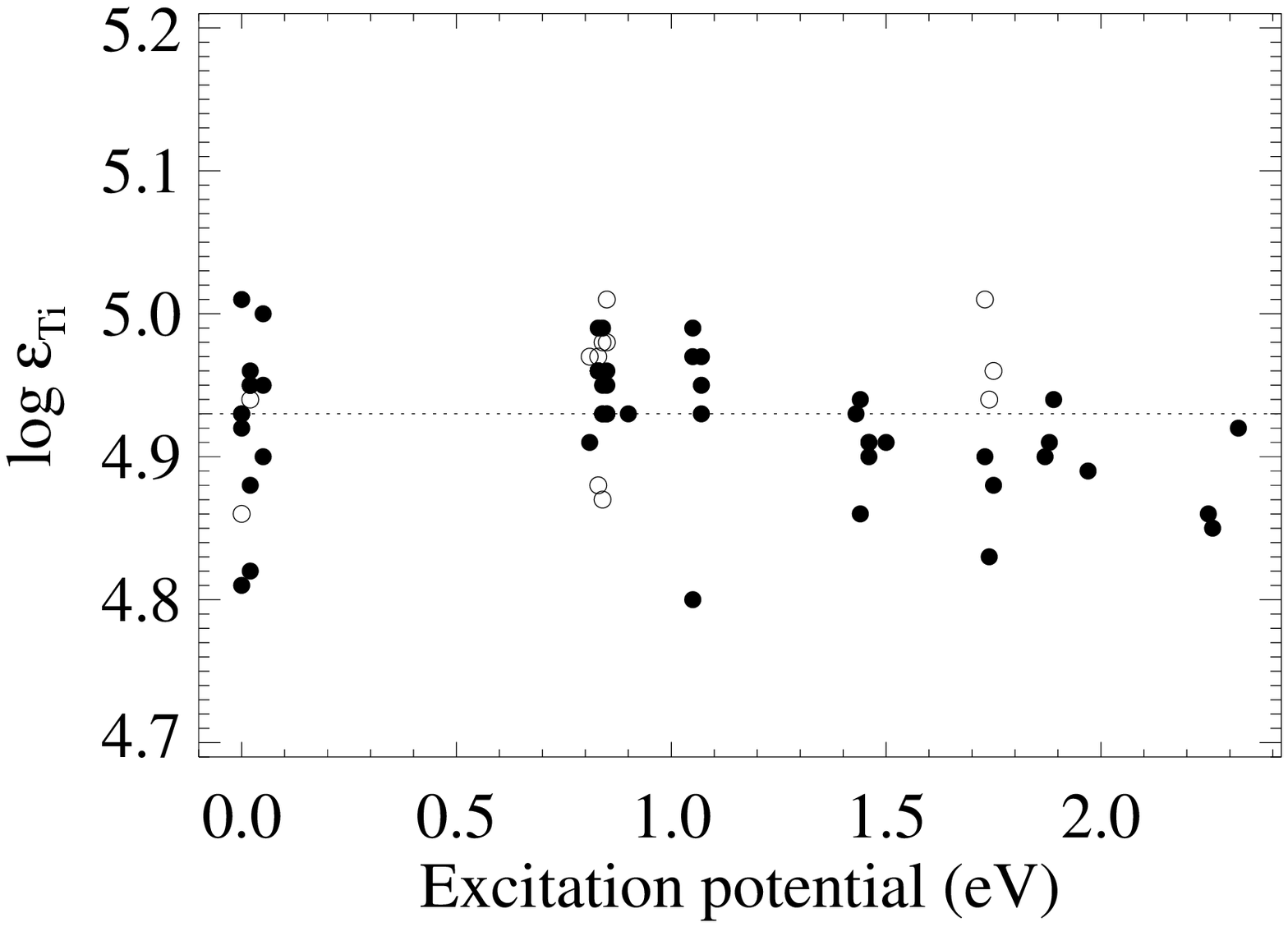}\hfill}
\vspace{-4mm}
\hbox{\includegraphics[scale=0.3]{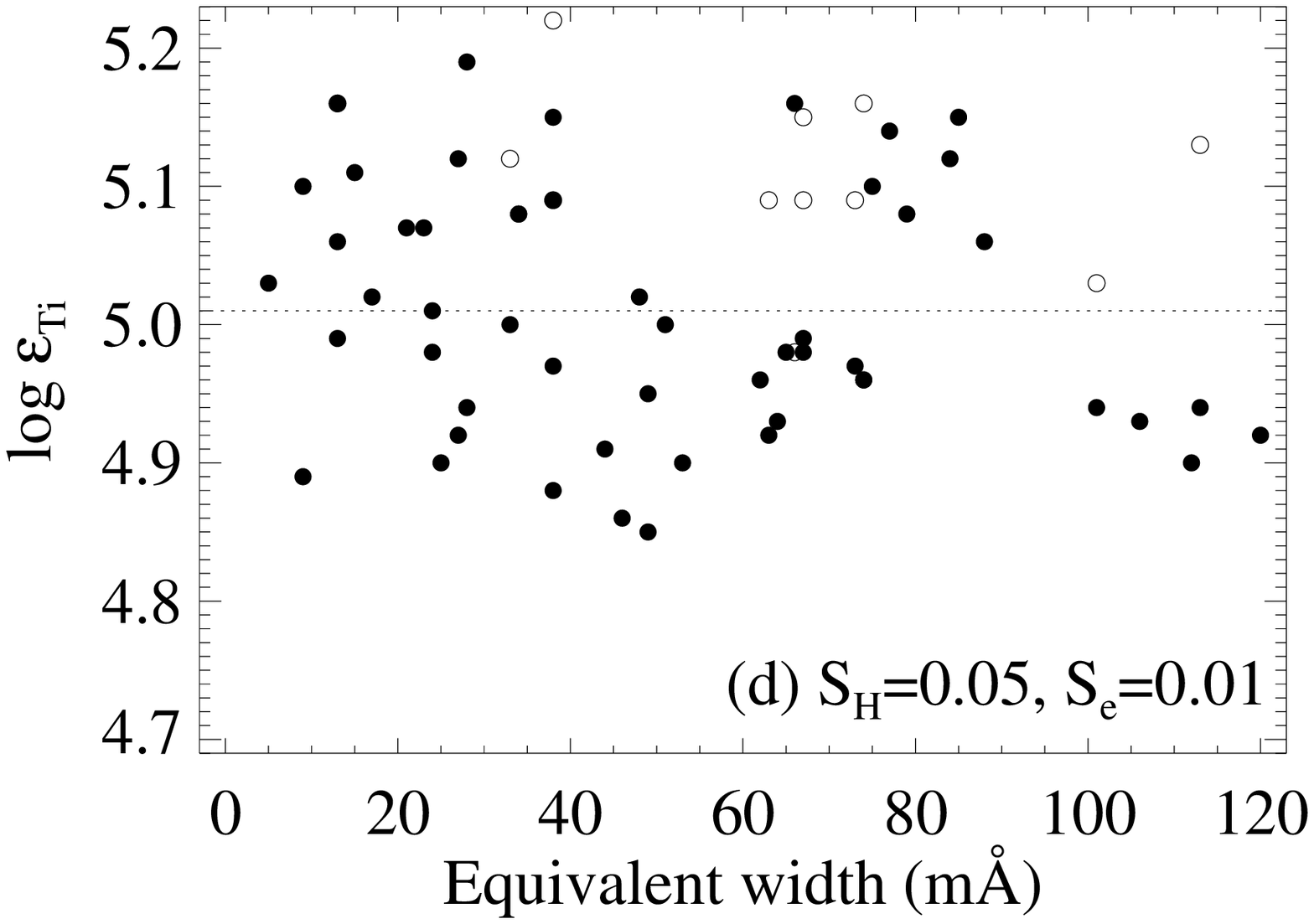}\hfill
\hspace{-6mm}
\includegraphics[scale=0.3]{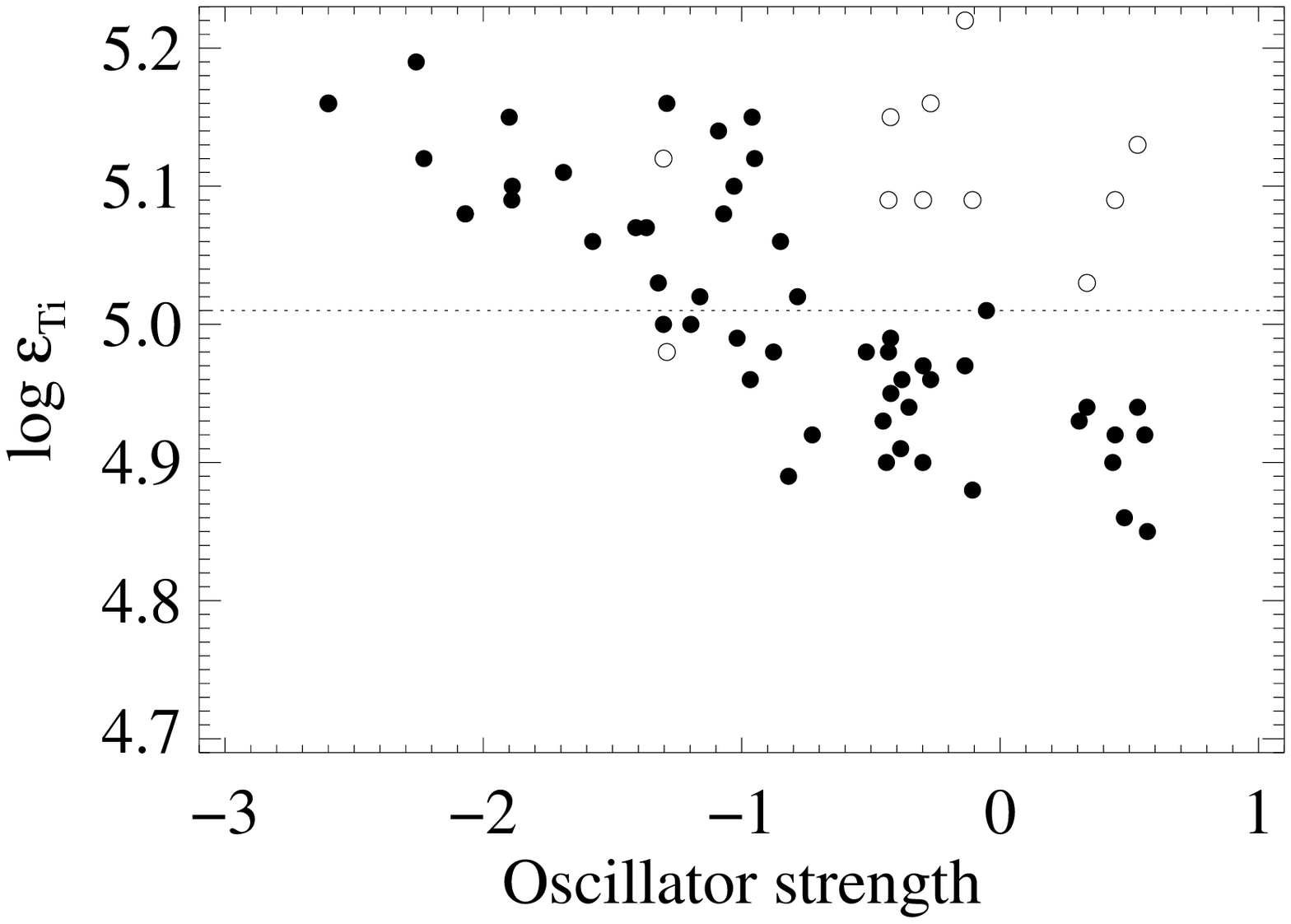}\hfill
\hspace{-6mm}
\includegraphics[scale=0.3]{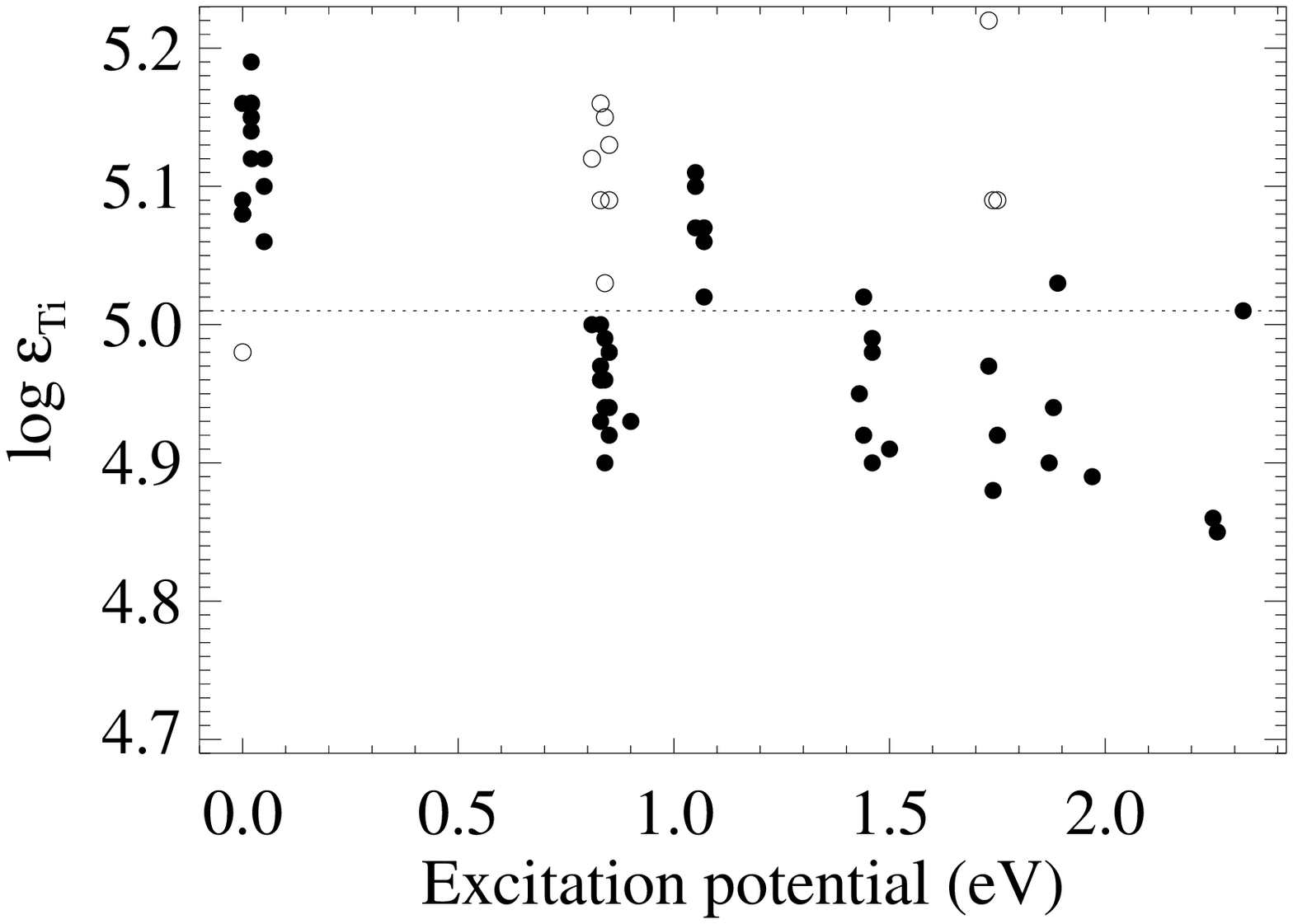}\hfill}
\vspace{-4mm}
\hbox{\includegraphics[scale=0.3]{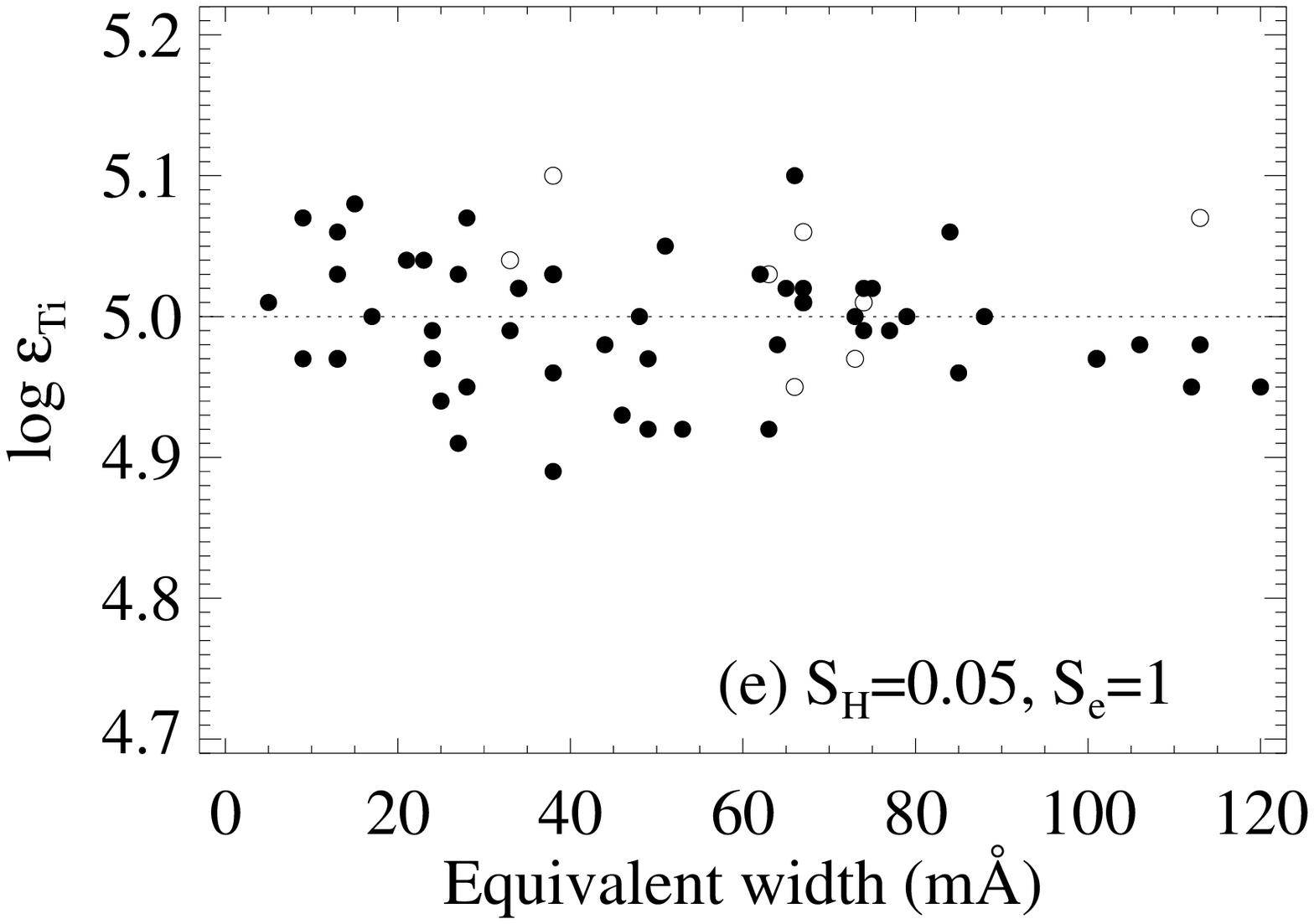}\hfill
\hspace{-6mm}
\includegraphics[scale=0.3]{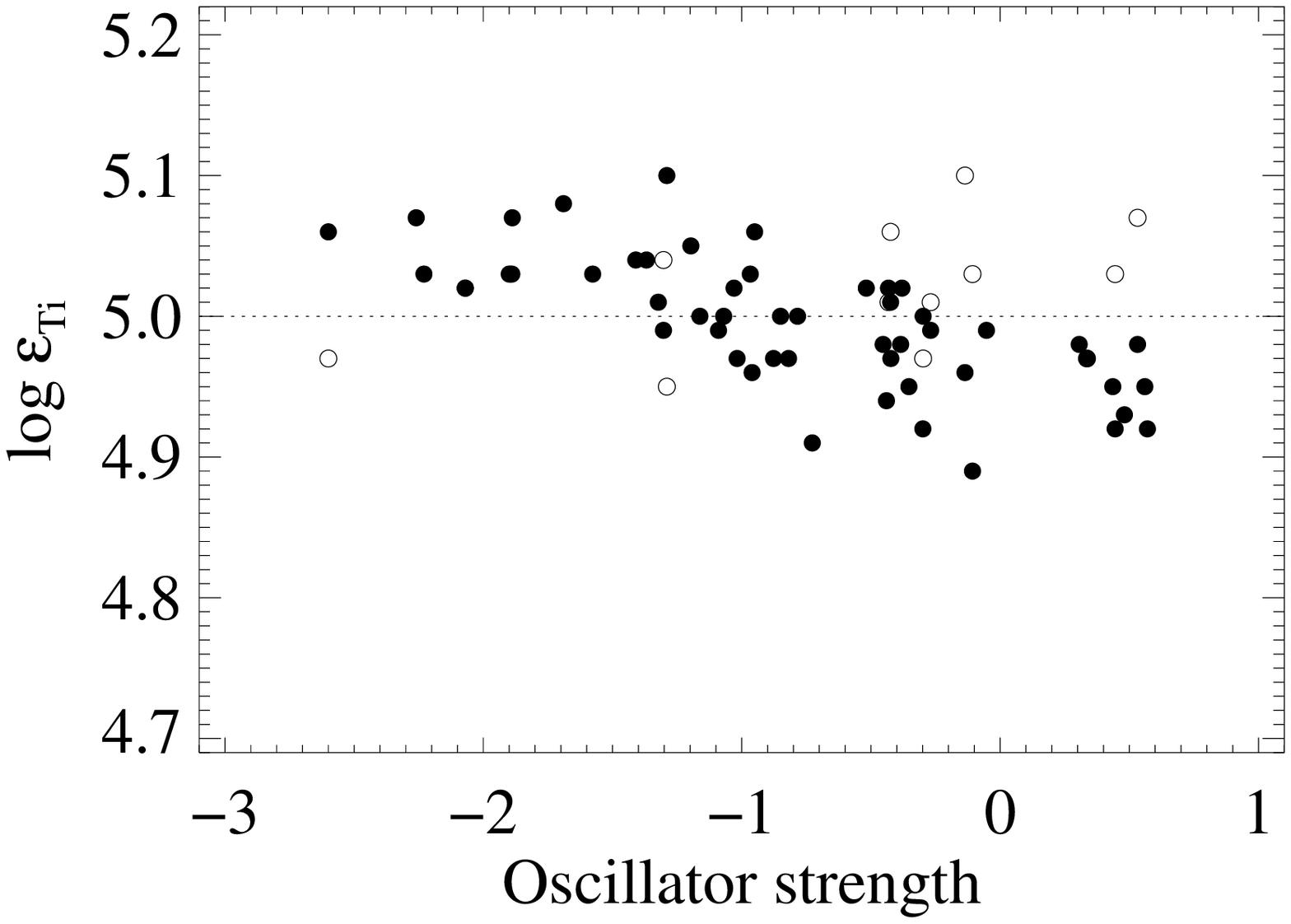}\hfill
\hspace{-6mm}
\includegraphics[scale=0.3]{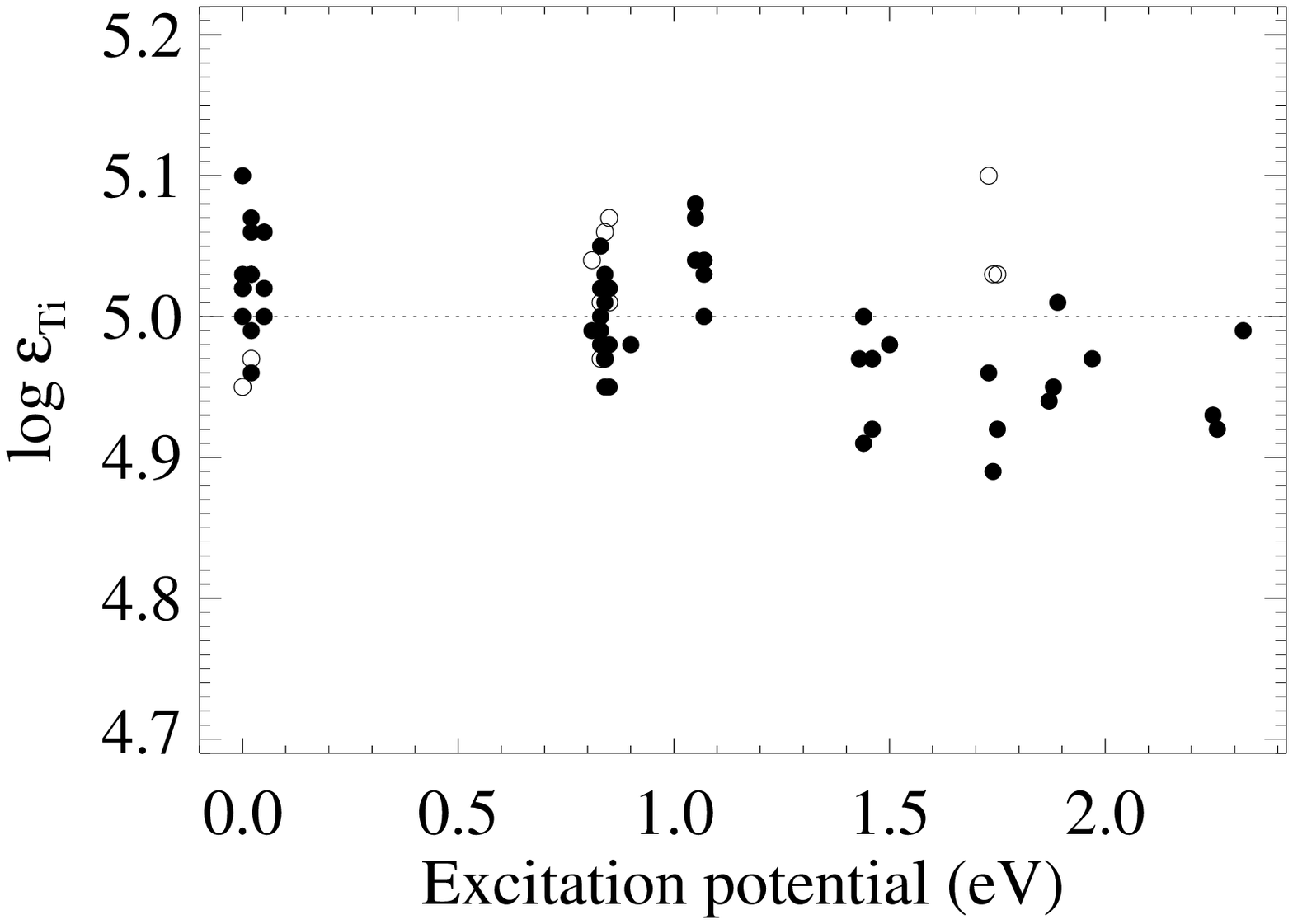}\hfill}
\vspace{-4mm}
\hbox{\includegraphics[scale=0.3]{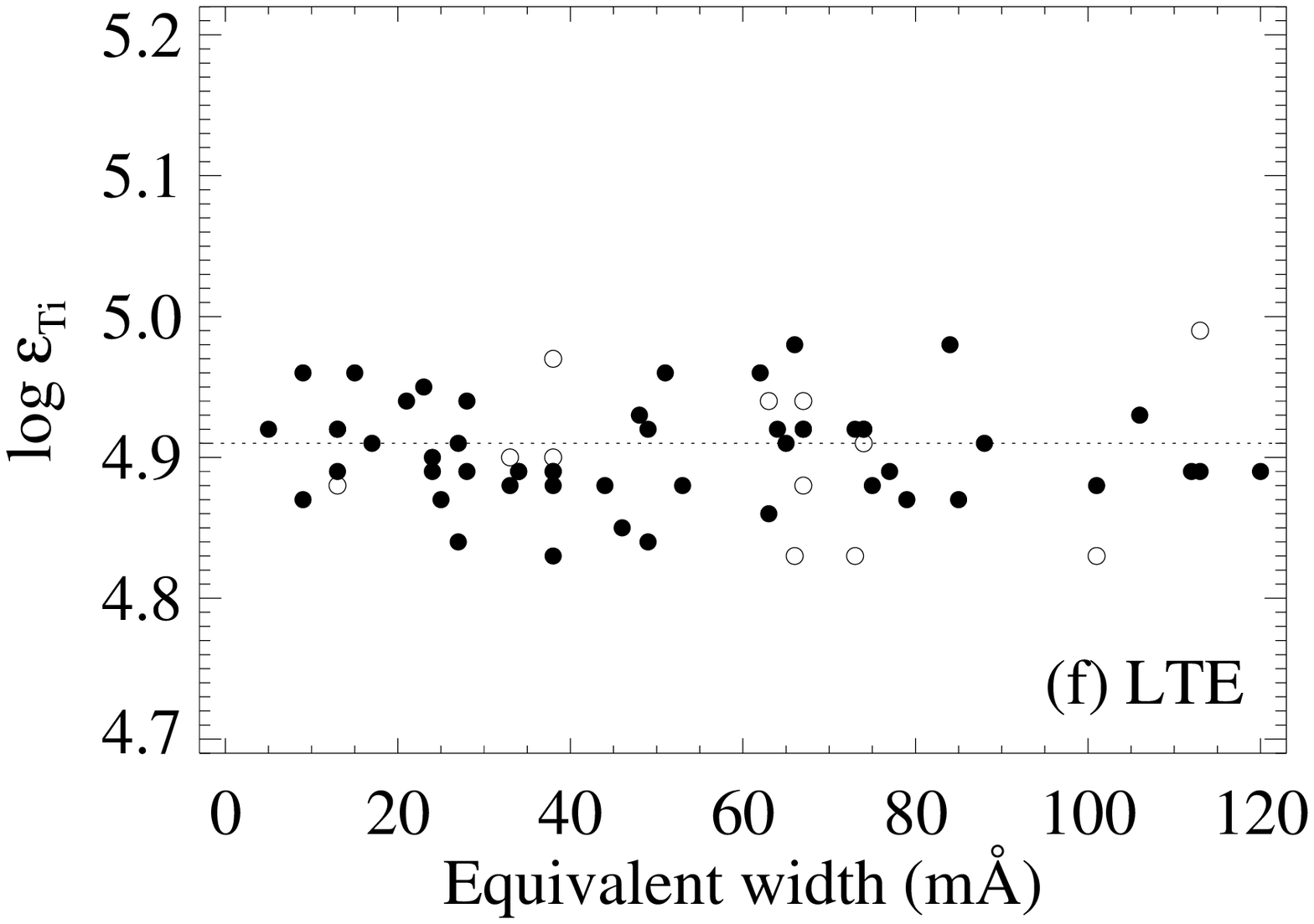}\hfill
\hspace{-6mm}
\includegraphics[scale=0.3]{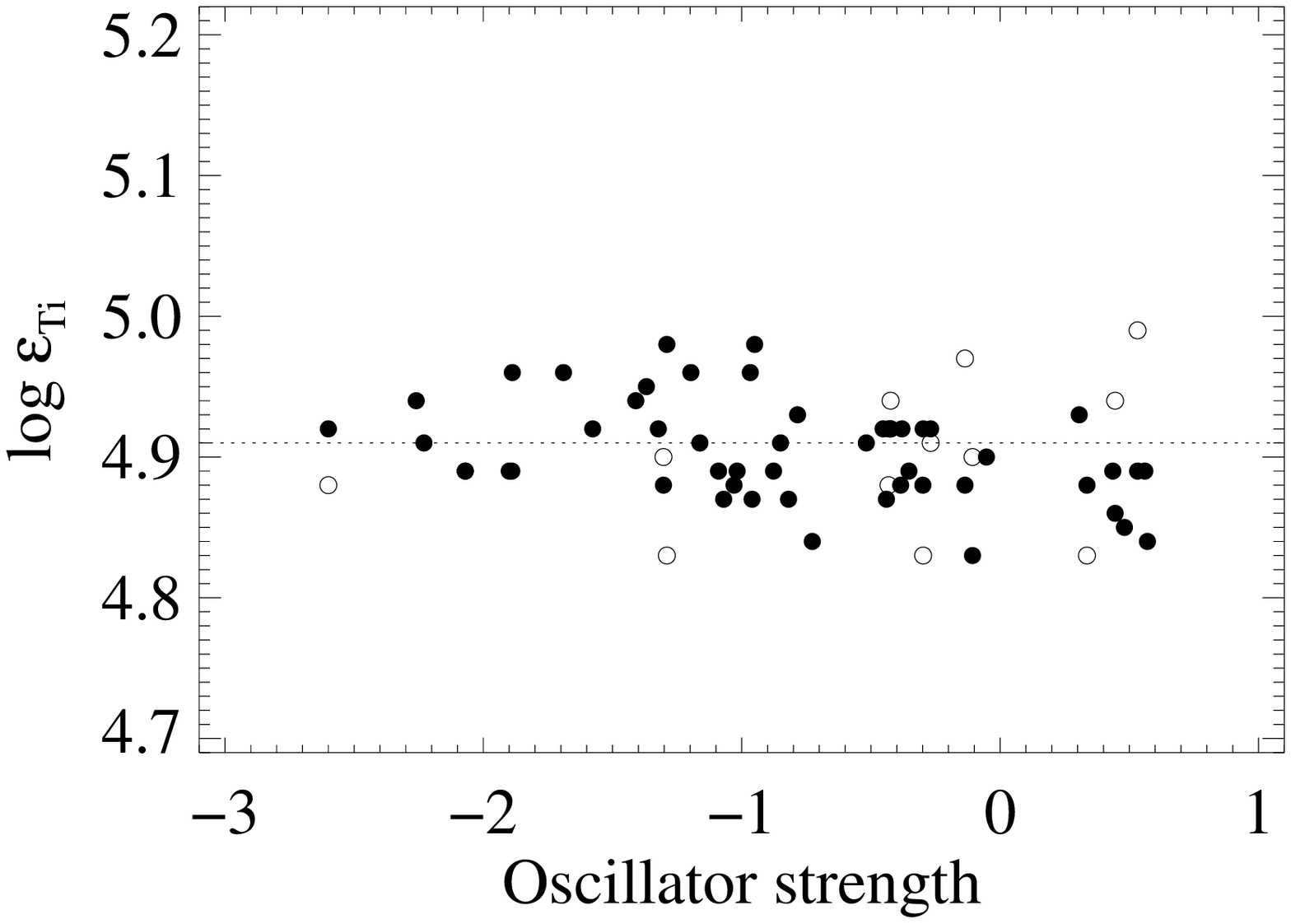}\hfill
\hspace{-6mm}
\includegraphics[scale=0.3]{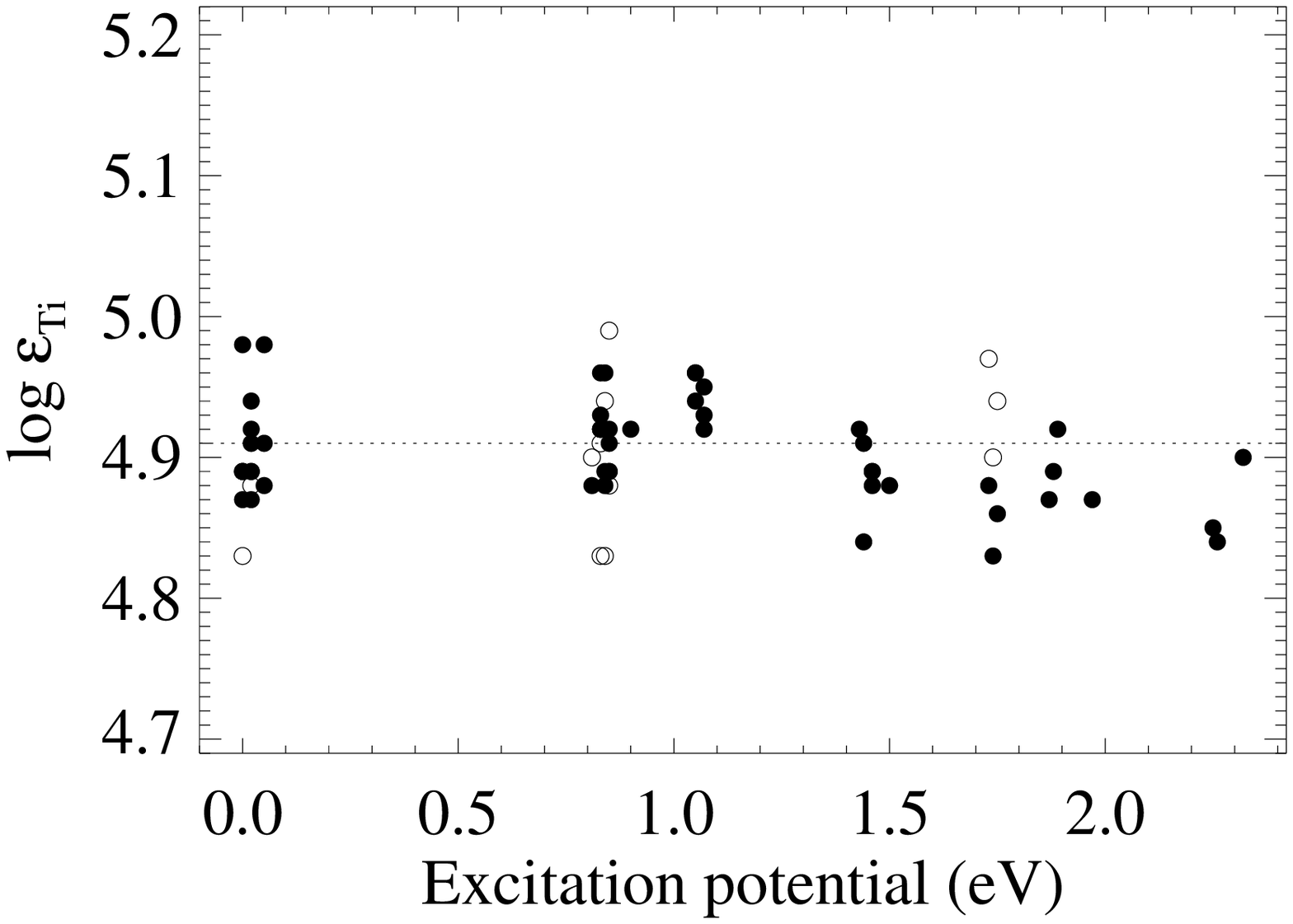}\hfill}
\vspace{-4mm}
\caption[]{NLTE Ti abundances computed with the MAFAGS-OS solar model and
different model atoms. Filled symbols denote the values computed with the $\log
gf$ set from Blackwell et al. (1982a, 1982b, 1983, 1986), and open symbols
correspond to the abundances derived with the data from
\citet{2006MNRAS.373.1603B}.}
\label{solar-abs}
\end{figure*}

NLTE abundances for each Ti I line derived with the MAFAGS-OS model atmosphere
and different atomic models (described in Sect. \ref{sec:modelatom}) are shown
in Fig. \ref{solar-abs} as a function of line equivalent width, oscillator
strength, and excitation potential of the lower level of the transition. The
combination of the scaling factors $\SH$ and $\Se$ used for each model atom is
indicated in the first panel of each row. The case of ($0.05, 10$) is omitted
because the results are almost identical to the case ($0.05, 1$) (Fig.
\ref{solar-abs}e). LTE results are also presented.
Filled symbols denote the values computed with the $\log gf$ set from Blackwell
et al. (1982a, 1982b, 1983, 1986), and open symbols correspond to the individual
abundances derived with the data from \citet{2006MNRAS.373.1603B}. There are
$12$ lines in common between the two experimental studies. Dashed lines show the
average NLTE abundance determined for each atomic model with the $gf$-values of
Blackwell et al., (\ref{abundances}). We do not present analogous plots for the
Ti II-based abundances because NLTE effects on Ti II in the solar case are
negligibly small, which is evident from Table \ref{abundances}.

Inspection of NLTE $\log \epsilon_{\rm Ti}$ values in Fig. \ref{solar-abs}
shows that NLTE abundances are very sensitive to the atomic data used in
statistical equilibrium calculations. For some combinations of $\SH$ and $\Se$,
e.g. ($3,0.01$)(Fig. \ref{solar-abs}a) and ($0.05,0.01$) (Fig. \ref{solar-abs}d)
the scatter between the Ti I lines is very large. The individual abundances
differ by a factor of two, although there are no physically significant trends
with the equivalent width. In addition, abundance seems to be correlated with
the oscillator strength and excitation potential of the lower level, if only the
$gf$-values of Blackwell et al. (1982a, 1982b, 1983, 1986) are considered. The
$gf$-$\log \epsilon$ and $\Elow$-$\log \epsilon$ slopes are larger for the cases
(a) and (d) that clearly gives rise to the large spread of abundance with $\EW$.
On the one side, this is an irrefutable argument against small scaling factors
to $e^-$ collision rates, e.g. $\Se = 0.01$.
One could also argue that small $\SH$ values are inappropriate, because they
lead to uncomfortably large solar Ti abundances (Table \ref{abundances}), $\geq
5.00$, which are not consistent with the meteoritic value\footnote{$\loge = 4.91
\pm 0.03$ dex \citep{2009ARA&A..47..481A}}. However, the latter depends on the
solar abundance of the reference element Si, which may
be subject to usual problems of spectroscopic methods as well
\citep{2000A&A...359..755A,2008A&A...486..303S}. A second possibility is that
the Ti model atom is incomplete in terms of number of levels and transitions.
\citet{2010IAUS..265..197M} have recently demonstrated that it is important to
include highly-excited \emph{predicted} levels and transitions between them in
statistical equilibrium calculations for Fe. Although these data also exist for
Ti, the number of predicted levels and transitions exceeds $14\,000$ and few
millions, respectively. Even more levels and radiative transitions are predicted
for Fe I: $\sim 37\,500$, respectively, $\sim 6$ millions. Such atomic models
are not tractable even with 1D NLTE codes and require efficient algorithms to
reduce them by combining the levels and lines. We are currently working on this.

It has to be kept in mind that photoionization cross-sections are also not
appropriate for Ti. However, in the absence of quantum-mechanical calculations,
it is impossible to predict whether the hydrogenic approximation over- or
underestimates the cross-sections. We conclude this in analogy to the similar
atom Cr I, for which the calculated photoionization cross-sections of
\citet{2009JQSRT.110.2148N} differ from the hydrogenic approximation by
orders of magnitude, being significantly larger for some levels and smaller for
others. In principle, the same is true for the electron and H I collision rates.
Comparison of R-matrix calculations for e$^-$ collisions for Ca II
\citep*{2007A&A...469.1203M} with the formula of \citet{1962ApJ...136..906V}
also reveals order of magnitude differences. What concerns inelastic H I
collision, compared to the Drawin's formulas, \textit{ab initio}
quantum-mechanical calculations predict significantly lower collision rates for
certain transitions of simple alkali atoms
\citep{2003PhRvA..68f2703B,2010A&A...519A..20B}, and they show that, in
addition to excitation, other effects like ion-pair formation become important.

There are also other explanations for the apparent trends of individual
abundances with the parameters of Ti I lines and/or large scatter, which are not
related to the deficiencies of NLTE modelling. First, one can question the
reliability of statistics based on small number of lines, i.e. there are just
few lines in our list with small $\log gf$'s. Also, the range of low-level
excitation potentials is limited to $0 - 2.5$ eV. Second, the uncertainties of
$gf$-values from Blackwell et al. (1982a, 1982b, 1983, 1986) might be
underestimated. It is interesting that the same regularity, i.e. variation of
abundance with multiplet, was also found in the LTE analysis of solar Ti I lines
by \citet[][their Fig. 4]{1987A&A...180..229B}, and it is present for the
individual Ti abundances based on the LTE assumption (Fig. \ref{solar-abs}f)
and/or the MAFAGS-ODF model atmosphere.
In addition to uncertainties in the $gf$ values measured with the Oxford
furnace, \citet{1987A&A...180..229B} also suggested that some Ti I lines are
affected by photospheric temperature and velocity fluctuations, which are very
pronounced in 3D hydrodynamical solar model atmospheres
\citep{2005ARA&A..43..481A}. Although this is most likely the case for
low-excitation temperature-sensitive lines, we do not see any trend of abundance
with line strength (Fig. \ref{solar-abs}a). The latter probably indicates that a
single depth-independent microturbulence is not a bad approximation. Otherwise,
the medium-strong lines ($\EW \approx 60 \ldots 100$) being very sensitive to
$\Vmic$ would show systematically higher (or lower) abundances.


The abundances derived from the NLTE and LTE profile fitting of the Ti II lines
are given in Table \ref{abundances}. Although the standard deviations of the
mean abundances, especially for the $gf$-values of \citet{2001ApJS..132..403P},
are quite large, there is no trend of individual line abundances with $\Elow$,
$\EW$ and $\log gf$ (not shown). As expected from the results of Sect.
\ref{sec:nlte_effect}, LTE abundances are slightly larger than that computed
under NLTE. The effect of the isotopic shift on the abundance determinations is
not negligible. For example, a single component used to fit the observed Ti II
line at $4488$ \AA, overestimates the abundance by $0.06$ dex compared to the
result obtained with $5$ isotopic components (Fig. \ref{profiles}d). To
demonstrate that the profile composed of several components is also slightly
asymmetric, the Ti II line was plotted excluding the contribution of blends
in the wings.

Some Ti II lines require very large abundances, $\loge \geq 5.05$ dex, to fit
their profiles in the solar flux spectrum. Few of them also demonstrate broad
asymmetric cores in the solar disk-center intensity spectrum, which can not be
reproduced even taking isotopic structure into account. One of these lines,
$4443$ \AA, with an isotopic shift of $\sim 380$ MHz \citep{2010PhyS...81f5301N}
is shown in Fig. \ref{profiles}e. Very specific profile shapes are probably not
due to convective motions \citep{2000A&A...359..729A}. Thus, we neglect such
lines in the solar analysis assuming that they contain unresolved blends in the
inner cores\footnote{The presence of blends can be revealed either by searching
for asymmetries in the observed line profiles or by inspecting theoretical line
lists. One problem is that often weak blends overlap with the core of a stronger
line under investigation that leaves the profile of the latter fully symmetric.
Another problem is that theoretical line lists are compiled from transitions
observed in laboratory conditions, which do not correspond to astrophysical
conditions. A comparison of observed and \emph{synthetic} spectra proves that
many observed lines still lack identification, which may well contribute to the
well-known ``missing UV opacity'' problem in the UV. It is natural to expect
missing opacity in other spectral regions as well.}. The synthetic profiles of
other discrepant lines systematically underestimate depths of both wings, e.g.
$4395$ \AA\ (Fig. \ref{profiles}f). Thus, the abundance estimate inferred from
such profile fits is rather subjective and depends, among other parameters, on
the adopted value for the macroturbulence, which otherwise would not affect
determination of the abundance being an external line broadening mechanism. We
refrain from using these lines in the solar analysis, also because they are
very sensitive to $\Vmic$.

In Table \ref{abundances}, it can be seen that MAFAGS-OS model and NLTE line
formation with $\SH = 3$ achieve good ionization equilibrium of Ti I/Ti II for
the Sun. For the given $\SH$, the mean abundances determined with $\Se = 0.01,
1, 10$ are also consistent with the Ti II-based abundances within their
respective uncertainties, however $\Se = 1$ provides the smallest
abundance spread for the solar case. With respect to the Ti I lines, almost
equal mean abundances were obtained for the $gf$-values from
\citet{2006MNRAS.373.1603B} and Blackwell et al. (1982a, 1982b, 1983, 1986), the
latter data scaled by $+0.056$ dex. However, our results for the Ti II lines
favour the $gf$-values of \citet{2001ApJS..132..403P}, which lead to a better
agreement between the two ionization states. The MAFAGS-OS NLTE abundances,
whether based on Ti I or Ti II lines, also agree well with the Ti abundance in C
I meteorites, $\loge = 4.91 \pm 0.03$ dex \citep{2009ARA&A..47..481A}.

\subsection{Ti ionization equilibrium for the metal-poor
stars}{\label{sec:stars}}
\subsubsection{Observations and stellar parameters}
To test ionization equilibrium of Ti in the atmospheres with reduced metal
content, we have chosen four stars from the sample of
\citet{2008A&A...492..823B}. The reader is referred to this paper for a detailed
description of observations. The spectra of HD 84937, HD 140283, and HD 122563
were taken from the UVESPOP survey \citep{2003Msngr.114...10B}. The spectra of
HD 102200 were taken by T. Gehren and colleagues on the ESO UVES echelle
spectrograph at the VLT UT2 in Paranal, Chile. The UVES spectra have a
slit-determined resolution of $\lambda/\Delta\lambda \sim 50\,000$ and a
signal-to-noise ratio better than $S/N \sim 300$ near $5000$ \AA.

Stellar parameters were adopted from
\citet{2004A&A...413.1045G,2006A&A...451.1065G} for HD 84937, HD 140283, and HD
102200, and from \citet{2008A&A...478..529M} for HD 122563. These studies used
the same observed spectra, model atmospheres (MAFAGS-ODF), and the codes as
adopted in this work that secures consistency of our abundance analysis. The
effective temperatures $\Teff$ were determined from Balmer line profile fits,
and gravities $\log g$ were based on \emph{Hipparcos} parallaxes from the
\citet{1997yCat.1239....0E} catalogue. The spectroscopic temperatures are
consistent with $\Teff$'s determined by \citet{2010A&A...512A..54C} using the
Infrared Flux method. \citet{2010A&A...512A..54C} derived $\Teff = 6155$ K for
HD 102200, $\Teff = 5777$ K for HD 140283, and $\Teff = 6408$ K for HD 84937.
The iron abundance and microturbulence velocities were derived from LTE fitting
of Fe II lines, which do not suffer from NLTE effects for the range of stellar
parameters investigated here \citep{2010IAUS..265..197M}.

MAFAGS-OS model atmospheres are available only for HD 84937 and HD 122563 (L.
Mashonkina, private communication). The gravities adopted for the MAFAGS-OS
models are on average $\sim 0.1$ dex larger than that used to compute the
MAFAGS-ODF model atmospheres. The reason is that new reduction of
\emph{Hipparcos} data \citep*{2007A&A...474..653V} revealed some errors in the
earlier version of the catalogue \citep{1997yCat.1239....0E}. However, for our
stars the parallax differences between both catalogue versions are within the
parallax error itself, which propagates into $0.1$ dex error in $\log g$.

Stellar parameters are given in Table \ref{stel_param}. The estimated errors are
$100$ K for $\Teff$, $0.1$ dex for $\log g$, $0.1$ dex for [Fe/H], and $0.2$
km/s for $\Vmic$.

\begin{table}
\caption{Stellar parameters for the selected sample. See text.}
\renewcommand{\footnoterule}{} 
\tabcolsep1.0mm \small
\label{stel_param}
\begin{tabular}{lcccccccc}
\hline\noalign{\smallskip}
Object & \multicolumn{3}{c}{MAFAGS-ODF} & \multicolumn{3}{c}{MAFAGS-OS} &
 $\xi_{\rm t}$ & [$\alpha$/Fe] \\ 
~~ & $\Teff$ & $\log g$ & [Fe/H] & $\Teff$ & $\log g$ & [Fe/H] & km/s & \\
\noalign{\smallskip}\hline\noalign{\smallskip}
HD 84937  & 6350 & 4.00 & $-2.16$ & 6350 & 4.09 & $-2.15$ &  1.8 & 0.4\\
HD 102200 & 6120 & 4.17 & $-1.28$ &      &      &         &  1.4 & 0.4\\
HD 122563 & 4600 & 1.50 & $-2.51$ & 4600 & 1.60 & $-2.50$ &  1.9 & 0.2\\
HD 140283 & 5773 & 3.66 & $-2.38$ &      &      &         &  1.5 & 0.4\\
\noalign{\smallskip}\hline\noalign{\smallskip}
\end{tabular}
\end{table}

\subsubsection{Ti abundances in the metal-poor stars}{\label{sec:stars}}

The analysis of the metal-poor stars is differential with respect to the Sun. An
individual abundance derived for each line in a stellar spectrum is related
to the solar abundance from that line computed with the same parameters and
assumptions in the spectrum synthesis, including model atmosphere type (OS or
ODF). The Ti abundance in a star relative to the Sun is given by:
\[
\mathrm{[Ti/H]}= \loggfestar - \loggfesun 
\]
We also used several Ti II lines ($4394.05$, $4395.85$, $4443.8$ \AA), which
have been excluded from the solar abundance analysis because of profile
distortions, which are likely caused by blends (Fig. \ref{profiles}e) and
sensitivity of profile wings to microturbulence (Fig. \ref{profiles}f).
However, in spectra of very metal-poor stars the lines are much weaker and
appear to be symmetric showing no evidence for blends. Here, we discuss the
stellar Ti abundances normalized to the Fe abundance of a star, [Ti/Fe]; this
parameter is relevant to Galactic chemical evolution studies.
\begin{figure}
\begin{center}
\includegraphics[scale=0.3]{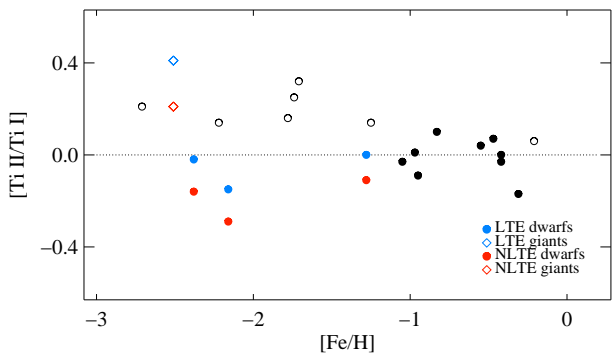}
\hspace{-5mm}
\caption[]{NLTE (red symbols) and LTE (blue symbols) abundances of Ti in the
metal-poor stars. For comparison, the LTE abundance ratios of
\citet{1991A&A...241..501G} are shown with open and filled black symbols,
respectively.}
\label{ab_comp}
\end{center}
\end{figure}

The Ti abundances determined for four metal-poor stars are given in Table
\ref{ab_stars}. Fig. \ref{ab_comp} displays the difference between [Ti/Fe]
ratios computed from the lines of two ionization stages, [Ti$_{\rm II}$/Ti$_{\rm
I}$] $=$ [Ti$_{\rm II}$/Fe] - [Ti$_{\rm I}$/Fe], as a function of stellar
metallicity. NLTE and LTE ratios for the metal-poor stars are shown with
red and blue symbols, respectively. The figure also includes LTE Ti abundances
in metal-poor giants and dwarfs from the study of  \citet{1991A&A...241..501G}.
They define stars with $\log g < 3$ giants.

Our LTE abundances based on Ti I and Ti II lines agree within their respective
uncertainties for three stars, except for the very metal-poor giant HD 122563.
In contrast, ionization equilibrium is not satisfied under NLTE with $\SH = 3$
and $\Se = 1$. Except for the metal-poor giant, the Ti I lines give
systematically higher abundances than Ti II lines. Assuming any other
combination of collision scaling factors makes the discrepancy for the three
stars worse because NLTE effects on Ti I increase more for the metal-poor stars
than for the Sun. Variation of $\Se$ has almost no effect on the Ti abundances,
whereas decreasing the efficiency of hydrogen collisions by few orders of
magnitude, $\SH = 0.05$, leads to extreme overionization from the lowest Ti I
states. This is expected because concentration of free electrons in metal-poor
atmospheres is very low and thermalization of levels can only occur due to H I -
atom collisions. As a result, the NLTE abundances determined with $\SH = 0.05$
are on average $0.2$ dex larger than those computed with $\SH = 3$, and the
difference between the latter and the LTE-based abundances is of the same order.

For HD 84937, our Ti II-based LTE abundance is $0.15 $ dex lower than the LTE
abundance determined from the Ti I lines. This discrepancy has no relation to
NLTE effects, because there is no physical mechanism to produce overpopulation
of low-lying Ti I levels at expense of Ti II in the atmospheres, where Ti I is
the minor ionization stage. For HD 122563, we encounter a different problem: the
NLTE abundance determined from the Ti I lines is $0.2$ dex lower than that from
the Ti II, although the discrepancy is twice as small compared to LTE.

Thus, it appears that our reference NLTE model overestimates NLTE effects on Ti
in high-gravity atmospheres, corresponding to dwarfs, however it underestimates
overionization in the atmospheres of giants. For metal-poor dwarfs and
subgiants, nearly thermalized occupation numbers for Ti I levels are
necessary to satisfy ionization balance. As seen in Fig. \ref{ab_comp}, the
results of \citet{1991A&A...241..501G} support this conclusion. In their LTE
analysis of metal-poor stars, the mean difference between ionized and neutral
species is $+0.07 \pm 0.03$ dex for dwarfs and $+0.2 \pm 0.03$ dex for giants. 
As noted above (Sect. \ref{sec:intro}), similar offsets for metal-poor giants
were also reported in the LTE studies by \citet{1983ApJ...265L..93B},
\citet{2002ApJS..139..219J}, and \citet{2010arXiv1008.3721T}.

\subsubsection{Discussion}{\label{sec:discussion}}

The analysis of four metal-poor stars showed that the LTE approximation in Ti
line formation calculations is not adequate for evolved stars. However our NLTE
models with available atomic data for Ti do not perform better if both evolved
and unevolved stars are considered. In fact, there are few reasons why the
problem can still be attributed to deficiencies of the NLTE model.

All lines detected in the stellar spectra originate from the metastable levels
\Ti{a}{5}{F}{}{3,4,5} with the excitation energy $\sim 0.8$ eV. These levels,
among many other low-lying states, dominate ionization balance in Ti I. Thus,
inadequate results may reflect erroneous photoionization cross-sections, which
are computed in the hydrogenic approximation. We have performed a test
decreasing the cross-sections for several low-excitation Ti I states, including
\Ti{a}{5}{F}{}{}, by three orders of magnitude. Although the individual line
abundances decrease by $\sim 0.03$ dex, this is by far insufficient to bring two
ionization stages in agreement. We conclude that it is a cumulative effect of
non-hydrogenic photoionization in Ti I, rather than cross-sections for
individual atomic levels, that may lead to a quantitatively different
distribution of atomic level populations. In the absence of quantum-mechanical
calculations, it is impossible to predict whether the hydrogenic approximation
over- or underestimates the cross-sections for each Ti I level. Therefore, we
can not explore this possibility further.

If the problem is in NLTE, then it is likely that NLTE modelling of a similar
atom, Fe, suffers from similar deficiencies as Ti. For example, accurate
cross-sections for transitions caused by inelastic H I and e$^-$ collisions are
missing for both atoms, and calibration of collision efficiency on observed
spectra is necessary. In this case, one would expect the problem to be reduced
or even eliminated if one compares the abundances determined from lines of equal
ionization stages, i.e. Ti I and Fe I. The NLTE abundances based on Fe I lines
and MAFAGS-OS model
atmospheres\footnote{\citet{2004A&A...413.1045G,2006A&A...451.1065G} do not
provide estimates of NLTE Fe abundances based on Fe I lines and MAFAGS-ODF model
atmospheres for the stars in our sample.} are available only for HD 84937 and HD
122563, [Fe$_{\rm I}$/H]$ = -2.00 \pm 0.07$, respectively, [Fe${\rm _I}$/H]$ =
-2.61 \pm 0.09$ (L. Mashonkina, private communication). The former value is
$0.15$ dex higher than our standard Fe II-based value for the metal-poor
subdwarf, [Fe/H]$ = -2.15$. For the metal-poor giant, the NLTE Fe
abundance determined from Fe I lines is $0.11$ dex lower than the reference
Fe II-based value [Fe/H]$ = -2.50$. It is interesting that Fe
abundances derived from the NLTE analysis of Fe I lines are slightly discrepant
with that derived from Fe II lines, and the sign of discrepancy is the same
as for Ti both for the dwarf and giant stars. As a result, the [Ti/Fe] ratios
for the two metal-poor stars determined from the NLTE analysis of Ti I and Fe I
lines agree much better with that obtained from Ti II and Fe II lines. For a
subdwarf HD 84937, we obtain [Ti${\rm _I}$/Fe$_{\rm I}$]$ = +0.44 \pm 0.02$,
which is consistent with [Ti$_{\rm II}$/Fe$_{\rm II}$]$ = +0.35 \pm 0.07$ within
the respective uncertainties of both values. For a giant HD 122563, [Ti${\rm
_I}$/Fe$_{\rm I}$]$ = +0.23 \pm 0.15$ and [Ti$_{\rm II}$/Fe$_{\rm II}$]$ = +0.22
\pm 0.02$. Although the standard deviations of abundances are still large, the
NLTE approach in this case has the advantage that it minimizes offsets between
two ionization stages of Ti for both evolved and main sequence stars.

There is also some similarity between our results, i.e. an apparent
need for the high degree of thermalization of Ti I levels in the atmospheres of
dwarfs, and that of \citet*{2003A&A...407..691K}, who investigated kinetic
equilibrium of Fe in metal-poor stars with the same methods, model atmospheres,
and codes, as we do. They find that a very large scaling factor to Drawin's H I
collision rates, $\SH = 3$, in addition to the enforced thermalization of the
upper Fe I levels \citep{2001A&A...366..981G}, is necessary to fullfill
ionization equilibrium of Fe in their sample of dwarfs and subgiants with [Fe/H]
$< -1$. Their results also indicate that NLTE effects on Fe II are negligible,
which is supported by the results of Mashonkina et al. (2010) for Fe and is also
true for Ti II (Table \ref{ab_stars}).

We conclude that the ratios [Ti/Fe] in a stellar atmosphere computed exclusively
using Ti II and Fe II lines should be robust, at least in the framework of 1D
modelling, and should be used for Galactic chemical evolution studies,
whenever possible. It is strongly recommended to avoid Ti I lines, as well as to
take average of two ionization stages, in abundance analyses and in
determination of surface gravity for evolved stars. Otherwise, systematic
offsets between giants and dwarfs are unavoidable and will produce spurious
abundance trends with metallicity.

\begin{table*}
\caption{Ti abundances for the selected sample.}
\small
\tabcolsep1.0mm
\label{ab_stars}
\begin{center}
\begin{tabular}{lcl|rr@{$\,\pm\,$}lr@{$\,\pm\,$}l|rr@{$\,\pm\,$}lr@{$\,\pm\,$}
l|r@{$\,\pm\,$}lr@{$\,\pm\,$}l|r@{$\,\pm\,$}lr@{$\,\pm\,$}l}
\hline\noalign{\smallskip}
Object & [Fe/H] & [Mg/Fe] & $N_{\rm{Ti I}}$ & \multicolumn{4}{c}{[Ti$_{\rm
I}$/Fe] ODF} & $N_{\rm{Ti II}}$ & \multicolumn{4}{c}{[Ti$_{\rm II}$/Fe] ODF} &
\multicolumn{4}{c}{[Ti$_{\rm I}$/Fe] OS} & \multicolumn{4}{c}{[Ti$_{\rm
II}$/Fe] OS} \\
  &  &  &  & \multicolumn{2}{c}{LTE} & \multicolumn{2}{c}{NLTE} & & 
\multicolumn{2}{c}{LTE} & \multicolumn{2}{c}{NLTE} & \multicolumn{2}{c}{LTE} &
\multicolumn{2}{c}{NLTE} & \multicolumn{2}{c}{LTE} & \multicolumn{2}{c}{NLTE} \\
\noalign{\smallskip}\hline\noalign{\smallskip}
HD 84937  & $-2.16$ & 0.32 & 7 & $0.49$ & $0.02$ & $0.63$ & $0.02$  & 6 & $0.34$
& $0.08$ & $0.34$ & $0.09$ & $0.46$ & $0.02$ & $0.59$ & $0.02$ & $0.35$ & $0.06$
& $0.35$ & $0.07$ \\
HD 102200 & $-1.28$ & 0.34 & 16 & $0.28$ & $0.03$ & $0.39$ & $0.05$ & 11 &
$0.28$ & $0.05$ & $0.28$ & $ 0.04$ \\
HD 140283 & $-2.38$ & 0.43 & 5 & $0.19$ & $0.03$ & $0.34$ & $0.02$  & 6 & $0.17$
& $0.07$ & $0.18$ & $0.08$ \\
HD 122563 & $-2.51$ & 0.45 & 5 & $-0.12$ & $0.07$ & $0.13$ & $0.11$ & 3 & $0.29$
& $0.03$ & $0.34$ & $0.03$ & $-0.19$ & $0.6$ & $0.12$ & $0.15$ & $0.21$ & $0.02$
& $0.22$ & $0.02$ \\
\noalign{\smallskip}\hline\noalign{\smallskip}
\end{tabular}
\end{center}
\end{table*}

\section{Summary}{\label{sec:summary}}

Statistical equilibrium of Ti was computed for a restricted range of stellar
parameters, focussing on late-type stars. The Ti model atom was constructed
with available experimental atomic data for levels and radiative transitions.
Photoionization was assumed hydrogenic, except for the few Ti I levels, for
which we adopted experimental cross-sections from \citet{2009ChJCP..22..615Y}.
Inelastic H I collision rates were computed following the classical
\citet{1968ZPhy..211..404D} formalism in the version of
\citet{1984A&A...130..319S}. The Drawin's cross-sections were increased by a
factor of three, $\SH =3$, because this value produced the smallest abundance
scatter and satisfied ionization equilibrium of Ti I/Ti II for the Sun.

NLTE effects on Ti I level populations and lines are significant for any
combination of $\Teff$, $\log g$, and [Fe/H]. The dominant effect is
overionization that makes NLTE particularly important at low metallicity. 
In the solar case, LTE approach underestimates the abundances by $0.05 - 0.1$
dex. 
The solar Ti abundances derived from NLTE profile fitting of Ti I and Ti II
lines with MAFAGS-OS theoretical model atmosphere are $\log \epsilon = 4.94 \pm
0.05$ dex \citep[$\log gf$'s from][]{2006MNRAS.373.1603B} and $\log
\epsilon = 4.95 \pm 0.06$ dex \citep[$\log gf$'s from][]{2001ApJS..132..403P}.
Both values agree well with the Ti abundance in C I meteorites, $\loge = 4.91
\pm 0.03$ dex \citep{2009ARA&A..47..481A}. The results obtained with $gf$-values
for Ti I transitions measured with Oxford furnace
\citep[][and earlier references]{1986MNRAS.220..289B} are identical, $\log
\epsilon = 4.93 \pm 0.04$ dex. The $gf$-values for Ti II transitions from
\citet{1993A&A...273..707B} lead to $\sim 0.04$ dex larger solar Ti abundances.
MAFAGS-ODF models and/or LTE approach do not perform as good as MAFAGS-OS and
NLTE with calibrated collision efficiency.

Ti abundances were computed for four metal-poor stars: HD 102200, HD 84937, HD
140283, HD 122563. LTE approach leads to an agreement between Ti I and Ti
II-based abundances for unevolved stars, however it produces large discrepancy
for giants. The same problem was also reported in other LTE studies of Ti
abundances in metal-poor stars. Under NLTE, the discrepancy between
the two ionization stages is significantly reduced for giants. However, now
the Ti I lines give higher abundances than the Ti II lines for dwarfs and
subgiants. This failure should by no means be seen as invalidity of the NLTE
approach, but rather as a manifestation of inaccurate or missing atomic data. It
is probably important to include highly-excited predicted levels and transitions
between them in the model atom. In addition, accurate photoionization
cross-sections and cross-sections for inelastic collisions with H I atoms are
urgently needed.

One of 'temporary' solutions might be to relate NLTE abundances determined
from lines of equal ionization stages, i.e. Ti I and Fe I. In our preliminary
tests, the [Ti/Fe] ratios for the metal-poor dwarf HD 84937 and giant
HD 122563 determined from the NLTE analysis of Ti I and Fe I lines agree with
those obtained from Ti II and Fe II lines. However, the standard deviations of
abundances are not small. Also, one has to keep in mind the remaining
discrepancy between Fe I and Fe II lines.

At present, we strongly recommend to disregard Ti I lines in abundance analyses
of giants, as well as in calibration of their surface gravities. Since the NLTE
effects on Ti II and Fe II are small for the range of stellar parameters
investigated in this work, the ratios [Ti/Fe] computed exclusively
using Ti II and Fe II lines should be robust and should be used for Galactic
chemical evolution studies.

\section*{Acknowledgments}
This work is based on observations collected at the European Southern
Observatory, Chile, 67.D-0086A, and the Calar Alto Observatory, Spain. We thank
Dr. Lyudmila Mashonkina for the MAFAGS-OS model atmospheres for selected stars.
We thank Dr. Aldo Serenelli and Prof. Martin Asplund for revision of the
manuscript and interesting suggestions. The detailed review by an anonymous
referee helped to significantly improve the paper.

\bibliographystyle{mn2e}
\bibliography{references}
\bsp

\label{lastpage}
\end{document}